\documentclass[preprint,rmp,aps]{revtex4}

\usepackage{psfrag}
\usepackage{bm}% bold math

\usepackage{amsmath}
\usepackage{amssymb}
\usepackage{latexsym}

\ifx\pdfoutput\undefined
\usepackage{graphicx}
\else
\usepackage[pdftex]{graphicx}
\usepackage{epstopdf}
\fi
\newcommand{\Order}[2][\epsilon]{\ensuremath{\mathcal{O}\left(#1^{#2}\right)}}

\begin{document}

\begin{abstract}

The seemingly simple problem of determining the drag on a body moving
through a very viscous fluid has, for over 150 years, been a source of
theoretical confusion, mathematical paradoxes, and experimental
artifacts, primarily arising from the complex boundary layer structure
of the flow near the body and at infinity.  We review the extensive
experimental and theoretical literature on this problem, with special
emphasis on the logical relationship between different approaches.  The
survey begins with the developments of matched asymptotic expansions,
and concludes with a discussion of perturbative renormalization group
techniques, adapted from quantum field theory to differential
equations.  The renormalization group calculations lead to a new
prediction for the drag coefficient, one which can both reproduce and
surpass the results of matched asymptotics.

\end{abstract}

\title{Simple Viscous Flows: From Boundary Layers to the Renormalization Group}
\author{John Veysey II}
%\email{veysey@uiuc.edu}
\author{Nigel Goldenfeld}
%\email{nigel@uiuc.edu}
\affiliation{Department of Physics, University
of Illinois at Urbana-Champaign, 1110 W. Green St., Urbana, IL 61801,
USA} \maketitle

\tableofcontents

\section{INTRODUCTION TO LOW $R$ FLOW}
\label{chap:Intro Low R}

\subsection{Overview}

In 1851, shortly after writing down the Navier-Stokes
equations, Sir George Gabriel Stokes turned his attention to what
modern researchers might whimsically refer to as \lq\lq the hydrogen
atom" of fluid mechanics: the determination of the drag on a sphere or
an infinite cylinder moving at fixed speed in a highly viscous fluid
\cite{Stokes51}. Just as the quantum theory of the hydrogen atom
entailed enormous mathematical difficulties, ultimately leading to the
development of quantum field theory, the problem posed by Stokes has
turned out to be much harder than anyone could reasonably have
expected: it took over 100 years to obtain a justifiable lowest order
approximate solution, and that achievement required the invention of a
new branch of applied mathematics, \emph{matched asymptotic
expansions}.  And just as the fine structure of the hydrogen atom's
spectral lines eventually required renormalization theory to resolve
the problems of \lq\lq infinities" arising in the theory, so too,
Stokes' problem is plagued by divergences that are, to a physicist,
most naturally resolved by renormalization group theory
\cite{Feynman48, Schwinger48, Tomonaga48, Stuckelberg1953,
Gellmann1954, Wilson1971a, Wilson1971b, Wilson1983, CGO96}.

In order to appreciate the fundamental difficulty of such problems, and
to expose the similarity with familiar problems in quantum
electrodynamics, we need to explain how perturbation theory is used in
fluid dynamics.  Every flow that is governed by the Navier-Stokes
equations only (i.e. the transport of passive scalars, such as
temperature, is not considered; there are no rotating frames of
reference or other complications) is governed by a single dimensionless
parameter, known as the Reynolds number, which we designate as $R$. The
Reynolds number is a dimensionless number made up of a characteristic
length scale $L$, a characteristic velocity of the flow $U$, and the
kinematic viscosity $\nu\equiv \eta/\rho$, where $\eta$ is the
viscosity and $\rho$ is the density of the fluid.  In the problems at
hand, defined precisely below, the velocity scale is the input fluid
velocity at infinity, $u_\infty$, and the length scale is the radius
$a$ of the body immersed in the fluid.  Then the Reynolds number is
given by:
\begin{equation}
R\equiv \frac{u_\infty a}{\nu}
\end{equation}
The Reynolds number is frequently interpreted as the ratio of the
inertial to viscous terms in the Navier-Stokes equations. For very
viscous flows, $R\rightarrow 0$, and so we anticipate that a sensible
way to proceed is perturbation theory in $R$ about the problems with
infinite viscosity, i.e. $R=0$.  In this respect, the unwary reader
might regard this as an example very similar to quantum
electrodynamics, where the small parameter is the fine structure
constant.  However, as we will see in detail below, there is a
qualitative difference between a flow with $R=0$ and a flow with
$R\rightarrow 0$.  The fundamental reason is that by virtue of the
circular or spherical geometry, the ratio of inertial to viscous forces
in the Navier-Stokes equations is not a constant everywhere in space:
it varies as a function of radial distance $r$ from the body, scaling
as \Order[R r/a]{}.  Thus, when $R=0$, this term is everywhere zero; but for
any non-zero $R$, as $r/a\rightarrow\infty$ the ratio of inertial to
viscous forces becomes arbitrarily large.  Thus, inertial forces can
not legitimately be regarded as negligible with respect to viscous
forces everywhere: the basic premise of perturbation theory is not valid.

Perturbation theory has to somehow express, or manifest, this fact, and
it registers its objection by generating divergent terms in its
expansion.  These divergences are not physical, but are the
perturbation theory's way of indicating that the zeroth order
solution---the point about which perturbation theory proceeds---is not
a correct starting point.  The reader might wonder if the precise
nature of the breakdown of perturbation theory, signified by the
divergences, can be used to deduce what starting point would be a valid
one.  The answer is yes: this procedure is known as the perturbative
renormalization group (RG), and we will devote a significant fraction
of this article to expounding this strategy.  As most readers will know,
renormalization \cite{Feynman48, Schwinger48, Tomonaga48} and
renormalization group \cite{Stuckelberg1953, Gellmann1954, Wilson1971a,
Wilson1971b, Wilson1983} techniques in quantum field theories have been
stunningly successful. In the most well-controlled case, that of quantum
electrodynamics, the smallness of the fine structure constant allows
agreement of perturbative calculations with high-precision measurements
to 12 significant figures \cite{GABR06}. Do corresponding
techniques work as well in low Reynolds fluid dynamics, where one wishes to
calculate and measure the drag $C_D$ (defined precisely below)? Note
that in this case, it is the {\it functional form\/} in $R$ for the
drag that is of interest, rather than the drag at {\it one\/} particular
value of $R$, so the measure of success is rather more involved.
Nevertheless, we will see that calculations can be compared with
experiments, but there too will require careful interpretation.

Historically a different strategy was followed, leading to a set of
techniques known generically as singular perturbation theory, in
particular encompassing boundary layer theory and the method of matched
asymptotic expansions.  We will explain these techniques, developed by
mathematicians starting in the 1950's, and show their connection with
renormalization group methods.

\begin{figure}
\psfrag{Re}{$R = U_\infty a/\nu$}
\psfrag{CD}{$C_D R/4\pi$}
\psfrag{Jayaweera XXXXXX}{Jayaweera}
\psfrag{Tritton}{Tritton}
\psfrag{Kaplun}{Eqn. \ref{cylinder:KaplunCD}}
\begin{center}
\includegraphics[width=.494 \textwidth]{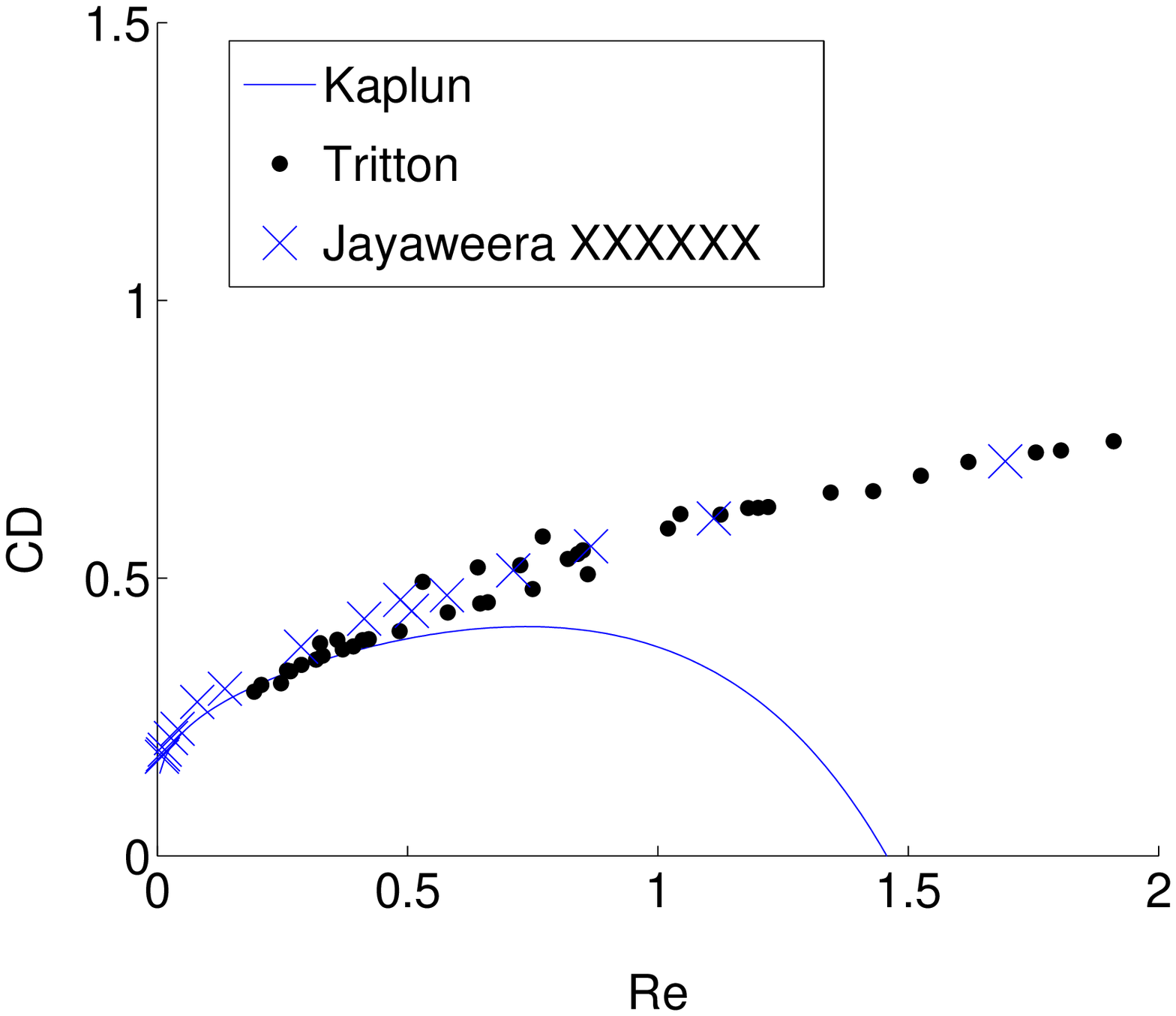}
\hfill
\psfrag{Re}{$R = U_\infty a/\nu$}
\psfrag{CD}{$C_D R/6\pi - 1$}
\psfrag{Maxworthy XXXXXX}{Maxworthy}
\psfrag{Le Clair}{Le Clair}
\psfrag{Dennis}{Dennis}
\psfrag{Chester and}{Eqn. \ref{matchedcd}}
\includegraphics[width=.494 \textwidth]{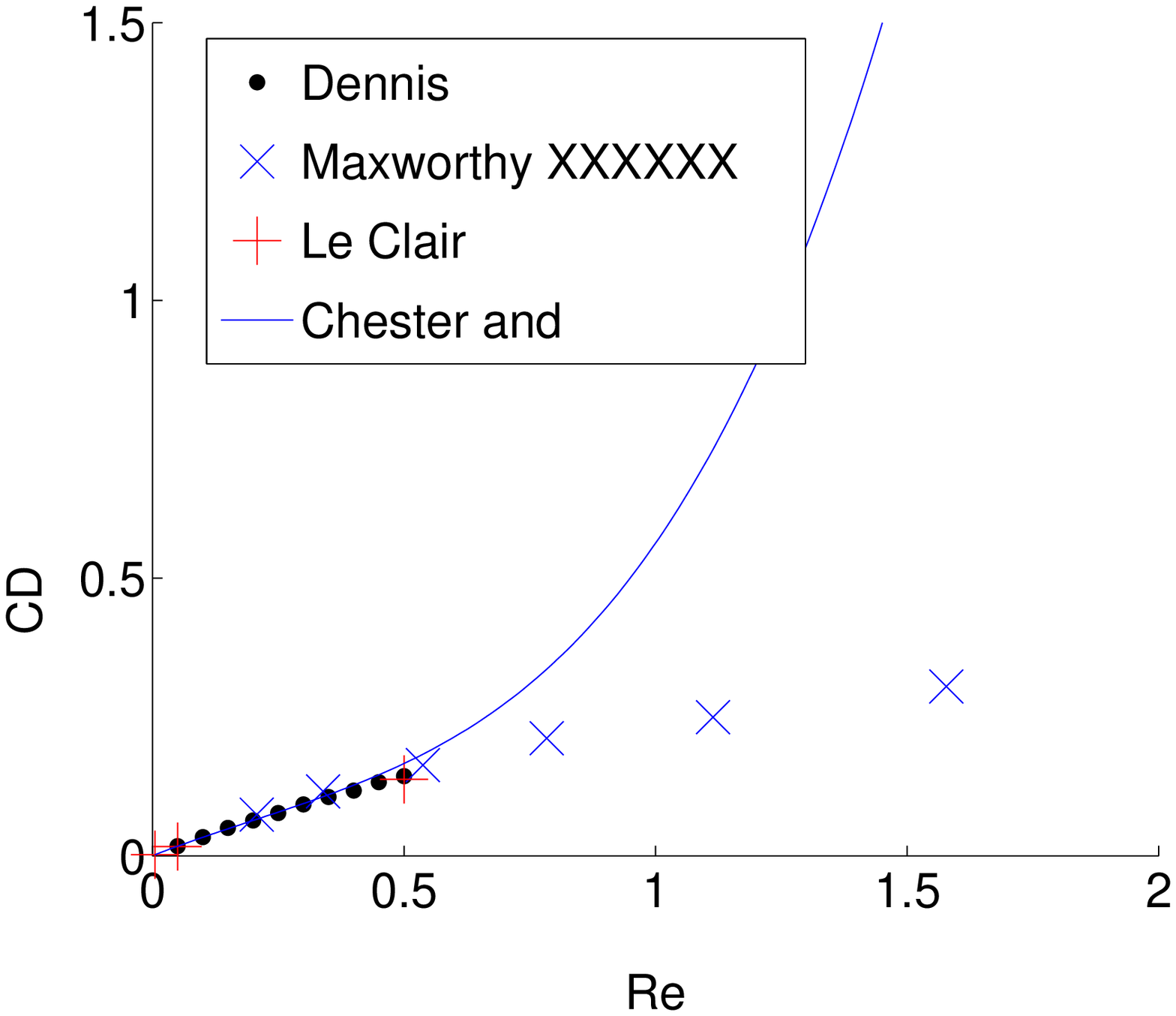}
\end{center}
\caption{(Color online) Comparing experiment with ``state of the art'' theoretical
predictions for a sphere \cite{Jay65,Tritton59} (right) and a cylinder
\cite{Den71,Maxworthy65,LC70} (left).} \label{bestinitialcomparision}
\end{figure}

Although the calculational techniques of matched asymptotic expansions are
widely regarded as representing a systematically firm footing, their best
results apply only to infinitesimally small Reynolds number. As shown
in Figure \ref{bestinitialcomparision}, the large deviations between
theory and experiment for $R \sim 0.5$ demonstrate the need for theoretical
predictions which are more robust for small but non-infinitesimal Reynolds
numbers.  Ian Proudman, who, in a {\it tour de force\/} helped obtain the
first matched asymptotics result for a sphere \cite{Proudman57}, expressed
it this way: ``It is therefore particularly disappointing that the
numerical `convergence' of the expansion is so poor.'' \cite{Chester69} In
spite of its failings, Proudman's solution from 1959 was the first
mathematically rigorous one for flow past a sphere; all preceding
theoretical efforts were worse.

Further complicating matters, the literature surrounding these
problems is rife with ``paradoxes'', revisions, ad-hoc justifications,
disagreements over attribution, mysterious factors of two, conflicting
terminology, non-standard definitions, and language barriers. Even a recent
article attempting to resolve this quagmire \cite{Lindgren99} contains an
inaccuracy regarding publication dates and scientific priority. This
tortured history has left a wake of experiments and numerical calculations
which are of widely varying quality, although they can appear to agree when
not examined closely. For example, it turns out that the finite size of
experimental systems has a dramatic effect on measurements and simulations,
a problem not appreciated by early workers.

Although in principle the matched asymptotics results can be
systematically extended by working to higher order, this is not
practical. The complexity of the governing equations prohibits further
improvement. We will show here that techniques based on the renormalization
group ameliorate some of the technical difficulties, and result in a
more accurate drag coefficient at small but non-infinitesimal Reynolds
numbers. Given the historical importance of the techniques developed to
solve these problems, we hope that our solutions will be of
general methodological interest.

We anticipate that some of our readers will be fluid dynamicists interested
in assessing the potential value of renormalization group techniques. We
hope that this community will see that our use of the renormalization group
is quite distinct from applications to stochastic problems, such as
turbulence, and can serve a different purpose. The second group of readers
may be physicists with a field theoretic background, encountering fluids
problems for the first time, perhaps in unconventional settings, such as
heavy ion collisions and QCD \cite{PhysRevLett.86.402, CMT05, Hen05, HG06,
BRW06, CKM06} or 2D electron gases \cite{stone1990sdf,
1998PhyB..256...47E}. We hope that this review will expose them to the
mathematical richness of even the simplest flow settings, and introduce a
familiar conceptual tool in a non-traditional context.

This review has two main purposes.  The first purpose of the present
article is to attempt a review and synthesis of the literature,
sufficiently detailed that the subtle differences between different
approaches are exposed, and can be evaluated by the reader.  This is
especially important, because this is one of those problems so detested
by students, in which there are a myriad of ways to achieve the right
answer for the wrong reasons.  This article highlights all of these.

A second purpose of this article is to review the use of
renormalization group techniques in the context of singular
perturbation theory, as applied to low Reynolds number flows. These
techniques generate a non-trivial estimate for the functional form of $C_D
(R)$ that can be sensibly used at moderate values of $R\sim \Order[1]{}$,
not just infinitesimal values of $R$. As $R \rightarrow 0$, these new
results reduce to those previously obtained by matched asymptotic
expansions, in particular accounting for the nature of the mathematical
singularities that must be assumed to be present for the asymptotic
matching procedure to work.

Renormalization group techniques were originally developed in the
1950's to extend and improve the perturbation theory for quantum
electrodynamics.  During the late 1960's and 1970's, renormalization
group techniques famously found application in the problem of phase
transitions \cite{Wilson1971a,Kadanoff1966,Widom1963}. During the
1990's, renormalization group techniques were developed for ordinary
and partial differential equations, at first for the analysis of
nonequilibrium (but deterministic) problems which exhibited anomalous
scaling exponents \cite{GMO+90, CGO91} and subsequently for the related
problem of travelling wave selection \cite{CGO+94, CGO94b, CG95}.  The
most recent significant development of the renormalization group---and
the one that concerns us here---was the application to singular
perturbation problems \cite{CGO94, CGO96}.  The scope of \cite{CGO96}
encompasses boundary layer theory, matched asymptotic expansions,
multiple scales analysis, WKB theory, and reductive perturbation theory
for spatially-extended dynamical systems. We do not review all these
developments here, but focus only on the issues arising in the highly
pathological singularities characteristic of low Reynolds number flows.
For a pedagogical introduction to renormalization group techniques, we
refer the reader to \cite{GoldenfeldBook}, in particular Chapter 10
which explains the connection between anomalous dimensions in field
theory and similarity solutions of partial differential equations.  We
mention also that the RG techniques discussed here have also been the
subject of rigorous analysis \cite{bricmont:rpd, bricmont1994rga,
moise1998nsd, ziane2000crg, moise2000rgm, moise2001rgm, blomker2002sns,
lan2004abc, wirosoetisno59fgw, petcu2004rgm} in other contexts of fluid
dynamics, and have also found application in cavitation
\cite{josserand1999cie} and cosmological fluid dynamics
\cite{iguchi1998rga, nambu1999rlw, nambu2000rga, belinchon2002rga,
nambu2002bra}.

This review is organized as follows.  After precisely posing the
mathematical problem, we review all prior theoretical and experimental
results. We identify the five calculations and measurements which are
accurate enough, and which extend to sufficiently small Reynolds number, to
be useful for evaluating theoretical predictions. Furthermore, we review
the history of all theoretical contributions, and clearly present the
methodologies and approximations behind previous solutions. In doing so, we
eliminate prior confusion over chronology and attribution. We conclude by
comparing the best experimental results with our new,
RG-based, theoretical prediction. This exercise makes the shortcomings that
Proudman lamented clear.

% ------------------------------------------------------------------------------
% Mathematical Problem
% ------------------------------------------------------------------------------

\subsection{Mathematical formulation}

The goal of these calculations is to determine the drag force exerted
on a sphere and on an infinite cylinder by steady, incompressible, viscous
flows. The actual physical problem concerns a body moving at constant
velocity in an infinite fluid, where the fluid is at rest in the laboratory
frame. In practice, it is more convenient to analyze the problem using an
inertial frame moving with the fixed body, an approach which is entirely
equivalent.\footnote{Nearly all workers, beginning with Stokes
\cite{Stokes51}, use this approach, which Lindgren \cite{Lindgren99} refers
to as the ``steady'' flow problem.}

\begin{figure}
\psfrag{UInfinity}{$\vec{u}_{\infty}$}
\psfrag{a}{$a$}
\psfrag{r}{$r$}
\psfrag{q}{$\theta$}
%\psfrag{x}{$x$}
%\psfrag{y}{$y$}
\begin{center}
\includegraphics*[width=.35 \textwidth]{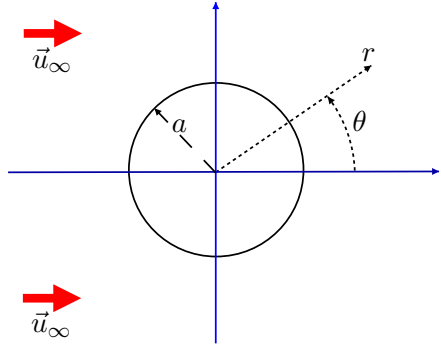}
\caption{(Color online) Schematic for flow past a sphere or cylinder.}
\label{fig:Flowschematic}
\end{center}
\end{figure}

Flow past a sphere or circle is shown schematically in Figure
\ref{fig:Flowschematic}. The body has a characteristic length scale,
which we have chosen to be the radius ($a$), and it is immersed in uniform
stream of fluid. At large distances, the undisturbed fluid moves
with velocity $\vec{u}_{\infty}$.

The quantities shown in Table \ref{tab:gov1} characterize the problem.
We assume incompressible flow, so $\rho =$ const.
\begin{table}
\begin{center}
  \begin{tabular}{|c|l|} \hline
   Quantity & Description\\
   \hline
   $\vec{r}$ & Coordinate Vector \\
   $\vec{u}(\vec{r})$ & Velocity Field \\
   $\rho$ & Fluid Density   \\
   $p(\vec{r})$ & Pressure \\
   $\nu$ & Kinematic Viscosity \\
   $ a $ & Characteristic Length of Fixed Body \\
   $\vec{u}_{\infty}$ & The Uniform Stream Velocity  \\
\hline
  \end{tabular}
  \caption{Quantities needed to characterize low $R$ flow past a rigid body.}
  \label{tab:gov1}
\end{center}
\end{table}
The continuity equation (Eqn. \ref{Continuity2}) and the time-independent
Navier-Stokes equations (Eqn. \ref{NS1}) govern steady-state, incompressible flow.
\begin{equation}
\label{Continuity2}
\nabla\cdot\vec{u} = 0
\end{equation}
\begin{equation}
\label{NS1}
(\vec{u} \cdot \nabla
\vec{u}) = - \frac{\nabla p}{\rho} + \nu \nabla^{2}\vec{u}
\end{equation}
These equations must be solved subject to two boundary conditions,
given in Eqn. \ref{boundaryconditions}. First, the \emph{no-slip}
conditions are imposed on the surface of the fixed body (Eqn.
\ref{no-slip}). Secondly, the flow must be a uniform stream far from
the body (Eqn. \ref{uniform}). To calculate the pressure, one also
needs to specify an appropriate boundary condition (Eqn.
\ref{pressurebc}), although as a matter of practice this is immaterial,
as only pressure differences matter when calculating the drag
coefficient.
\begin{subequations}
\label{boundaryconditions}
\begin{eqnarray}
\vec{u}(\vec{r}) &=& 0 \quad \vec{r}
\in \textrm{\{Surface of Fixed Body\}} \label{no-slip} \\
\lim_{|\vec{r}| \to \infty} \vec{u}(\vec{r}) &=& \vec{u}_{\infty} \label{uniform} \\
\lim_{|\vec{r}| \to \infty} p(\vec{r}) &=& p_{\infty} \label{pressurebc}
\end{eqnarray}
\end{subequations}

It is convenient to analyze the problem using non-dimensional quantities, which
are defined in Table \ref{tab:gov2}.
\begin{table}
\begin{center}
  \begin{tabular}{|c|l|} \hline
   Dimensionless Quantity & Definition\\
   \hline
$\vec{r}^{*}$ & $\vec{r}/a$ \\
$\vec{u}^{*}(\vec{r})$ &  $\vec{u}(\vec{r})/|\vec{u}_\infty|$ \\
$p^{*}(\vec{r})$ &  $a\ p(\vec{r})/\rho \ \nu \ |\vec{u}_\infty|$\\
$\vec{\nabla}^{*}$ & $a \ \vec{\nabla}$ \\
\hline
  \end{tabular}
  \caption{Dimensionless variables.}
  \label{tab:gov2}
\end{center}
\end{table}
When using dimensionless variables, the governing equations assume the
forms given in Eqns. \ref{Continuity2ND} and \ref{NonDNS}, where we
have introduced the \emph{Reynolds Number}, $ R  = |\vec{u}_{\infty}|
a/\nu$, and denoted scaled quantities by an asterisk.
\begin{equation}
\label{Continuity2ND}
\nabla^{*} \cdot \vec{u^{*}} = 0
\end{equation}
\begin{equation}
R (\vec{u}^{*} \cdot
\nabla^{*})\vec{u}^{*} = - \nabla^{*} p^{*} + \nabla^{*2} \vec{u}^{*}
\label{NonDNS}
\end{equation}

The boundary conditions also transform, and will later be given
separately for both the sphere and the cylinder (Eqns. \ref{Sphere BC},
\ref{Cylinder BC}). Henceforth, the $^{*}$ will be omitted from our
notation, except when dimensional quantities are explicitly introduced.
It is useful to eliminate pressure from Eqn. \ref{NonDNS} by taking the
curl and using the identity $\nabla \times \nabla p = 0$, leading to
\begin{equation}
\label{NS3}
 (\vec{u} \cdot \nabla)(\nabla \times \vec{u}) - ((\nabla \times \vec{u}) \cdot
\vec{u})= \frac{1}{R} \nabla^{2}(\nabla \times \vec{u})
\end{equation}

% ------------------------------------------------------------------------------
% Mathematical Problem: Cylinder
% ------------------------------------------------------------------------------

\subsubsection{Flow past a cylinder}

For the problem of the infinite cylinder, it is natural to use
cylindrical coordinates, $\vec{r}=(r, \theta, z)$.  We examine the
problem where the uniform flow is in the $\hat{x}$ direction (see
Figure \ref{fig:Flowschematic}). We will look for 2-d solutions, which
satisfy $\partial_z{\vec{u}} = 0$.

Since the problem is two dimensional, one may reduce the set of
governing equations (Eqns. \ref{Continuity2ND} and \ref{NonDNS}) to a
single equation involving a scalar quantity, the \emph{Lagrangian}
stream function, usually denoted $\psi(r,\theta)$. It is defined by
Eqn. \ref{cylinderstreamfunction}.\footnote{Although many authors
prefer to solve the vector equations, we follow Proudman and Pearson
\cite{Proudman57}.}
\begin{equation}
u_r =\frac{1}{r} \frac{\partial \psi}{\partial \theta} \qquad u_\theta =
-\frac{\partial \psi}{\partial r} \qquad u_z = 0
\label{cylinderstreamfunction}
\end{equation}

This definition guarantees that equation (\ref{Continuity2ND}) will be
satisfied \cite{Goldstein29}. Substituting the stream function into
equation (\ref{NS3}), one obtains the governing equation (Eqn.
\ref{CylinderEqn}). Here we follow the compact notation of Proudman and
Pearson \cite{Proudman57,Hinch91}.
\begin{equation}
\label{CylinderEqn}
\nabla_r^4 \psi(r,\theta) = - \frac{R}{r} \frac{\partial (\psi,
\nabla_r^2)}{\partial(r,\theta)}
\end{equation}
where
\begin{displaymath}
\nabla_r^2 \equiv \frac{\partial^2}{\partial r^2} + \frac{1}{r}
\frac{\partial}{\partial r} + \frac{1}{r^2} \frac{\partial^2}{\partial \theta^2}
\end{displaymath}

The boundary conditions which fix $\vec{u}(\vec{r})$ (Eqns.
\ref{no-slip}, \ref{uniform}) also determine $\psi(r,\theta)$ up to an
irrelevant additive constant.\footnote{The constant is irrelevant
because it vanishes when the derivatives are taken in Eqn.
\ref{cylinderstreamfunction}.} Eqn. \ref{Cylinder BC} gives the
boundary conditions expressed in terms of stream functions.
\begin{subequations}
\label{Cylinder BC}
\begin{eqnarray}
\psi(r = 1, \theta) &=& 0 \\
\frac{\partial \psi(r,\theta)}{\partial r} \bigg|_{r=1} &=& 0 \\
\lim_{r \to \infty} \frac{\psi(r,\theta)}{r} &=& \sin(\theta)
\end{eqnarray}
\end{subequations}
To calculate the drag for on a cylinder, we must first solve Equation
\ref{CylinderEqn} subject to the boundary conditions given in Eqn. \ref{Cylinder
BC}.

% ------------------------------------------------------------------------------
% Mathematical Problem: Sphere
% ------------------------------------------------------------------------------

\subsubsection{Flow past a sphere}

To study flow past a sphere, we use spherical coordinates: $\vec{r}=(r,
\theta, \phi)$.  We take the uniform flow to be in the $\hat{z}$
direction. Consequently, we are interested in solutions which are
independent of $\phi$, because there can be no circulation about the
$\hat{z}$ axis.

Since the problem has axial symmetry, one can use the \emph{Stokes'
stream function} (or Stokes' current function) to reduce Eqns.
\ref{Continuity2ND} and \ref{NonDNS} to a single equation. This stream
function is defined through the following relations:
\begin{equation}
v_r = \frac{1}{r^2 \sin{\theta}}\psi_{\theta} \qquad
v_{\theta} = -\frac{1}{r \sin{\theta}} \psi_{r} \qquad
v_{\phi} = 0
\label{spherestreamfunction}
\end{equation}
These definitions guarantee that Eqn. \ref{Continuity2ND} will be
satisfied. Substituting Eqn. \ref{spherestreamfunction} into
Eqn. \ref{NS3}, one obtains the governing equation for $\psi(r,\theta)$ \cite{Proudman57}:
\begin{equation}
\label{SphereEqn}
D^4 \psi = R \left( \frac{1}{r^2} \frac{\partial(\psi, D^2 \psi)}{\partial(r,\mu)} +
\frac{2}{r^2} D^2 \psi L \psi \right)
\end{equation}
In this equation,
\begin{subequations}
\nonumber
\begin{eqnarray}
\nonumber \mu &\equiv& \cos{\theta} \\
\nonumber D^2 &\equiv& \frac{\partial^2}{\partial r^2} + \frac{1-\mu^2}{r^2}
\frac{\partial^2}{\partial \mu^2} \\
\nonumber L &\equiv& \frac{\mu}{1-\mu^2}\frac{\partial}{\partial r} + \frac{1}{r} \frac{\partial}{\partial \mu}
\end{eqnarray}
\end{subequations}
Here we follow the notation of Proudman and Pearson \cite{Proudman57}.
Other authors, such as Van Dyke \cite{VanDyke1975} and Hinch
\cite{Hinch91}, write their stream function equations in an equivalent,
albeit less compact, notation.

As in the case of the cylinder, the boundary conditions which fix
$\vec{u}(\vec{r})$ (Eqns. \ref{no-slip}, \ref{uniform}) determine
$\psi$ up to an irrelevant additive constant. The transformed boundary
conditions are given by Eqn. \ref{Sphere BC}.
\begin{subequations}
\label{Sphere BC}
\begin{eqnarray}
\psi(r = 1, \mu) &=& 0 \\
\frac{\partial \psi(r,\mu)}{\partial r} \bigg|_{r=1} &=& 0 \\
\lim_{r \to \infty} \frac{\psi(r,\mu)}{r^2} &=& \frac{1}{2} \left( 1 -
  \mu^2 \right)
\end{eqnarray}
\end{subequations}

In this paper, we obtain approximate solutions for
Eqn. \ref{CylinderEqn} (subject to Eqn. \ref{Cylinder BC}), and
Eqn. \ref{SphereEqn} (subject to Eqn. \ref{Sphere BC}). These solutions
are then used to calculate drag coefficients, which we compare to
experimental results.

% ------------------------------------------------------------------------------
% Mathematical Problem: Drag Coefficient
% ------------------------------------------------------------------------------

\subsubsection{Calculating the drag coefficient}

Once the Navier-Stokes equations have been solved, and the stream
function is known, calculating the drag coefficient, $C_D$, is a
mechanical procedure. We follow the methodology described by Chester
and Breach \cite{Chester69}. This analysis is consistent with the work
done by Kaplun \cite{Kaplun57c} and Proudman \cite{Proudman57},
although these authors do not detail their calculations.

This methodology is significantly different from that employed by other
workers, such as Tomotika \cite{Tomotika50, Oseen10}. Tomotika
calculates $C_D$ approximately, based on a linearized calculation of
pressure. Although these approximations are consistent with the
approximations inherent in their solution of the Navier-Stokes
equations, they are inadequate for the purposes of obtaining a
systematic approximation to any desired order of accuracy.

Calculating the drag on the body begins by determining the force
exerted on the body by the moving fluid. Using dimensional variables,
the force per unit area is given by \cite{LandauLifschitz}:
\begin{equation}
P_i = -\sigma_{ik} n_k
\label{forceperunitarea}
\end{equation}
Here $\sigma_{ik}$ is the stress tensor, and $\vec{n}$ is a unit vector
normal to the surface. For an incompressible fluid, the stress tensor
takes the form \cite{LandauLifschitz}:
\begin{equation}
\sigma_{ik} = -p\delta_{ik} + \eta\left(\frac{\partial
v_{i}}{\partial{x_k}}+\frac{\partial v_k}{\partial
x_i}\right)
\label{Stress2}
\end{equation}
$\eta$ is the \emph{dynamic viscosity}, related to the kinematic
viscosity by $\eta = \nu \rho$. The total force is found by integrating
Eqn. \ref{forceperunitarea} over the surface of the solid body. We now
use these relations to derive explicit formula, expressed in terms of
stream functions, for both the sphere and the cylinder.

{\it a. Cylinder} In the case of the cylinder, the components of the
velocity field are given through the definition of the Lagrangian
stream function (Eqn. \ref{cylinderstreamfunction}). Symmetry requires
that the net force on the cylinder must be in the same direction as the
uniform stream. Because the uniform stream is in the $\hat{x}$
direction, if follows from Eqns. \ref{forceperunitarea} and
\ref{Stress2} that the force\footnote{The form of $\sigma_{ik}$ in
cylindrical coordinates is given is Landau \cite{LandauLifschitz}.} on
the cylinder per unit length is given by:
\begin{eqnarray}
\label{cylinderforceeqn}
F_{\hat{x}} &=& \oint (\sigma_{rr} \cos{\theta} - \sigma_{r\theta}
\sin{\theta}) \text{d}s \\
\nonumber &=& \left[ \int_0^{2\pi} \left( \sigma_{rr} \cos{\theta} - \sigma_{r\theta}
\sin{\theta} \right) r\ \text{d}\theta \right]_{r=a} \\
\nonumber &=& \bigg[ \int_0^{2\pi} \bigg( \left(-p+2\eta \frac{\partial
        v_r}{\partial r} \right) \cos{\theta} - \eta \left(\frac{1}{r}
      \frac{\partial v_r}{\partial \theta} + \frac{\partial
        v_\theta}{\partial r} - \frac{v_\theta}{r} \right) \sin{\theta} \bigg) r\ \text{d}\theta\bigg]_{r=a}
\end{eqnarray}

The drag coefficient for an infinite cylinder is \emph{defined} as $C_D
= F_{\text{Net}}/\rho |\vec{u}_\infty|^2 a$. Note that authors (e.g.,
\cite{Kaplun67, Tritton59}) who define the Reynolds number based on
diameter nonetheless use the same definition of $C_D$, which is based
on the radius.  For this problem, $F_{\text{Net}} = F_{\hat{x}}$, as
given by  Eqn. \ref{cylinderforceeqn}. Introducing the dimensionless
variables defined in Table \ref{tab:gov2} into Eqn.
\ref{cylinderforceeqn}, we obtain Eqn. \ref{finalcylforce}. Combining
this with the definition of $C_D$, we obtain Eqn. \ref{CDcylinder}.
\begin{equation}
\label{finalcylforce}
F_{\hat{x}} = \frac{\rho  |\vec{u}_\infty|^2 a}{R} \bigg[ \int_0^{2\pi} \bigg( \left(-p(r,\theta)+2
      \frac{\partial u_r}{\partial r} \right) \cos{\theta} - \left(\frac{1}{r}
      \frac{\partial u_r}{\partial \theta} + \frac{\partial
        u_\theta}{\partial r} - \frac{u_\theta}{r} \right) \sin{\theta} \bigg) r\ \text{d}\theta\bigg]_{r=1}
\end{equation}
\begin{equation}
\label{CDcylinder}
C_D = \frac{1}{R}  \bigg[ \int_0^{2\pi} \bigg( \left(-p(r,\theta)+2
      \frac{\partial u_r}{\partial r} \right) \cos{\theta} - \left(\frac{1}{r}
      \frac{\partial u_r}{\partial \theta} + \frac{\partial
        u_\theta}{\partial r} - \frac{u_\theta}{r} \right) \sin{\theta}
      \bigg) r\ \text{d}\theta \bigg]_{r=1}
\end{equation}

To evaluate this expression, we must first derive $p(r,\theta)$ from
the stream function. The pressure can be determined to within an
irrelevant additive constant by integrating the $\hat{\theta}$
component of the Navier-Stokes equations (Eqn. \ref{NonDNS})
\cite{Chester69, LandauLifschitz}. The constant is irrelevant because,
in Eqn. \ref{CDcylinder}, $\int_0^{2\pi} C \cos{\theta} \text{d}
\theta=0$. Note that all gradient terms involving $z$ vanish by
construction.
\begin{equation}
\label{cylpressure}
p(r,\theta) = r \ \int \left[ -R \left( \left(\vec{u} \cdot \nabla \right)
  u_\theta + \frac{u_r u_\theta}{r} \right)  + \nabla^2 u_\theta + \frac{2}{r^2}
\frac{\partial u_r}{\partial \theta} - \frac{u_\theta}{r^2} \right] \text{d} \theta
\end{equation}

Given a solution for the stream function $\psi$, the set of
dimensionless Eqns. \ref{cylinderstreamfunction}, \ref{CDcylinder}, and
\ref{cylpressure} uniquely determine $C_D$ for a cylinder. However, because the
velocity field satisfies no-slip boundary conditions, these general
formula often simplify considerably.

For instance, consider the class of stream functions which meets the
boundary conditions (Eqn. \ref{Cylinder BC}) and can be expressed as a
Fourier sine series: $\psi(r,\theta) = \sum_{n=1}^\infty f_n(r)
\sin{n\theta}$. Using the boundary conditions it can be shown that, for
these stream functions, Eqn. \ref{CDcylinder} reduces to the simple
expression given by Eqn. \ref{cylinder:convenientdrag}.
\begin{equation}
\label{cylinder:convenientdrag}
C_D = -\frac{\pi}{R} \left(\frac{\textrm{d}^3}{\textrm{d}r^3} f_1(r)\right)_{r=1}
\end{equation}

{\it b. Sphere}

 The procedure for calculating $C_D$ in the case of the sphere is nearly
 identical to that for the cylinder. The components of the velocity field are given through the definition of
the Stokes' stream function (Eqn. \ref{spherestreamfunction}). As
before, symmetry requires that any net force on the cylinder must be in
the direction of the uniform stream, in this case the $\hat{z}$
direction.

From Eqn. \ref{forceperunitarea}, the net force on the sphere is given by Eqn. \ref{sphereforceeqn}.
\begin{eqnarray}
\label{sphereforceeqn}
F_{\hat{z}} &=& \oint (\sigma_{rr} \cos{\theta} - \sigma_{r\theta}
\sin{\theta}) \text{d}s \\
\nonumber &=& 2 \pi \left[ \int_0^{\pi} \left( \sigma_{rr} \cos{\theta} - \sigma_{r\theta}
\sin{\theta} \right) r^2 \sin{\theta} \ \text{d}\theta \right]_{r=a}
\end{eqnarray}

For the sphere, the drag coefficient is \emph{defined} as $C_D \equiv
F_{\text{Net}}/\rho |\vec{u}_\infty|^2 a^2$. Often the drag coefficient
is given in terms of the \emph{Stokes' Drag}, $D_S \equiv 6 \pi \rho
|\vec{u}_\infty| a \nu =  6 \pi \rho |\vec{u}_\infty|^2 a^2/R$. In
these terms, $C_D = F_{\text{Net}} 6\pi/D_S R$. If $F_\text{Net} =
D_S$, $C_D = 6 \pi/R$, the famous result of Stokes \cite{Stokes51}.

Not all authors follow Stokes' original definition of $C_D$. For
instance, S. Goldstein \cite{Goldstein29,Goldstein38} and H. Liebster
\cite{Liebster26,Liebster24} define $C_D$ using a factor based on
cross-sectional areas: $C_D^{\textrm{Goldstein}} = C_D 2/\pi$. These
authors also define $R$ using the diameter of the sphere rather than
the radius. S. Dennis, defines $C_D$ similarly to Goldstein, but
without the factor of two: $C_D^{\textrm{Dennis}} = C_D/\pi$
\cite{Den71}.

Using the form of Eqn. \ref{Stress2} given in Landau
\cite{LandauLifschitz} and introducing the dimensionless variables
defined in Table \ref{tab:gov2} into Eqn. \ref{sphereforceeqn}, we
obtain Eqn. \ref{finalsphforce}. Combining this with the definition of
$C_D$, we obtain Eqn. \ref{CDsphere}.
\begin{equation}
\label{finalsphforce}
F_{\hat{z}} = \frac{D_s}{3} \bigg[ \int_0^{\pi} \bigg( \left(-p(r,\theta)+2
      \frac{\partial u_r}{\partial r} \right) \cos{\theta} - \left(\frac{1}{r}
      \frac{\partial u_r}{\partial \theta} + \frac{\partial
        u_\theta}{\partial r} - \frac{u_\theta}{r} \right) \sin{\theta}
      \bigg) r^2 \sin{\theta} \ \text{d}\theta \bigg]_{r=1}
\end{equation}
\begin{equation}
\label{CDsphere}
C_D = \frac{2 \pi}{R} \bigg[ \int_0^{\pi} \bigg( \left(-p(r,\theta)+2
      \frac{\partial u_r}{\partial r} \right) \cos{\theta} - \left(\frac{1}{r}
      \frac{\partial u_r}{\partial \theta} + \frac{\partial
        u_\theta}{\partial r} - \frac{u_\theta}{r} \right) \sin{\theta}
      \bigg) r^2 \sin{\theta} \ \text{d}\theta \bigg]_{r=1}
\end{equation}

As with the cylinder, the pressure can be determined to within an
irrelevant additive constant by integrating the $\hat{\theta}$
component of the Navier-Stokes equations (Eqn. \ref{NonDNS})
\cite{Chester69, LandauLifschitz}. Note that gradient terms involving
$\phi$ must vanish.
\begin{equation}
\label{sphpressure}
p(r,\theta) = r \ \int \left[ -R \left( \left(\vec{u} \cdot \nabla \right)
u_\theta + \frac{u_r u_\theta}{r} \right)  + \nabla^2 u_\theta + \frac{2}{r^2}
\frac{\partial u_r}{\partial \theta} - \frac{u_\theta}{r^2 \sin{\theta}^2} \right] \text{d} \theta
\end{equation}
Given a solution for the stream function $\psi$, the set of
dimensionless Eqns. \ref{spherestreamfunction}, \ref{CDsphere}, and
\ref{sphpressure} uniquely determine $C_D$ for a sphere.

As with the cylinder, the imposition of no-slip boundary conditions
considerably simplifies these general formula. In particular, consider
stream functions of the form $\psi(r,\theta) = \sum_{n=1}^\infty
f_n(r)Q_n(\cos{\theta})$, where $Q_n(x)$ is defined as in Eqn.
\ref{Oseen:Goldstein}. If these stream functions satisfy the boundary
conditions, the drag is given by Eqn. \ref{sphere:simpledrag}:
\begin{equation}
\label{sphere:simpledrag}
C_D = \frac{2\pi}{3 R} \left( -2 f_1^{''}(r) + f_1^{'''}(r) \right)_{r=1}
\end{equation}

{\it c. A subtle point}

When applicable, Eqns. \ref{cylinder:convenientdrag} and
\ref{sphere:simpledrag} are the most convenient way to calculate the
drag given a stream function. They simply require differentiation of a
single angular term's radial coefficient. However, they only apply to
functions that can be expressed as a series of harmonic functions.
Moreover, for these simple formula to apply, the series expansions
\emph{must} meet the boundary conditions exactly. This requirement
implies that \emph{each} of the functions $f_i(r)$ independently meets
the boundary conditions.

The goal of our work is to derive and understand approximate solutions
to the Navier-Stokes' equations. These approximate solutions generally
will not satisfy the boundary conditions exactly. What --- if any ---
applicability do Eqns. \ref{cylinder:convenientdrag} and
\ref{sphere:simpledrag} have if the stream function does not exactly
meet the boundary conditions?

In some rare cases, the stream function of interest can be expressed in
a convenient closed form. In these cases, it is natural to calculate
the drag coefficient using the full set of equations. However we will
see that the solution to these problems is generally only expressible
as a series in harmonic functions. In these cases, it actually
preferable  to use the simplified equations
\ref{cylinder:convenientdrag} and \ref{sphere:simpledrag}.

First, these equations reflect the essential symmetry of the problem,
the symmetry imposed by the uniform flow. Eqns.
\ref{cylinder:convenientdrag} and \ref{sphere:simpledrag} explicitly
demonstrate that, given an exact solution, only the lowest harmonic
will matter: Only terms which have the same angular dependence as the
uniform stream will contribute to the drag. By utilizing the simplified
formula for $C_D$ as opposed to the general procedure, we effectively
discard contributions from higher harmonics. This is exactly what we
want, since these contributions are artifacts of our approximations,
and would not be present in an exact solution.

The contributions from inaccuracies in how the lowest harmonic meets
the boundary conditions are more subtle. As long as the boundary
conditions are satisfied to the accuracy of the overall approximation,
it does not matter whether one uses the full-blown or simplified drag
formula. The drag coefficients will agree to within the accuracy of the
original approximation.

In general, we will use the simplified formula. This is
the approach taken explicitly by many matched asymptotics workers
 \cite{Chester69,Ski75}, and implicitly by other workers \cite{Proudman57,VanDyke1975}.
It should be noted that these workers only use the portion\footnote{To be precise, they use only the Stokes' expansion,  rather than a uniform expansion.} of their solutions which can exactly meet the assumptions of the simplified drag formula. However, as we will subsequently discuss, this is an oversimplification.

% ------------------------------------------------------------------------------
% Brief History
% ------------------------------------------------------------------------------

\section{HISTORY OF LOW $R$ FLOW STUDIES}
\label{chap:history}

\subsection{Experiments and numerical calculations}

Theoretical attempts to determine the drag by solving the
Navier-Stokes' equations have been paralleled by an equally intricate
set of experiments. In the case of the sphere, experiments usually
measured the terminal velocity of small falling spheres in a
homogeneous fluid. In the case of the cylinder, workers measured the
force exerted on thin wires or fibers immersed in a uniformly flowing
viscous fluid.

These experiments, while simple in concept, were difficult
undertakings. The regime of interest necessitates some combination of
small objects, slow motion, and viscous fluid. Precise measurements are
not easy, and neither is insuring that the experiment actually examines
the same quantities that the theory predicts. All theoretical drag
coefficients concern objects in an infinite fluid, which asymptotically
tends to a uniform stream. Any real drag coefficient measurements must
take care to avoid affects due to the finite size of the experiment.
Due to the wide variety of reported results in the literature, we found
it necessary to make a complete survey, as presented in this section.

\subsubsection{Measuring the drag on a sphere}

As mentioned, experiments measuring the drag on a sphere at low
Reynolds number were intertwined with theoretical developments. Early
experiments, which essentially confirmed Stokes' law as a reasonable
approximation, include those of Allen \cite{Allen00}, Arnold
\cite{Arnold11}, Williams \cite{Williams15}, and Wieselsberger
\cite{Wie21}.

\begin{figure}[tb]
\begin{center}
\includegraphics[width=.8 \textwidth]{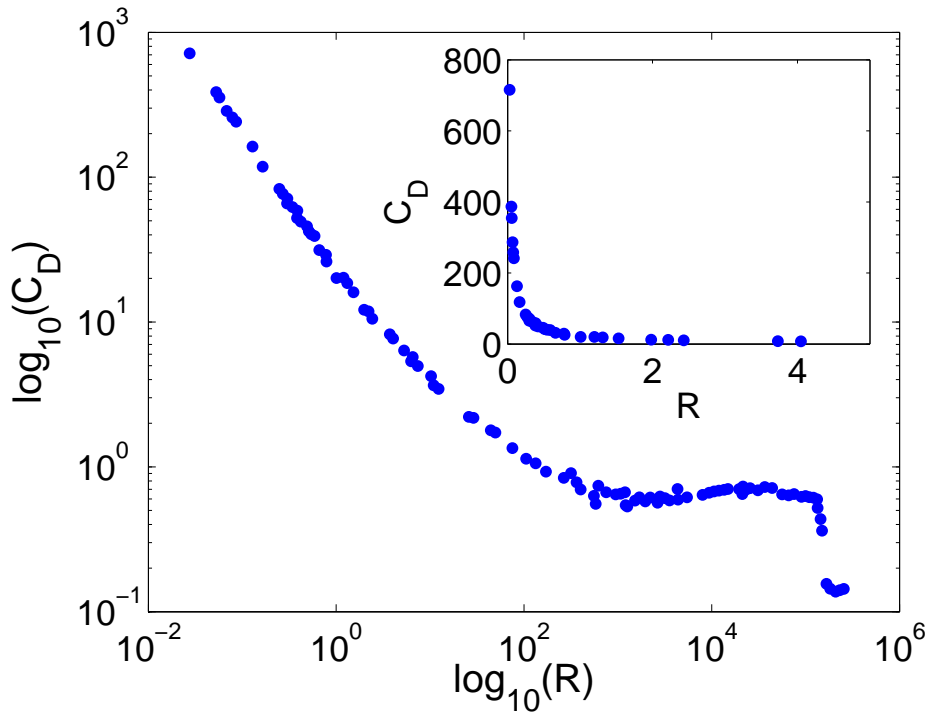}
\caption{(Color online) Early measurements of the drag on a sphere \cite{Goldstein38}.}
\label{Goldstein Sphere}
\end{center}
\end{figure}

The next round of experiments were done in the 1920s, motivated by the
theoretical advances begun by C. W. Oseen \cite{Oseen10}. These
experimentalists included Schmeidel \cite{schmiedel28} and Liebster
\cite{Liebster26,Liebster24}. The results of Allen, Liebster, and
Arnold were analyzed, collated, and averaged by Castleman
\cite{Castleman25}, whose paper is often cited as a summary of prior
experiments. The state of affairs after this work is well summarized in
plots given by Goldstein (p. 16) \cite{Goldstein38}, and Perry
\cite{Perry50}. Figure \ref{Goldstein Sphere} shows Goldstein's plot,
digitized and re-expressed in terms of the conventional definitions of
$C_D$ and $R$.

Figure \ref{Goldstein Sphere} shows the experimental data at this
point, prior to the next theoretical development, matched asymptotics.
Although the experimental data seem to paint a consistent portrait of
the function $C_D(R)$, in reality they are not good enough to
discriminate between different theoretical predictions.

Finite geometries cause the most significant experimental errors for
these measurements \cite{Tri88,Maxworthy65,Lindgren99}. Tritton notes
that ``the container diameter must be more than one hundred times the
sphere diameter for the error to be less than 2 percent'', and Lindgren
estimates that a ratio of 50 between the container and sphere diameters
will result in a 4\% change in drag force.

In 1961, Fidleris et al. experimentally studied the effects of finite
container size on drag coefficient measurements \cite{Fid61}. They
concluded that there were significant finite size effects in previous
experiments, but also proposed corrections to compensate for earlier
experimental limitations. Lindgren also conducted some related
experiments \cite{Lindgren99}.

T. Maxworthy also realized this problem, and undertook experiments
which could be used to evaluate the more precise predictions of matched
asymptotics theories. In his own words,
\begin{quote}
From the data plotted in Goldstein or Perry, it would appear that the
presently available data is sufficient to accurately answer any
reasonable question. However, when the data is plotted `correctly';
that is, the drag is non-dimensionalized with respect to the Stokes
drag, startling inaccuracies appear. It is in fact impossible to be
sure of the drag to better than $\pm 20\%$ ... The difficulties faced
by previous investigators seemed to be mainly due to an inability to
accurately compensate for wall effects \cite{Maxworthy65}.
\end{quote}

\begin{figure}[tb]
\psfrag{Re}{\mbox{\Large $R$}}
\psfrag{CD}{\mbox{\Large $C_D R/6\pi - 1$}}
\psfrag{Maxworthy}{Maxworthy}
\psfrag{Previous Expts.}{Previous Expts.}
\begin{center}
\includegraphics[width=.8 \textwidth]{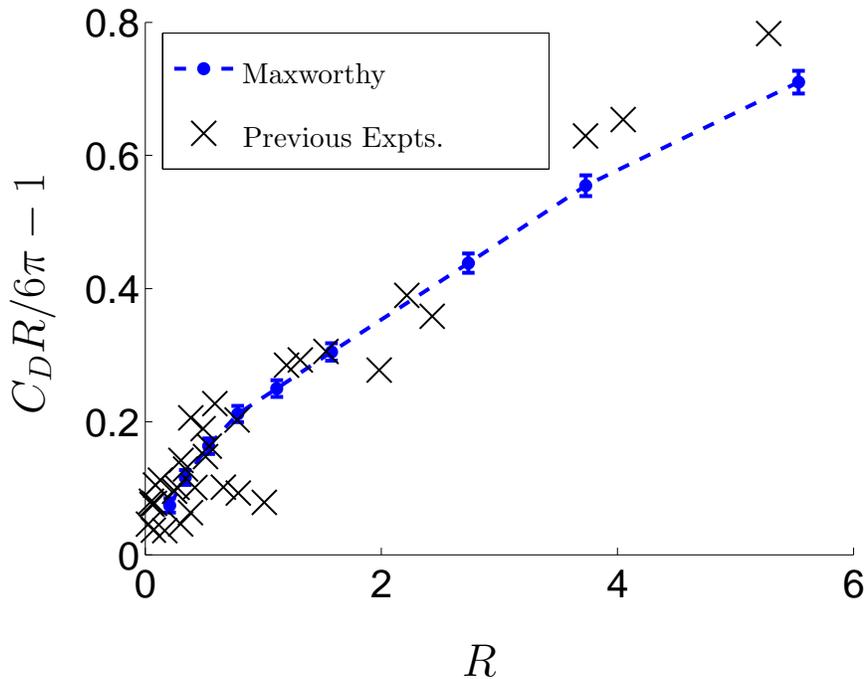}
\caption{(Color online) Maxworthy's accurate measurements of the drag on a sphere
\cite{Maxworthy65} contrasted with previous experiments
\cite{Goldstein38}.} \label{Maxworthy Sphere}
\end{center}
\end{figure}

Maxworthy refined the falling sphere technique to produce the best
experimental measurements yet --- $2\%$ error. He also proposed a new
way of plotting the data, which removes the $R^{-1}$ divergence in Eqn.
\ref{CDsphere} (as $R\rightarrow 0$). His approach makes clear the
failings of earlier measurements, as can be seen in Figure
\ref{Maxworthy Sphere}, where the drag measurements are normalized by
the Stokes drag, $C_D^{\textrm{Stokes}} = 6 \pi/R$.

In Maxworthy's apparatus, the container diameter is over 700 times the
sphere diameter, and does not contribute significantly to experimental
error, which he estimates at better than 2 percent. Note that the data
in Figure \ref{Maxworthy Sphere} are digitized from his paper, as raw
data are not available.

This problem also attracted the attention of atmospheric scientists,
who realized its significance in cloud physics, where ``cloud drops may
well be approximated by rigid spheres.''\cite{Pru70} In a series of
papers (e.g.,  \cite{Pru70,Pru68,Bea69,LC70}), H.R. Pruppacher and
others undertook numerical and experimental studies of the drag on the
sphere. They were motivated by many of the same reasons as Maxworthy,
because his experiments covered only Reynolds numbers between 0.4 and
11, and because ``Maxworthy's experimental setup and procedure left
considerable room for improvement'' \cite{Pru68}.

Their results included over 220 measurements, which they binned and
averaged. They presented their results in the form of a set of linear
fits. Adopting Maxworthy's normalization, we collate and summarize
their findings in Eqn. \ref{Pruppacher1}.
\begin{equation}
C_D \frac{R}{6\pi} - 1  = \left\{ \begin{array}{ll} 0.102\left(2 R\right)^{0.955} & 0.005 < R \leq 1.0 \\ 0.115 \left(2 R\right)^{0.802} & 1.0 < R \leq 20 \\ 0.189\left(2 R\right)^{0.632} & 20 < R \leq 200 \end{array} \right.
\label{Pruppacher1}
\end{equation}

Unfortunately, one of their later papers includes the following
footnote (in our notation): ``At $R < 1$ the most recent values of $C_D
R/6\pi - 1$ (Pruppacher, 1969, unpublished) tended to be somewhat
higher than those of Pruppacher and Steinberger.'' \cite{LC70} Their
subsequent papers plot these unpublished data as ``experimental
scatter.'' As the unpublished data are in much better agreement with
both Maxworthy's measurements and their own numerical analysis
\cite{LC70}, it makes us question the accuracy of the results given in
Eqn. \ref{Pruppacher1}.

There are many other numerical calculations of the drag coefficient for a
sphere, including: Dennis \cite{Den71}, Le Clair  \cite{LC70,Pru70},
Hamielec  \cite{Ham67}, Rimon  \cite{Rim69}, Jenson  \cite{Jen59}, and
Kawaguti  \cite{Kaw50}. Most of these results are not useful either
because of large errors (e.g., Jenson), or because they study ranges of
Reynolds number which do not include $R < 1$.  Many numerical studies
examine only a few (or even just a single) Reynolds numbers. For the
purposes of comparing theoretical predictions of $C_D$ at low Reynolds
number, only Dennis  \cite{Den71} and Le Clair  \cite{LC70} have useful
calculations. Both of these papers report tabulated results which are
in very good agreement with both each other and Maxworthy; at $R=0.5$,
the three sets of results agree to within 1\% in $C_D$, and to within
10\% in the transformed variable, $C_D R/6\pi -1$. The agreement is
even better for $R < 0.5$.

\begin{figure}[tb]
\psfrag{Re}{\mbox{\Large $R$}}
\psfrag{CD}{\mbox{\Large $C_D R/6\pi - 1$}}
\psfrag{CD2 XXXXXXX}{\tiny{$C_D R/6\pi - 1$}}
\psfrag{Maxworthy XXXXXX}{Maxworthy}
\psfrag{Pruppacher}{Eqn. \ref{Pruppacher1}}
\psfrag{Le Clair}{Le Clair}
\psfrag{Dennis}{Dennis}
\begin{center}
\includegraphics[width=.8 \textwidth]{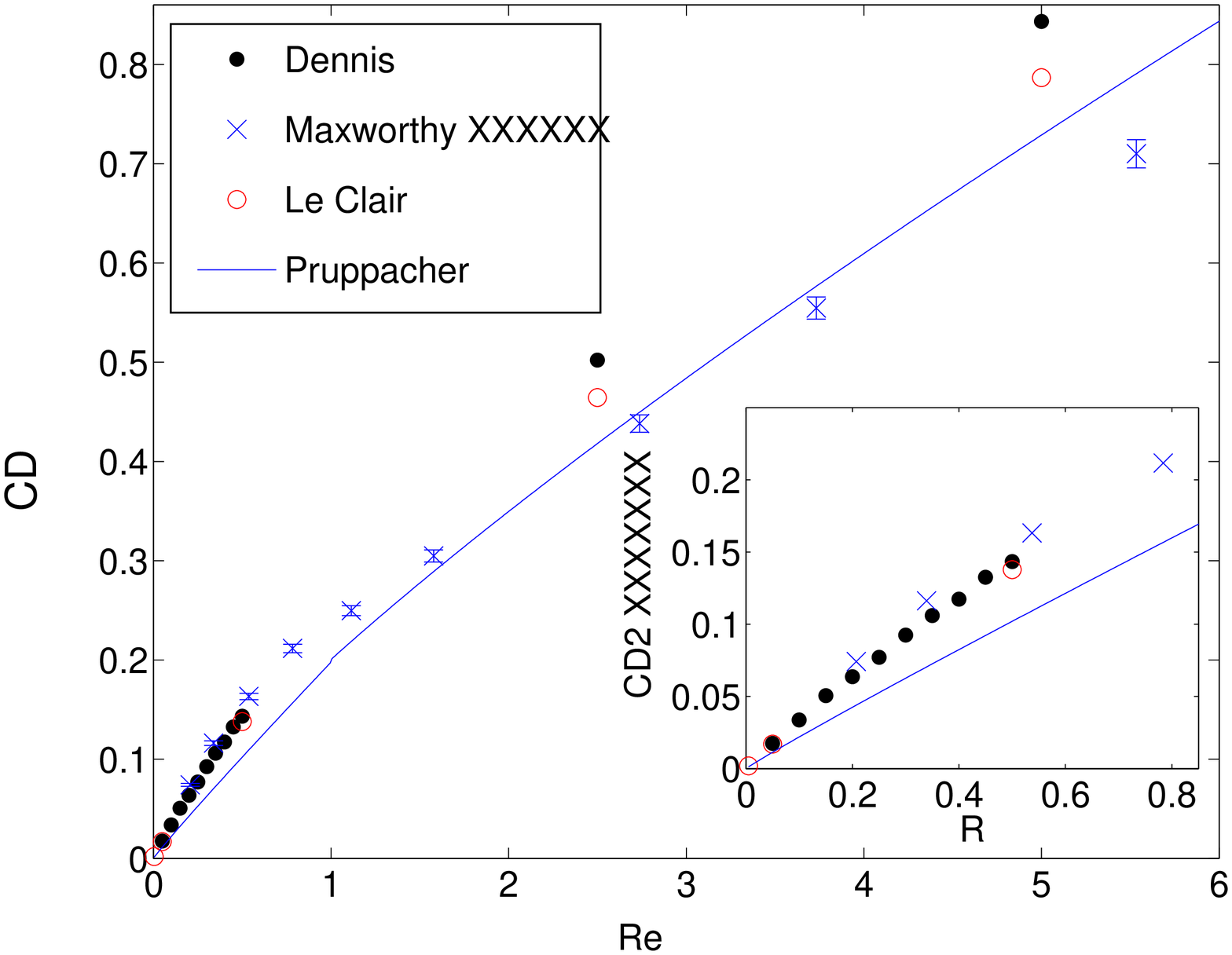}
\caption{(Color online) A summary of experimental and numerical studies of $C_D$ for a sphere \cite{Maxworthy65,LC70,Den71}.}
\label{TotSphere1}
\end{center}
\end{figure}

Figure \ref{TotSphere1} shows all relevant experimental and numerical
results for the drag on a sphere. Note the clear disagreement between
Pruppacher's results (Eqn. \ref{Pruppacher1}), and all of the other
results for $R < 1$ --- including Le Clair and Pruppacher's numerical
results \cite{LC70}.  This can be clearly seen in the inset graph.
Although Pruppacher's experiment results do agree very well with other
data for larger values of $R$ ($R \gtrsim 20$), we will disregard them
for the purposes of evaluating theoretical predictions at low Reynolds
number.

It should also be noted that there is a community of researchers
interested in sedimentation and settling velocities who have studied
the drag on a sphere. In a contribution to this literature, Brown
reviews all of the authors discussed here, as he tabulates
$C_D$ for $R < 5000$ \cite{Bro03}. His report addresses a larger range
of Reynolds numbers and he summarizes a number of experiments not
treated here. His methodology is to apply the Fidleris' correction
\cite{Fid61} to previous experiments where tabulated experimental data
was published.\footnote{Brown incorrectly reports Dennis' work
\cite{Den71} as experimental.} While this yields a reasonably
well-behaved drag coefficient for a wide range of Reynolds numbers, it
is not particularly useful for our purposes, as less accurate work
obfuscates the results of the most precise experiments near $R=0$. It
also does not include numerical work or important results which are
only available graphically (e.g., Maxworthy \cite{Maxworthy65}).

\subsubsection{Measuring the drag on a cylinder}

Experiments designed to measure the drag on an infinite cylinder in a
uniform fluid came later than those for spheres. In addition to being a
more difficult experiment --- theoretical calculations assume the
cylinder is infinite --- there were no theoretical predictions to test
before Lamb's result in 1911 \cite{Lam11}.

In 1914, E. F. Relf conducted the first experiments \cite{Rel14}. These
looked at the force exerted on long wires in a fluid. Relf measured the
drag down to a Reynolds number of about ten. In 1921, Wieselberger
measured the drag at still lower Reynods number, reaching $R=2.11$ by
looking at the deflection of a weight suspended on a wire in an air
stream \cite{Wie21b}.

These experiments, combined with others \cite{Lin31,Goldstein38} at
higher Reynolds number, characterize the drag over a range of Reynolds
numbers (see Goldstein, pg. 15). However, they do not probe truly small
Reynolds numbers ($R\ll 1$), and are of little use for evaluating
theories which are only valid in that range. Curiously, there are no
shortage of claims otherwise, such as Lamb, who says ``The formula is
stated to be in good agreement with experiment for sufficiently small
values of $U_\infty a/\nu$; see Wieselsberger'' \cite{Lamb1932}.

In 1933, Thom measured the ``pressure drag'', extending observations
down to $R=1.75$. Thom also notes that this Reynolds number is still
too high to compare with calculations: ``Actually, Lamb's solution only
applies to values of $R$ less than those shown, in fact to values much
less than unity, but evidently in most cases the experimental results
are converging with them.'' \cite{Tho33}

\begin{figure}[tb]
\psfrag{Re}{\mbox{\Large $R$}}
\psfrag{CD}{\mbox{\Large $C_D R/4\pi$}}
\psfrag{CD2 XXXXXXX}{\tiny{$C_D R/4\pi$}}
\psfrag{Jayaweera XXXXX}{Jayaweera}
\psfrag{Finn}{Finn}
\psfrag{Tritton}{Tritton }
\begin{center}
\includegraphics[width=.8 \textwidth]{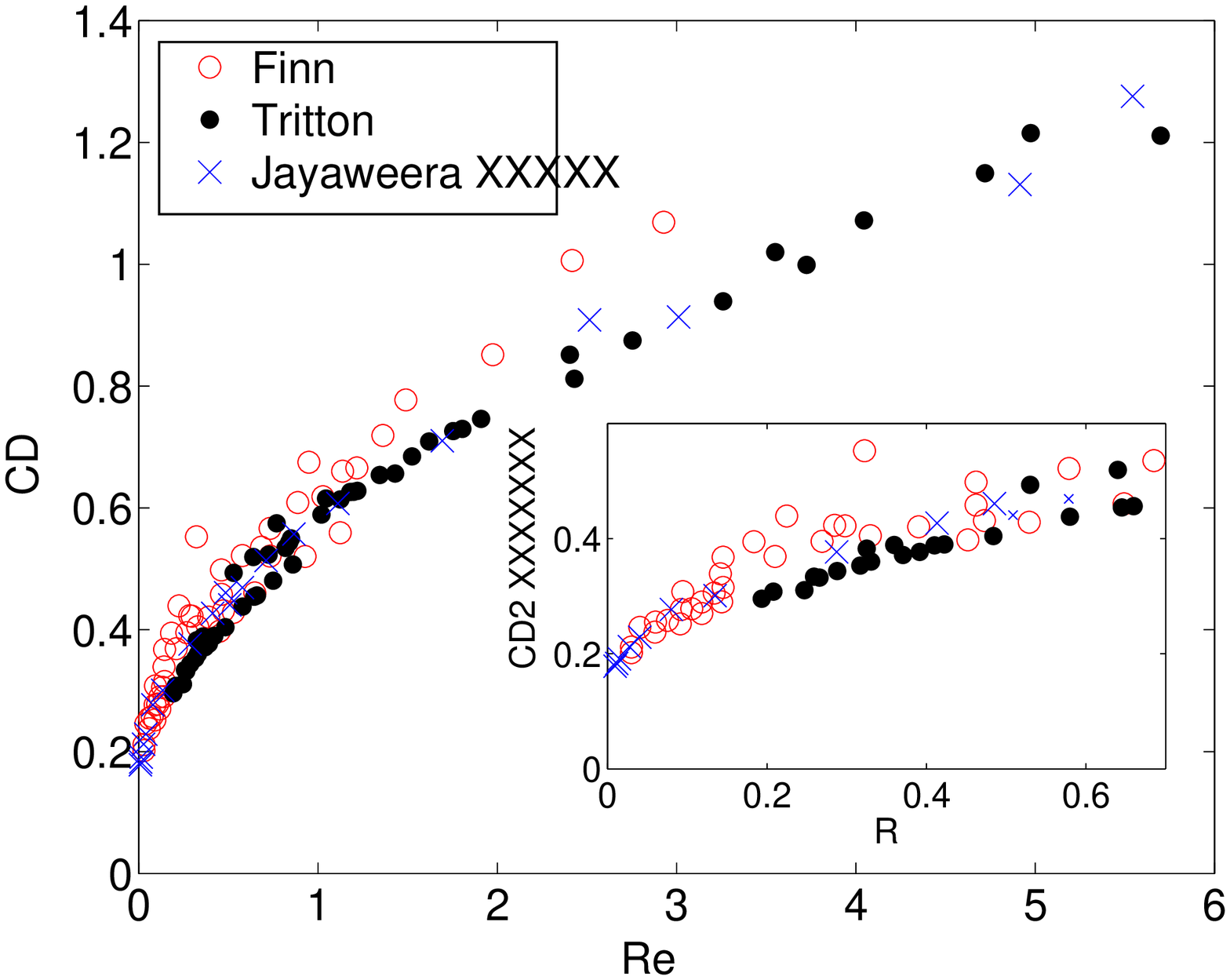}
\caption{(Color online) Summary of measurements of the drag on a cylinder \cite{Jay65,Fin53,Tritton59}.}
\label{TotCyl1}
\end{center}
\end{figure}

In 1946, White undertook a series of measurements, which were flawed
due to wall effects \cite{Whi46}. The first high quality experiments
which measured the drag at low Reynolds number were done by R. K. Finn
\cite{Fin53}. His results, available only in graphical form, are
reproduced in Figure \ref{TotCyl1}. While vastly superior to any previous
results, there is considerable scatter in Finn's measurements, and they
have largely been surpassed by later experiments.

\begin{figure}[tb]
\psfrag{Re}{$R = \frac{U_\infty a}{\nu}$}
\psfrag{Log}{$\log_{10}$}
\psfrag{Previous Experiments and}{Previous Expts. \cite{Goldstein38}}
\begin{center}
\includegraphics[width=.8 \textwidth]{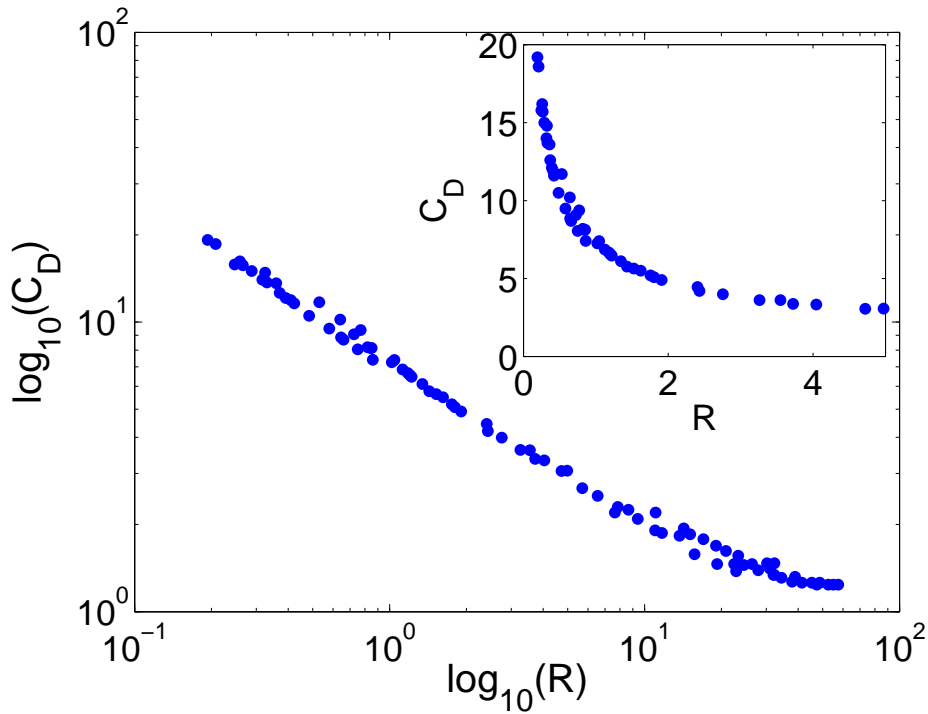}
\caption{(Color online) Tritton's measurements of the drag on a cylinder \cite{Tritton59}.}
\label{Tritton Data 1}
\end{center}
\end{figure}

Tritton, in 1959, conducted experiments which reached a Reynolds number
of $R=0.2$, and also filled in some gaps in the $R-C_D$ curve
\cite{Tritton59}. Tritton estimates his accuracy at $\pm 6\%$, and
compares his results favorably to previous work, commenting that,
``Probably the lowest R points of the other workers were stretching
their techniques a little beyond their limits.'' Tritton is also the
first author to give a discussion of systematic
errors.\footnote{Tritton does caution that his measurements may be
negatively biased at higher Reynolds number ($R \gtrsim 30$).} Tritton's
results are shown in Figure \ref{Tritton Data 1}. All of his data are
available in tabular form.

Maxworthy improved plots of the drag on a sphere (Fig. \ref{Goldstein
Sphere}), by arguing that the leading divergence must be removed to
better compare experiments and predictions (Fig. \ref{Maxworthy
Sphere}). This same criticism applies to plots of the drag on a
cylinder. In the case of the cylinder, $C_D$ goes as $R^{-1}$ (with
logarithmic corrections) as $R \rightarrow 0$ (Eqn. \ref{CDcylinder}).
This means we ought to plot $C_D R/4\pi$. This function tends to zero
as $R\rightarrow 0$, so it is not necessary to plot $C_D R/4\pi-1$, as
in the case of the sphere. Figure \ref{TotCyl1} shows both Finn's and
Tritton's data re-plotted with the leading divergence removed.

In 1965, K. O. L. F. Jayaweera  \cite{Jay65} undertook drag
measurements of the drag on very long (but finite) cylinders. At very
low Reynolds number ($R \leq 0.135$), his data are available in tabular
form. At higher Reynolds number, they had to be digitized. His data,
plotted with the leading divergence removed, are also shown in Figure
\ref{TotCyl1}.

The agreement amongst these experiments is excellent. Henceforth,
Finn's data will not be plotted, as it exhibits larger experimental
variations, and is surpassed by the experiments of Jayaweera and
Tritton. Jayaweera's data exhibit the least scatter, and may be
slightly better than Tritton's. However, both experiments have
comparable, large ratios of cylinder length to width (the principle
source of experimental error), and there is no a priori reason to favor
one experimental design over the other. We consider these two
experiments to be equivalent for the purposes of evaluating theoretical
predictions.

As with the sphere, there are numerical calculations, including:
Underwood \cite{Und69}, Son \cite{Son69}, Kawaguti \cite{Kaw66}, Dennis
\cite{Den64}, Thom \cite{Tho33}, Apelt \cite{Ape61}, and Allen
\cite{All55}. Of these, most treat only a few Reynolds numbers, none of
which are sufficiently small. Others, such as Allen and Dennis, have
had their results subsequently questioned \cite{Und69}. The only
applicable studies are Kawaguti \cite{Kaw66}, and Underwood
\cite{Und69}. Kawaguti has a calculation only for $R=0.5$, and is
omitted. Underwood's results are in principle important and useful, but
are only available in a coarse plot, which cannot be digitized with
sufficient accuracy. Consequently, no numerical results will be used
for evaluating analytical predictions.

There are many different experimental and numerical drag coefficient
measurements.  We will subsequently use only the best as
benchmarks for evaluating the performance of theoretical predictions.
In the case of the sphere, the experimental measurements of Maxworthy
\cite{Maxworthy65} as well as the numerical calculations of Dennis
\cite{Den71} and Le Clair \cite{LC70} all extend to sufficiently small
$R$ and possess sufficient accuracy. For the cylinder the experiments
of both Tritton \cite{Tritton59} and Jayaweera \cite{Jay65} are both
excellent. Although they exhibit small differences, we cannot judge
either to be superior, and we will compare both with theoretical
results.

\subsection{Theoretical history}

Since these problems were posed by Stokes in 1851, there have been many
attempts to solve them. All of these methods involve approximations,
which are not always rigorous (or even explicitly stated). There is
also considerable historical confusion over contributions and
attribution.\footnote{For an explanation of confusion over early work,
see Lindgren \cite{Lindgren99}. Proudman and Pearson \cite{Proudman57}
also begin their article with an insightful, nuanced discussion,
although there are some errors \cite{Lindgren99}.} Here we review and
summarize the substantial contributions to the literature, focusing on
what approximations are used, in both deriving governing equations and
in their subsequent solution. We discuss the validity and utility of
important results. Finally, we emphasize methodological shortcomings
and how they have been surmounted.

\subsubsection{Stokes and paradoxes}

In the first paper on the subject, Stokes approximated $R=0$ in Eqn.
\ref{NonDNS} and solved the resulting equation (a problem equivalent to
solving Eqn. \ref{SphereEqn} with $R=0$) \cite{Stokes51}. After
applying the boundary conditions (Eqn. \ref{Sphere BC}), his solution
is given in terms of a stream function by Eqn. \ref{stokessol}.
\begin{equation}
\label{stokessol}
\psi(r,\mu) = \frac{1}{4}\left( 2 r^2 - 3 r + \frac{1}{r} \right) \left(1-\mu^2
\right)
\end{equation}
By substituting $\psi(r,\mu)$ into Eqns. \ref{spherestreamfunction},
\ref{CDsphere}, and \ref{sphpressure} (or by using Eqn.
\ref{sphere:simpledrag}), we reproduce the famous result of Stokes,
given by Eqn. \ref{stokes:famoussol}.
\begin{equation}
\label{stokes:famoussol}
C_D = \frac{6 \pi}{R}
\end{equation}

Stokes also tackled the two dimensional cylinder problem in a similar
fashion, but could not obtain a solution. The reason for his failure
can be seen by setting $R=0$ in Eqn. \ref{CylinderEqn}, and attempting a
direct solution. Enforcing the $\sin\theta$ angular dependence results
in a solution of the form $\psi(r,\theta) = \left( C_1 r^3 + C_2 r
\ln{r}  + C_3 r + C_4/r \right) \sin\theta$. Here $C_i$ are integration
constants. No choice of $C_i$ will meet the boundary conditions Eqn.
(\ref{Cylinder BC}), as this solution cannot match the uniform flow at
large $r$. The best one can do is to set $C_1 = 0$, resulting in a
partial solution:
\begin{equation}
\psi(r,\theta) = C \left(2 r \ln{r}  - r + \frac{1}{r} \right) \sin\theta
\label{stokes:cylindersol}
\end{equation}
Nonetheless, this solution is \emph{not} a description of fluid flow
which is valid everywhere. Moreover, due to the indeterminable constant
$C$, Eqn. \ref{stokes:cylindersol} cannot be used to estimate the drag
on the cylinder.

A more elegant way to see that no solution may exist is through
dimensional analysis \cite{LandauLifschitz,Happel73}. The force per
unit length may only depend on the cylinder radius, fluid viscosity,
fluid density, and uniform stream velocity. These quantities are given
in Table \ref{tab:dimanal1}, with $M$ denoting a unit of mass, $T$ a
unit of time, and $L$ a unit of length. From these quantities, one may
form two dimensionless groups \cite{Buckingham04}: $\Pi_0 = R =
|\vec{u}_{\infty}| a/\nu$, $\Pi_1 = F_{\text{Net}}/(\rho \nu
|\vec{u}_{\infty}|)$. Buckingham's $\Pi$ Theorem \cite{Buckingham04}
then tells us that:
\begin{equation}
\Pi_0 = F(R)
\label{buckingham}
\end{equation}
If we make the assumption that the problem does not depend on $R$, as
Stokes did, then we obtain $\Pi_1 = \text{const}$, whence
\begin{equation}
F_{\text{Net}} \propto \rho \nu |\vec{u}_{\infty}|
\label{buckinghamresult}
\end{equation}

However, Eqn. \ref{buckinghamresult} does not depend on the cylinder
radius, $a$! This is physically absurd, and demonstrates that Stokes'
assumptions cannot yield a solution. The explanation is that when we
take the $R\rightarrow 0$ limit in Eqn. \ref{buckingham}, we made the
incorrect assumption that $F(R)$ tended toward a \emph{finite, non-zero
limit}. This is an example of \emph{incomplete similarity}, or
\emph{similarity of the second kind} (in the Reynolds number)
\cite{Barenblatt79}. Note that the problem of flow past a sphere
involves force, \emph{not} force per unit length, and therefore is not
subject to the same analysis.

\begin{table}
\begin{center}
  \begin{tabular}{|c|l|c|}
\hline
   Quantity & Description & Dimensions\\
   \hline
   $F_{\text{Net}}$ & Net Force per Unit Length & $MT^{-2}$ \\
   $\nu$ & Kinematic Viscosity & $L^2T^{-1}$\\
   $ a $ & Cylinder Radius & $L$ \\
   $ \rho $ & Fluid Density & $ML^{-3}$ \\
   $|\vec{u}_{\infty}|$ & The Uniform Stream Speed & $LT^{-1}$ \\
\hline
  \end{tabular}
\caption{Dimensional analysis of Stokes' problem.}
  \label{tab:dimanal1}
\end{center}
\end{table}

Stokes incorrectly took this nonexistence of a solution to mean that
steady-state flow past an infinite cylinder could not exist. This
problem, which is known as \emph{Stokes' paradox}, has been shown to
occur with any unbounded two-dimensional flow \cite{Kra53}.  But such flows
really do exist, and this mathematical problem has since been resolved by
the recognition of the existence of boundary layers.

In 1888, Whitehead, attempted to find higher approximations for flow
past a sphere, ones which would be valid for small but non-negligible
Reynolds numbers \cite{Whitehead89}. He used Stokes' solution (Eqn.
\ref{stokessol}) to approximate viscous contributions (the LHS of Eqn.
\ref{SphereEqn}), aiming to iteratively obtain higher approximations
for the inertial terms. In principle, this approach can be repeated
indefinitely, always using a linear governing equation to obtain higher
order approximations. Unfortunately, Whitehead found that his next
order solution could not meet all of the boundary conditions (Eqn.
\ref{Sphere BC}), because he could not match the uniform stream at
infinity \cite{VanDyke1975}. These difficulties are analogous to the
problems encountered in Stokes' analysis of the infinite cylinder.

Whitehead's approach is equivalent to a perturbative expansion in the
Reynolds number, an approach which is ``never valid in problems of
uniform streaming'' \cite{Proudman57}. This mathematical difficulty is
common to all three-dimensional uniform flow problems, and is known as
\emph{Whitehead's paradox}. Whitehead thought this was due to
discontinuities in the flow field (a ``dead-water wake''), but this is
incorrect, and his ``paradox'' has also since been resolved
\cite{VanDyke1975}.

% The reasons for Whitehead's failure can also be understood by examining Stokes' solution, Eqn. \ref{stokessol}. Stokes neglected inertial terms in the Navier-Stokes equations (\ref{NonDNS}), considering the flow to be dominated by viscous terms. Calculating a second approximation requires that we have a uniformly valid approximation to the neglected inertial terms. Subsitituting Stokes' solution (Eqn. \ref{stokessol} into the LHS of Eqn. \ref{NonDNS} results in a driving term which has a $\sin{\theta}^2 \cos \theta$ angular dependence. Any particular solution to the second order governing equations must also reflect this dependence, and will consequently be unable to meet the boundary conditions at infinity.

\subsubsection{Oseen's equation}

{\it a. Introduction}

In 1893, Rayleigh pointed out that Stokes' solution would be uniformly
applicable if certain inertial forces were included, and noted that the
ratio of those inertial forces to the viscous forces which Stokes
considered could be used to estimate the accuracy of Stokes'
approximations \cite{Ray93}.

Building on these ideas in 1910, C. W. Oseen proposed an ad hoc
approximation to the Navier-Stokes equations which resolved both
paradoxes. His linearized equations (the \emph{Oseen equations})
attempted to deal with the fact that the equations governing Stokes'
perturbative expansion are invalid at large $|\vec{r}|$, where they
neglect important inertial terms. In addition to Oseen, a number of
workers have applied his equations to a wide variety of problems,
including both the cylinder and the sphere.\footnote{Lamb
\cite{Lamb1932} solved the Oseen equations for the cylinder
approximately, as Oseen \cite{Oseen10} did for the sphere. The
Oseen equations have been solved exactly for a cylinder by Fax\'en
\cite{Faxen27}, as well as by Tomotika and Aoi \cite{Tomotika50}, and
those for the sphere were solved exactly by Goldstein \cite{Goldstein29}.}

Oseen's governing equation arises independently in several different
contexts. Oseen derived the equation in an attempt to obtain an
approximate equation which describes the flow everywhere. In modern
terminology, he sought a governing equation whose solution is a
uniformly valid approximation to the Navier-Stokes equations. Whether
he succeeded is a matter of some debate. The short answer is ``Yes, he
succeeded, but he got lucky.''

This story is further complicated by historical confusion. Oseen's
equations ``are valid but for the wrong reason'' \cite{Lindgren99};
Oseen originally objected to working in the inertial frame where the
solid body is at rest, and therefore undertook calculations in the rest
frame of uniform stream. This complication is overlooked largely
because many subsequent workers have only understood Oseen's intricate
three paper analysis through the lens of Lamb's later work
\cite{Lam11}. Lamb --- in addition to writing in English --- presents a
clearer, ``shorter way of arriving at his [Oseen's] results'', which he
characterizes as ``somewhat long and intricate.'' \cite{Lam11}

In 1913 Fritz Noether, using both Rayleigh's and Oseen's ideas,
analyzed the problem using stream functions \cite{Noe11}. Noether's
paper prompted criticisms from Oseen, who then revisited his own work.
A few months later, Oseen published another paper, which included a new
result for $C_D$ (Eqn. \ref{Oseen:dragsphere}) \cite{Ose13}. Burgess
also explains the development of Oseen's equation, and presents a clear
derivation of Oseen's principal results, particularly of Oseen's new
formula for $C_D$ \cite{BR16}.

Lindgren offers a detailed discussion of these historical developments
\cite{Lindgren99}. However, he incorrectly reports Noether's
publication date as 1911, rather than 1913. As a result, he incorrectly
concludes that Noether's work was independent of Oseen's, and
contradicts claims made in Burgess \cite{BR16}.

Although the theoretical justification for Oseen's approximations is
tenuous, its success at resolving the paradoxes of both Stokes and
Whitehead led to widespread use. Oseen's equation has been  fruitfully
substituted for the Navier-Stokes' equations in a broad array of low
Reynolds number problems. Happel and Brenner describe its application
to many problems in the dynamics of small particles where interactions
can be neglected \cite{Happel73}. Many workers have tried to explain
the utility and unexpected accuracy of Oseen's governing equations.

Finally, the Oseen equation, as a partial differential equation, arises
in both matched asymptotic calculations and in our new  work. In these
cases, however, its genesis and interpretation is entirely different,
and the similarity is purely formal. Due to its ubiquity and historical
significance, we now discuss both Oseen's equation and its \emph{many}
different solutions in detail.

{\it b. Why Stokes' approximation breaks down}

Oseen solved the paradoxes of Stokes and Whitehead by using Rayleigh's
insight: compare the magnitude of inertial and viscous forces
\cite{Oseen10,Ray93}. Stokes and Whitehead had completely neglected
inertial terms in the Navier-Stokes equations, working in the regime
where the Reynolds number is insignificantly small (so-called
``creeping flow''). However, this assumption can only be valid near the
surface of the fixed body. It is \emph{never} valid everywhere.

To explain why, we follow here the spirit of Lamb's analysis,
presenting Oseen's conclusions ``under a slightly different form.''
\cite{Lam11}

Consider first the case of the sphere. We can estimate the magnitude of
the neglected inertial terms by using Stokes' solution (Eqn.
\ref{stokessol}). Substituting this result into the RHS of Eqn.
\ref{SphereEqn}, we see that the dominant inertial components are
convective accelerations arising from the nonlinear terms in Eqn.
\ref{SphereEqn}. These terms reflect interactions between the uniform
stream and the perturbations described by Eqn. \ref{stokessol}. For
large values of $|\vec{r}|$, these terms are of \Order[R r]{-2}.

Estimating the magnitude of the relevant viscous forces is somewhat
trickier. If we substitute Eqn. \ref{stokessol} into the LHS of Eqn.
\ref{SphereEqn}, the LHS vanishes identically. To learn anything, we
must consider the terms individually. There are two kinds of terms
which arise far from the sphere. Firstly, there are components due
solely to the uniform stream. These are of \Order[r]{-2}. However, the
uniform stream satisfies Eqn. \ref{SphereEqn} independently, without
the new contributions in Stokes' solution. Mathematically, this means
that all of the terms of \Order[r]{-2} necessarily cancel amongst
themselves.\footnote{VanDyke \cite{VanDyke1975} does not treat this issue in detail, and we recommend Proudman
\cite{Proudman57} or Happel \cite{Happel73} for a more careful
discussion.} We are interested in the magnitude of the remaining terms,
perturbations which result from the other components of Stokes'
solution. These viscous terms (i.e. the $\partial_\theta^4$ term in
Eqn. \ref{SphereEqn}) are of \Order[r]{-3} as $r \to \infty$.

Combining these two results, the ratio of inertial to viscous terms, in
the $r \to \infty$ limit, is given by Eqn. \ref{oseen:stokesbreakdown}.
\begin{equation}
\frac{\textrm{inertial}}{\textrm{viscous}} = \Order[Rr]{}
\label{oseen:stokesbreakdown}
\end{equation}
This ratio is small near the body ($r$ is small) and justifies
neglecting inertial terms in that regime. However, Stokes' implicit
assumption that inertial terms are everywhere small compared to viscous
terms breaks down when $R r \sim \Order[1]{}$, and the two kinds of
forces are of the same magnitude. In this regime, Stokes' solution is
not valid, and therefore cannot be used to estimate the inertial terms
(as Whitehead had done). Technically speaking, Stokes' approximations
breaks down because of a singularity at infinity, an indication that
this is a \emph{singular perturbation} in the Reynolds' number. As
Oseen pointed out, this is the genesis of Whitehead's ``paradox''.

What does this analysis tell us about the utility of Stokes' solution?
Different opinions can be found in the literature. Happel, for
instance, claims that it ``is not uniformly valid'' \cite{Happel73},
while Proudman asserts ``Stokes' solution is therefore actually a
uniform approximation to the total velocity distribution.''
\cite{Proudman57} By a \emph{uniform approximation}, we mean that the
approximation asymptotically approaches the exact solution as the
Reynolds' number goes to zero \cite{Kaplun57a}; see Section
\ref{section:uniform} for further discussion.

Proudman and Pearson clarify their comment by noting that although
Stokes' solution is a uniform approximation to the total velocity
distribution, it does not adequately characterize the perturbation to
the uniform stream, or the \emph{derivatives} of the velocity. This is
a salient point, for the calculations leading to Eqn.
\ref{oseen:stokesbreakdown} examine components of the Navier-Stokes
equations, not the velocity field itself. These components are forces
--- derivatives of velocity.

However, Proudman and Pearson offer no proof that Stokes' solution is
actually a uniform approximation, and their claim that it is ``a valid
approximation to many bulk properties of the flow, such as the
resistance'' \cite{Proudman57} goes unsupported. In fact any
calculation of the drag necessitates utilizing derivatives of the
velocity field, so their argument is inconsistent.

We are forced to conclude that Stokes' solution is not a uniformly
valid approximation, and that his celebrated result, Eqn.
\ref{stokes:famoussol}, is the fortuitous result of uncontrolled
approximations. Remarkably, Stokes' drag formula is in fact the correct
zeroth order approximation, as can be shown using either matched
asymptotics or the Oseen equation! This coincidence is essentially due
to the fact that the drag is determined by  the velocity field and its
derivatives at the surface of the sphere, where $r=1$, and Eqn.
\ref{oseen:stokesbreakdown} is \Order[R]{1}. The drag coefficient
calculation uses Stokes' solution in the regime where his assumptions
are the most valid.

A similar analysis affords insight into the origin of Stokes' paradox
in the problem of the cylinder. Although we have seen previously that
Stokes' approach must fail for both algebraic and dimensional
considerations, examining the ratio between inertial and viscous forces
highlights the physical inconsistencies in his assumptions.

We can use the incomplete solution given by Eqn.
\ref{stokes:cylindersol} to estimate the relative contributions of
inertial and viscous forces in Eqn. \ref{CylinderEqn}. More
specifically, we examine the behavior of these forces at large values
of $r$. Substituting Eqn. \ref{stokes:cylindersol} into the RHS of Eqn.
\ref{CylinderEqn}, we find that the inertial forces are
$\mathcal{O}\left(R C^2 \log{r}/{r^2}\right)$ as $r \rightarrow
\infty$.

We estimate the viscous forces as in the case of the sphere, again
ignoring contributions due solely to the uniform stream. The result is
that the viscous forces are $\mathcal{O}\left(C \log{r}/{r^3}
\right)$.\footnote{This result disagrees with the results of Proudman
\cite{Proudman57} and VanDyke \cite{VanDyke1975}, who calculate that the
ratio of inertial to viscous forces $\sim R r \ln{r}$. However, both
results lead to the same conclusions.} Combining the two
estimates, we obtain the result given in Eqn.
\ref{oseen:stokesbreakdown2}.
\begin{equation}
\frac{\textrm{inertial}}{\textrm{viscous}} = \Order[Rr]{}
\label{oseen:stokesbreakdown2}
\end{equation}

This result demonstrates that the paradoxes of Stokes and Whitehead
are the result of the same failures in Stokes' uncontrolled
approximation. Far from the solid body, there is a regime where it is
incorrect to assume that the inertial terms are negligible in
comparison to viscous terms. Although these approximations happened to
lead to a solution in the case of the sphere, Stokes' approach is
invalid and technically inconsistent in both problems.

{\it c. How Oseen Resolved the Paradoxes}

Not only did Oseen identify the physical origin for the breakdowns in
previous approximations, but he also discovered a solution
\cite{Oseen10}. As explained above, the problems arise far from the
solid body, when inertial terms are no longer negligible. However, in
this region ($r \gg 1$), the flow field is nearly a uniform stream ---
it is almost unperturbed by the solid body. Oseen's inspiration was to
replace the inertial terms with linearized approximations far from the
body. Mathematically, the fluid velocity $\vec{u}$ in Eqn. \ref{NonDNS}
is replaced by the quantity $\vec{u}_\infty+\vec{u}$, where $\vec{u}$
represents the perturbation to the uniform stream, and is considered to
be small. Neglecting terms of \Order[|\vec{u}|]{2}, the viscous  forces
of the Navier-Stokes' equation --- $R \left(\vec{u} \cdot
\nabla\vec{u}\right)$ --- are approximated by $R \left(\vec{u}_\infty
\cdot\nabla\vec{u}\right)$.

This results in Oseen's equation:
\begin{equation}
\label{Oseen:Eqn}
R \left(\vec{u}_\infty \cdot \nabla \vec{u}\right) = - \nabla p + \nabla^2 \vec{u}
\end{equation}
The lefthand side of this equation is negligible in the region where
Stokes' solution applies. One way to see this is by explicitly
substituting Eqn. \ref{stokessol} or Eqn. \ref{stokes:cylindersol} into
the LHS of Eqn. \ref{Oseen:Eqn}. The result is of \Order[R]{}. This can
also be done self-consistently with any of the solutions of Eqn. \ref
{Oseen:Eqn}; it can thereby be explicitly shown that the LHS can only
becomes important when $r \gg 1$, and the ratios in Eqns.
\ref{oseen:stokesbreakdown} and \ref{oseen:stokesbreakdown2} are of
\Order[1]{}.

Coupled with the continuity equation (Eqn. \ref{Continuity2ND}), and
the usual boundary conditions, the Oseen equation determines the flow
field everywhere. The beautiful thing about Oseen's equation is that it
is \textit{linear}, and consequently is solvable in a wide range of
geometries. In terms of stream functions, the Oseen equation for a
sphere takes on the form given by Eqn. \ref{Oseen:SphereEqn}. The
boundary conditions for this equation are still given by Eqn.
\ref{Sphere BC}.
\begin{equation}
\label{Oseen:SphereEqn}
D^4 \psi = R \left(\frac{1-\mu^2}{r}
  \frac{\partial}{\partial \mu} + \mu \frac{\partial}{\partial r}
  \right) D^2 \psi(r,\mu)
\end{equation}
Here, $D$ is defined as in Eqn. \ref{SphereEqn}.

For the cylinder, where the boundary conditions are given by Eqn.
\ref{Cylinder BC}, Oseen's equation takes the form given by Eqn.
\ref{Oseen:CylinderEqn}.
\begin{equation}
\label{Oseen:CylinderEqn}
\nabla_r^4 \psi(r,\theta) = R \left(\cos(\theta)
\frac{\partial}{\partial r} -
\frac{\sin (\theta)}{r}\frac{\partial}{\partial \theta} \right) \nabla_r^2
\psi(r,\theta)
\end{equation}
Here $\nabla$ is defined as in Eqn. \ref{CylinderEqn}. This equation takes on a
particularly simple form in Cartesian coordinates (where $x=r\cos{\theta}$):
$\left( \nabla^2 - R\partial_x \right)\nabla^2\psi(r,\theta) =
0$.

A few historical remarks must be made. First, Oseen and Noether were
motivated to refine Stokes' work and include inertial terms because
they objected to the analysis being done in the rest frame of the solid
body. While their conclusions are valid, there is nothing wrong with
solving the problem in any inertial frame. Secondly, Oseen made no use
of stream functions; the above equations summarize results from several
workers, particularly Lamb.

There are many solutions to Oseen's equations, applying to different
geometries and configurations, including some exact solutions. However, for any
useful calculations, such as $C_D$, even the exact solutions need to be
compromised with approximations. There have been many years of
discussion about how to properly interpret Oseen's approximations, and
how to understand the limitations of both his approach and concomitant
solutions. Before embarking on this analysis, we summarize the
important solutions to Eqns. \ref{Oseen:SphereEqn} and
\ref{Oseen:CylinderEqn}.

{\it d. A plethora of solutions}

Oseen himself provided the first solution to Eqn. \ref{Oseen:SphereEqn}, solving
it exactly for flow past a sphere \cite{Oseen10}. Eqn. \ref{Oseen:1sol} reproduces this result
in terms of stream functions, a formula first given by Lamb \cite{Lamb1932}.
\begin{equation}
\label{Oseen:1sol}
\psi(r,\theta)=\frac{1}{4}\left(2 r^2 + \frac{1}{r}
\right)\sin^2{\theta} - \frac{3}{2 R} \left( 1+ \cos{\theta}
\right)\left(1 - e^{-\frac{1}{2}R r \left(1 -
  \cos{\theta}\right)}\right)
\end{equation}
This solution is reasonably behaved everywhere, and may be used to
obtain Oseen's improved approximation for the drag coefficient (Eqn.
\ref{Oseen:dragsphere}).
\begin{equation}
\label{Oseen:dragsphere}
C_D = \frac{6 \pi}{R}\left(1 + \frac{3}{8}R \right) + \Order[R]{2}
\end{equation}

Oseen obtained this prediction for $C_D$ after the prompting of
Noether, and only presented it in a later paper \cite{Ose13}. Burgess
also obtained this result \cite{BR16}. Oseen's work was hailed as a
resolution to Whitehead's paradox. While it \emph{did} resolve the
paradoxes (e.g., he explained how to deal with inertial terms), and his
solution is uniformly valid, it does \emph{not} posses sufficient
accuracy to justify the ``$3/8 R$'' term in Eqn.
\ref{Oseen:dragsphere}. What Oseen really did was to rigorously derive
the leading order term, proving the validity of Stokes' result (Eqn.
\ref{stokes:famoussol}). Remarkably, his new term is also correct! This
is a coincidence which will be carefully considered later.

This solution (Eqn. \ref{Oseen:1sol}) is exact in the sense that it
satisfies Eqn. \ref{Oseen:SphereEqn}. However, it does not exactly meet
the boundary conditions (Eqn. \ref{Sphere BC}) at the surface of the
sphere. It satisfies those requirements only approximately, to
\Order[R]{1}. This can readily be seen by expanding Eqn.
\ref{Oseen:1sol} about $r=1$:
\begin{equation}
\label{Oseen:SphereExpand}
\psi(r,\theta) = \frac{1}{4}\left( 2 r^2 - 3 r + \frac{1}{r} \right)\sin^2{\theta} + \Order[R]{1}
\end{equation}

Up to \Order[R]{} this is simply Stokes' solution (Eqn.
\ref{stokessol}), which vanishes identically at $r=1$. The new terms
fail to satisfy the boundary conditions at the surface, but are higher
order in $R$. Thus Oseen's solution is an exact solution to an
approximate governing equation which satisfies boundary conditions
approximately. The implications of this confounding hierarchy of
approximations will be discussed below.

Lamb contributed a simplified method for both deriving and solving
Oseen's equation \cite{Lam11}. His formulation was fruitfully used by
later workers (e.g.,  \cite{Faxen27,Goldstein29,Tomotika50}), and Lamb
himself used it both to reproduce Oseen's results and to obtain the
first result for the drag on an infinite cylinder.

Lamb's basic solution for flow around an infinite cylinder appears in a
number of guises. His original solution was given in terms of velocity
components, and relied on expansions of modified Bessel functions which
kept only the most important terms in the series. This truncation
results in a solution (Eqn. \ref{Oseen:Lambcyl1}) which only
approximately satisfies the governing equations (Eqn.
\ref{Oseen:CylinderEqn}), and is only valid near the surface.
\begin{subequations}
\label{Oseen:Lambcyl1}
\begin{eqnarray}
u_x &=& 1+ \delta \left(\gamma -\frac{1}{2} + \log{\frac{rR}{4}}+\frac{1}{2}\left(r^2-1\right)\frac{\partial^2}{\partial x^2} \log{r} \right) \\
u_y &=& \frac{\delta}{2} \left(r^2-1 \right) \frac{\partial^2}{\partial x \partial y} \log{r} \\
u_z &=& 0
\end{eqnarray}
\end{subequations}
In this equation, $\delta = \left(\frac{1}{2}-\gamma-\log{\frac{R}{4}}\right)^{-1}$.

Note that, although it only approximately satisfies Oseen's governing
equation, this result satisfies the boundary conditions (Eqn.
\ref{boundaryconditions}) exactly. Lamb used his solution to derive the
first result (Eqn. \ref{Oseen:LambDrag}) for the drag on an infinite
cylinder, ending Stokes' paradox:
\begin{equation}
\label{Oseen:LambDrag}
C_D = \frac{4\pi}{R} \left(\delta \right)
\end{equation}
In his own words, `` ... Stokes was
led to the conclusion that steady motion is impossible. It will appear
that when the inertia terms are partially taken into account ... that a
definite value for the resistance is obtained.'' \cite{Lam11} As with all
analysis based on the ad-hoc Oseen equation, it is difficult to quantify
either the accuracy or the limitations of Lamb's result.

Many authors formulate alternate expressions of Lamb's solution by
retaining the modified Bessel functions rather than replacing them with
expansions valid for small $R$ and $r$. This form is given by Eqn.
\ref{Oseen:LambcylClose}, and is related to the incomplete form given
by VanDyke (p. 162) \cite{VanDyke1975}.\footnote{Note that VanDyke
incorrectly attributes to this result to Oseen, rather than to Lamb.}
\begin{subequations}
\label{Oseen:LambcylClose}
\begin{eqnarray}
u_x &=& 1 + \delta \left(\frac{x^2}{r^4} - \frac{1}{2r^2} + \frac{2x}{Rr^2} - e^{Rx/2}K_0\left(\frac{Rr}{2}\right) - \frac{x}{r}e^{Rx/2} K_1\left(\frac{Rr}{2}\right)\right) \\
u_y &=& \delta \left( \frac{xy}{r^4} + \frac{2 y}{R r^2} - \frac{y}{r}e^{Rx/2}K_1\left(\frac{Rr}{2}\right)\right) \\
u_z &=& 0
\end{eqnarray}
\end{subequations}
Here $I_n$ and $K_n$ are modified Bessel functions.

In contrast to Eqn. \ref{Oseen:Lambcyl1}, this solution is an exact
solution to Oseen's equation (Eqn. \ref{Oseen:CylinderEqn}), but only
meets the boundary conditions to first approximation. In particular, it
breaks down for harmonics other than $\sin{\theta}$.  Whether Eqn.
\ref{Oseen:Lambcyl1} or Eqn. \ref{Oseen:LambcylClose} is preferred is a
matter of some debate, and ultimately depends on the problem one is
trying to solve.

Some workers prefer expressions like Eqn. \ref{Oseen:LambcylClose},
which are written in terms of $\vec{u}$. Unlike the solutions for the
stream function, these results can be written in closed form. This
motivation is somewhat misguided, as applying the boundary conditions
nonetheless requires a series expansion.

In terms of stream functions Eqn. \ref{Oseen:LambcylClose} transforms
into Eqn. \ref{Oseen:Lambcyl} \cite{Proudman57}.
\begin{equation}
\label{Oseen:Lambcyl}
\psi(r,\theta) = \left(r + \frac{\delta}{2 r} \right) \sin{\theta} -
\sum_{n=1}^\infty \delta\phi_n\left(\frac{R r}{2}\right) \frac{r
  \sin{n\theta}}{n}
\end{equation}
Here,
\begin{displaymath}
\phi_n(x) = 2 K_1(x)I_n(x) + K_0(x) \left(I_{n+1}(x) + I_{n-1}(x)\right)
\end{displaymath}

This result is most easily derived as a special case of Tomotika's
general solution (Eqn. \ref{cylinder:1sol}) \cite{Tomotika50}, although
Proudman et al. intimate that it can also be directly derived from
Lamb's solution (Eqn. \ref{Oseen:LambcylClose}) \cite{Proudman57}.

Bairstow et al. were the first to retain Bessel functions while solving
Oseen's Eqn. for flow past a cylinder \cite{Bai23}. They followed
Lamb's approach, but endeavored to extend it to larger Reynolds'
numbers, and obtained the drag coefficient given in Eqn.
\ref{Bairstowsol}. When expanded near $R=0$, this solution reproduces
Lamb's result for $C_D$ (Eqn. \ref{Oseen:LambDrag}). It can also be
obtained from Tomotika's more general solution (Eqn.
\ref{cylinder:1sol}).
\begin{equation}
C_D = \frac{4\pi}{R \left(I_0(R/2) K_0(R/2) + I_1(R/2)K_1(R/2)\right)}
\label{Bairstowsol}
\end{equation}
Bairstow also made extensive comparisons between experimental
measurements of $C_D$ and theoretical predictions \cite{Rel14}. He
concluded, ``For the moment it would appear that the maximum use has
been made of Oseen's approximation to the equations of viscous fluid
motion.''

At this point, the ``paradoxes'' were ``resolved'' but by an
approximate governing equation which had been solved approximately.
This unsatisfactory state of affairs was summarized by Lamb in the last
edition of his book: `` ... even if we accept the equations as adequate
the boundary-conditions have only been approximately satisfied.''
\cite{Lamb1932} His comment was prompted largely by the work of Hilding
Fax\'en, who initiated the next theoretical development, exact
solutions to Oseen's approximate governing equation (Eqn.
\ref{Oseen:Eqn}) which also exactly satisfy the boundary conditions.

Beginning with his thesis and spanning a number of papers Fax\'en
systematically investigated the application of boundary conditions to
solutions of Oseen's equations \cite{Fax23,Fax21}. Fax\'en initially
studied low Reynolds number flow around a sphere, and he began by
re-examining Oseen's analysis. He derived a formula for $C_D$ which
differed from Oseen's accepted result (Eqn. \ref{Oseen:dragsphere}).
Fax\'en realized that this was due to differences in the application of
approximate boundary conditions; within the limitations of their
respective analyses, the results actually agreed.

Fax\'en next solved Oseen's equation (Eqn. \ref{Oseen:SphereEqn}), but
in bounded, finite spaces where the boundary conditions could be
satisfied exactly. He initially studied flow near infinite parallel
planes, but ultimately focused on flow around a sphere within a
cylinder of finite radius. He aimed to calculate the drag force in a
finite geometry, and then take the limit of that solution as the radius
of the cylinder tends to infinity.

Unfortunately, in the low Reynolds number limit, the problem involves
incomplete similarity, and it is incorrect to assume that solutions
will be well behaved (e.g., tend to a finite value) as the boundary
conditions are moved to infinity.

The drag force which Fax\'en calculated involved a number of
undetermined coefficients, so he also calculated it using solutions to
Stokes' governing equations. This solution \emph{also} has unknown
coefficients, which he then calculated numerically. Arguing that the
two solutions ought to be the same, he matched coefficients between the
two results, substituted the numerical coefficients, and thereby
arrived at a drag force based on the Oseen governing equation.

This work is noteworthy for two reasons. First, the matching of
coefficients between solutions derived from the two different governing
equations is prescient, foreshadowing the development of matched
asymptotics 30 years later. Secondly, Fax\'en ultimately concluded
that Oseen's ``improvement'' (Eqn. \ref{Oseen:dragsphere}) on Stokes'
drag coefficient (Eqn. \ref{stokes:famoussol}) is invalid \cite{Fax23}.
Fax\'en's analysis demonstrates that --- when properly solved ---
Oseen's equation yields the same drag coefficient as Stokes', without
any additional terms \cite{Lindgren99}.

Studies by Bohlin and Haberman concur with Fax\'en's conclusions
\cite{Boh60,HS58,Lindgren99}. It is not surprising that his results
reject Oseen's new term ($3 R/8$). We previously explained that Oseen's
analysis, although it eliminates the ``paradoxes'', does not posses
sufficient accuracy to justify more than the lowest order term in Eqn.
\ref{Oseen:dragsphere}.

However, Fax\'en's results suffer from problems. First, they cannot be
systematically used to obtain better approximations. Secondly, Fax\'en
actually solves the problem for bounded flow, with the boundary
conditions prescribed by finite geometries. He uses a limiting
procedure to extend his solutions to unbounded flow (with boundary
conditions imposed on the uniform stream only at infinity, as in Eqn.
\ref{boundaryconditions}). In problems like this, which involve
incomplete similarity, it is preferable to work directly in the
infinite domain.

Fax\'en's meticulous devotion to properly applying boundary conditions
culminated in the first complete solution to Eqn.
\ref{Oseen:CylinderEqn}. In 1927, he published a general solution for
flow around an infinite cylinder which could exactly satisfy arbitrary
boundary conditions \cite{Faxen27}. Unfortunately, Fax\'en's solution
contains an infinite number of undetermined integration constants, and
approximations must be used to determine these constants. Although this
destroys the ``exact'' nature of the solution, these
approximations can be made in a controlled, systematic fashion --- an
improvement over the earlier results of Lamb and Oseen. Although
Fax\'en's heroic solution was the first of its kind, his real insight was
realizing that approximations in the application of boundary conditions
could be as important as the approximations in the governing equations.

His formal solutions are in essence a difficult extension of Lamb's
reformulation of Oseen's equations, and they inspired several similar
solutions. In 1929, Goldstein completed a similarly herculean calculation to
derive a general solution to Oseen's equation for flow around a sphere
\cite{Goldstein29}. Like Fax\'en's result for the cylinder, Goldstein's
solution can --- in principle ---  exactly satisfy the boundary
conditions. Unfortunately, it also suffers from the same problems: It
is impossible to determine all of the infinitely many integration
constants.

Goldstein's solution is summarized by Tomotika, who also translated it
into the language of stream functions \cite{Tomotika50}. We combine
elements from both papers in quoting the solution given in Eqn.
\ref{Oseen:Goldstein}.
\begin{equation}
\label{Oseen:Goldstein}
\psi(r,\theta) = -r^2 Q_1(\cos{\theta}) + \sum_{n=1}^\infty \left(B_n
r^{-n} + \sum_{m=0}^\infty X_m r^2 \Phi_{m,n}(r R /2) \right)
Q_n(\cos{\theta})
\end{equation}
In this equation,
\begin{subequations}
\begin{eqnarray}
Q_n(\mu) &=& \int_{-1}^\mu P_n(\mu) \textrm{d}\mu \\
\Phi_{m,n}(x) &=& - \left(\frac{m}{2m-1} \chi_{m-1}(x) + \frac{m+1}{2m+3}\chi_{m+1}(x)\right)f_{m,n}(x) \nonumber \\
& & - \left(\frac{m}{2m+1} f_{m-1,n}(x)+\frac{m+1}{2m+1} f_{m+1,n}(x) \right)\chi_m(x)  \\
\chi_m(x) &=& \left(2m+1\right) \left(\frac{\pi}{2x}\right)^{\left(\frac{1}{2}\right)}K_{m+\frac{1}{2}}(x) \\
f_{m,n}(x) &=& (2 n +1)\sum_{j=0}^m \frac{(2j)!(2m-2j)!(2n-2j)!}{(j!)^2(2m+2n-2j+1)!} \nonumber \\
& & \times \left(\frac{(m+n-j)!}{(m-j)!(n-j)!}\right)^2\phi_{m+n-2j}(x) \\
\phi_{n}(x)&=&(2n+1)\left(\frac{\pi}{2x}\right)^{\frac{1}{2}}I_{n+\frac{1}{2}}(x
)\nonumber \\f_{m,n}(x) &=&\sum_{j=0}^{m} C_m(k) \frac{\partial^j
\phi_n(x)}{\partial  x^j}
\end{eqnarray}
\end{subequations}
Here $K_n(x)$ and $I_n(x)$ are Bessel functions, $P_m(x)$ are Legendre
polynomials, and $C_m(k)$ is the coefficient of $x^k$ in $P_m(x)$. Note
that the second expression for  $f_{m,n}(x)$, written in terms of
derivatives, is computationally convenient \cite{Goldstein29}.

Eqn. \ref{Oseen:Goldstein} is given with undetermined constants of
integration, $B_n$ and $X_m$. Methods to determine these constants were
discussed by both Tomotika \cite{Tomotika50} and Goldstein
\cite{Goldstein29}. We will present our own analysis later.

There are many different results which have been obtained using the
above general solution. The exact formula for the stream function and
the drag coefficient depend on what terms in the solution are retained,
and how one meets the boundary conditions. In general, retaining $n$
angular terms in Eqn. \ref{Oseen:Goldstein} requires the retention of
$m=n-1$ terms in the second sum. In his original paper, Goldstein
retains three terms in each series, and thereby calculates the formula
for $C_D$ given in Eqn. \ref{GoldsteinCD}.
\begin{equation}
C_D = \frac{6\pi}{R}\left(1 + \frac{3}{8}R -\frac{19}{320} R^2 + \frac{71}{2560}R^3 - \frac{30179}{2150400} R^4 + \frac{122519}{17203200} R^5 + \Order[R]{6} \right)
\label{GoldsteinCD}
\end{equation}
The coefficient of the last term reflects a correction due to Shanks \cite{Sha55}.

To obtain the result in Eqn. \ref{GoldsteinCD}, Goldstein both
truncated his solution for the stream function and then expanded the
resulting $C_D$ about $R=0$. Van Dyke extended this result to include
an additional 24 terms, for purposes of studying the mathematical
structure of the series, but not because of any intrinsic physical
meaning \cite{VD70}. Van Dyke does not state whether he was
including more harmonics in the stream function solution or simply
increasing the length of the power series given in Eqn.
\ref{GoldsteinCD}.

In addition to expressing Goldstein's solution for the geometry of a
sphere in terms of stream functions, Tomotika derived his own exact
solution to Eqn. \ref{Oseen:CylinderEqn} for flow past a cylinder
\cite{Tomotika50}. Tomotika closely followed the spirit of Lamb
\cite{Lam11} and Goldstein \cite{Goldstein29}, and his resulting
``analysis is quite different from Fax\'en's.'' \cite{Tomotika50}. His
solution to Eqn. \ref{Oseen:CylinderEqn} is given in Eqn.
\ref{cylinder:1sol} below, conveniently expressed in terms of stream
functions. Note that Tomotika's result suffers from the same problems
as his predecessors: An infinite number of undetermined integration
constants.
\begin{equation}
\label{cylinder:1sol}
\psi(r,\theta) = r \sin{\theta} + \sum_{n=1}^\infty \left( B_n
r^{-n} + \sum_{m=0}^\infty X_m r \Phi_{m,n}(r R /2) \right) \sin{n \theta}
\end{equation}
Where
\begin{eqnarray}
\Phi_{m,n}(x) &=&
\left(K_{m+1}(x)+K_{m-1}(x)\right)\left(I_{m-n}(x)+I_{m+n}(x)\right) \nonumber \\
& & + K_m(x) \left(I_{m-n-1}(x)-I_{m-n+1}(x)-I_{m+n-1}(x)+I_{m+n+1}(x)\right)
\end{eqnarray}
As before, $B_n$ and $X_m$ are constants of integration which need to
be determined by the boundary conditions (Eqn. \ref{Cylinder BC}).

As with Goldstein's solution for the sphere, approximations are
necessary in order to actually calculate a drag coefficient. By
retaining the $m=0$ and $n=1$ terms, Tomotika reproduced Bairstow's
result for $C_D$ (Eqn. \ref{Bairstowsol}). He also numerically
calculated drag coefficients based on retaining more terms. As with the
Goldstein solution, keeping $n$ angular terms requires keeping
$m=n-1$ terms in the second sum.

The solutions given in Eqns. \ref{Oseen:Goldstein} and
\ref{cylinder:1sol} represent the culmination of years of efforts to solve
Oseen's equation for both the sphere and the cylinder. These general
solutions are also needed in both matched asymptotics and the new
techniques presented in this section \cite{Proudman57}.

There is a final noteworthy solution to Eqn. \ref{Oseen:CylinderEqn}.
In 1954, Imai  published a general method for solving the
problem of flow past an arbitrary cylindrical body \cite{II54}. His
elegant technique, based on analytic functions, applies to more general
geometries. Imai calculated a formula for $C_D$, approximating the
functions in his exact solution with power series about $R=0$. His
result (re-expressed in our notation) is given in Eqn. \ref{imaisol}.
\begin{equation}
C_D = \frac{4 \pi}{R} \delta + R \left(-\frac{\pi}{2} + \frac{\pi \delta}{4}-\frac{5 \pi \delta^2}{32} \right)
\label{imaisol}
\end{equation}
Note that Imai's result agrees with Eqn. \ref{Oseen:LambDrag} at lowest
order, the only order to which Oseen's equation really applies. A
priori, his result is neither better nor worse than any other solution
of Oseen's equation. It is simply different.

{\it e. Discussion}

We have presented Oseen's governing equations for low Reynold number
fluid flow. These equations are a linearized approximation to the
Navier-Stokes' equations. We have also presented a number of different
solutions, for both stream functions and drag coefficients; each of
these solutions comes from a unique set of approximations. The
approximations which have been made can be put into the following broad
categories:

\begin{itemize}
\item The governing equation --- Oseen's equation approximates the Navier-Stokes equations.
\item Solutions which only satisfy the Oseen's equation approximately.
\item Solutions which only satisfy the boundary conditions approximately.
\item Solutions where the stream function is expanded in a power series about $R=0$ after its derivation.
\item Approximations in the drag coefficient derivation.
\item Drag coefficients which were expanded in a power series about $R=0$ after their derivation.
\end{itemize}

The first approximation is in the governing equations. Oseen's
approximation is an \textit{ad hoc} approximation which, although it
can be shown to be self-consistent, requires unusual cleverness to
obtain. Because it is not derived systematically, it can be difficult
to understand either its applicability or the limitations of its
solutions. There have been years of discussion and confusion about both
the equation and its solutions. The short answer is this: Oseen's
governing equation is a zeroth order uniformly valid approximation to
the Navier Stokes equation; the equation and its solutions are valid
only at \Order[R]{0}.

It is not easy to prove this claim rigorously \cite{Fax23}. However, it
can be easily shown that Oseen's equations are self-consistent with its
solutions, and that the error in the solution is of \Order[R]{1}. One
way to explicitly demonstrate this is by substituting a solution of
Oseen's equation into the LHS of the Navier-Stokes equations (Eqn.
\ref{NonDNS}), thereby estimating the contribution of inertial terms
for the flow field characterized by the solution. By repeating that
substitution into the LHS of Oseen's equation (Eqn. \ref{Oseen:Eqn}),
one can estimate the contribution of inertial terms under Oseen's
approximations. Comparing the two results gives an estimate of the
inaccuracies in Oseen's governing equations.

Concretely, for the sphere, we substitute Eqn. \ref{Oseen:1sol} into
the RHS of Eqn. \ref{Oseen:SphereEqn}, and into the RHS of Eqn.
\ref{SphereEqn}. The difference between the two results is of
\Order[R]{1}.

For the cylinder, substitute Eqn. \ref{Oseen:Lambcyl} into the RHS of
Eqns. \ref{Oseen:CylinderEqn} and \ref{CylinderEqn}. The difference
between the exact and approximate inertial terms is of
\Order[R\delta]{}, where $\delta$ is defined as in Eqn.
\ref{Oseen:Lambcyl}.

These conclusions do not depend on the choice of solution (or on the
number of terms retained in Eqn. \ref{Oseen:Lambcyl}). They explicitly
show that the governing equation is only valid to \Order[R]{} (or
\Order[R\delta]{}). Consequently, the solutions can only be meaningful
to the same order, and the boundary conditions need only be satisfied
to that order. With these considerations, almost all of the solutions
in the preceding section are equivalent. The only ones which are not
--- such as Eqn. \ref{Oseen:Lambcyl1} --- are those in which the
solution itself has been further approximated.\footnote{In this case,
the Bessel functions have been expanded near $R=0$ and are no longer
well behaved as $R \rightarrow \infty$.}

Since the formulae for determining $C_D$ (Eqns. \ref{CDcylinder} and
\ref{CDsphere}) are of the form $1/R$ + terms linear in stream function
+ nonlinear terms, a stream function which is valid to \Order[R]{} will
result in a drag coefficient which is valid to \Order[R]{0}. Thus, in
all of the formula for $C_D$ which have been presented so far, only the
first term is meaningful. For a sphere, this is the Stokes' drag (Eqn.
\ref{stokes:famoussol}), and for the cylinder, Lamb's results (Eqn.
\ref{Oseen:LambDrag}).

We have concluded that it is only good fortune that Oseen's new
``$3/8R$'' term is actually correct. This concurs with the analysis of
Proudman et al., who wrote, ``Strictly, Oseen's method gives only the
leading term ... and is scarcely to be counted as superior to Stokes's
method for the purpose of obtaining the drag.'' \cite{Proudman57}
Proudman and Pearson also note that the vast effort expended finding
exact solutions to Oseen's equation is ``of limited value.''
Goldstein's formula for $C_D$, for instance, is expanded to
\Order[R]{5}, well beyond the accuracy of the original governing
equations. The reason for Oseen's good fortune is rooted in the
symmetry of the problem. Chester and Van Dyke both observe that the
non-linear terms which Oseen's calculation neglects, while needed for a
correct stream function, do not contribute to the drag because of
symmetry \cite{Che62,VanDyke1975}.

Lindgren argues that Fax\'en proved that, when the boundary conditions are
met properly and Oseen's equations solved exactly, the resulting $C_D$
is that obtained by Stokes (Eqn. \ref{stokes:famoussol})
\cite{Lindgren99}. Whether this argument is correct does not matter, as
Oseen's additional term is beyond the accuracy of his governing
equations.

There is another approximation which arises while computing $C_D$ in
the context of Oseen's equation. Many workers (e.g., \cite{Tomotika50})
compute the pressure in Eqns. \ref{CDcylinder} and \ref{CDsphere} by
integrating Oseen's equation (Eqn. \ref{Oseen:Eqn}, rather than the
Navier-Stokes equations (Eqn. \ref{NonDNS}). In Eqns. \ref{cylpressure}
and \ref{sphpressure}, we presented a pressure calculation based on the
Navier Stokes equations. Calculating pressure using the linearized
Oseen equation introduces an additional approximation into $C_D$. While
not necessarily problematic or inconsistent, this approximation can be
difficult to identify.

{\it f. Two different interpretations}
%Section from PnP here:, happel.
One criticism of the Oseen equation is that it may be obtained by
linearizing the Navier-Stokes equations, without regard to the
magnitude of inertial and viscous terms. By writing $\vec{u} =
\vec{U}_\infty + \delta\vec{u}$, treating $\delta\vec{u}$ as a small
perturbation, and expanding Eqn. \ref{NonDNS} one can formally
reproduce Oseen's equations. Clearly, the disturbance to the uniform
stream is not negligible near the surface of the solid body, and
therefore Oseen's equations  ``would appear to be a poor approximation
in the neighborhood of the body where the boundary condition
$\vec{u}=0$ requires that the true inertial term be small.''
\cite{Happel73}

This incorrect argument, put forth as a reason to use Stokes'
solutions, overlooks the origins of Oseen's equations. The point of
Oseen's approximation is that inertial terms are \emph{only}
significant at large values of $\vert r\vert$, where $R\vert r\vert$ is
no longer negligible. Near the surface of the solid, the approximate
inertial terms which Oseen introduced are negligible in comparison to
the viscous terms, because they are multiplied by the factor $R$ (in
the LHS of Eqn. \ref{Oseen:Eqn}). Hence the difference between Oseen's
and Stokes' equations in the neighborhood of the sphere will be of
\Order[R]{}, and is beyond the scope of either theory.

{\it g. Better approximations}

The approach of Whitehead was essentially to improve Stokes' solution
for the sphere in an iterative fashion \cite{Whitehead89}. By
substituting the first approximation into the governing equations, he
estimated the neglected terms. He then tried, and failed, to solve the
resulting governing equation. This approach fails because the Stokes'
equations are not uniformly valid to zeroth order.

Oseen's equations are uniformly valid, and, as Proudman remarked,
``there seems little reason to doubt that Whitehead's iterative method,
using Oseen's equation rather than Stokes's equation would yield an
expansion, each successive term of which would represent a uniformly
valid higher approximation to the flow. In each step of the iteration,
a lower-order approximation would be used to calculate those particular
inertia terms that are neglected ... the expansion generated in this
way would seem to be the most economic expansion possible.''
\cite{Proudman57}

Proudman did not follow through on this idea, instead developing a
solution based on matched asymptotics expansions (see below). In an
appendix, Van Dyke relates the unpublished work of C. R. Illingworth
(1947) \cite{VanDyke1975}. Illingworth carried through Whitehead's
program, deriving a new expression (Eqn. \ref{illingworth}) for $C_D$,
which agrees to \Order[R^2\ln{R}]{} with the later results of matched
asymptotic calculations (Eqn. \ref{matchedcd}).
\begin{equation}
C_D = \frac{6\pi}{R} \left( 1+ \frac{3}{8}R + \frac{9}{40}R^2 \log{R} + 0.1333 R^2 + \frac{81}{320}R^3\log{R} - 0.0034 R^3 + \ldots \right)
\label{illingworth}
\end{equation}
Although this result has since been subsumed by matched asymptotics, it
is nonetheless remarkable, substantially improving any previous drag
calculations, and rigorously justifying Oseen's $3/8R$ term.

There have also been efforts (e.g.,  \cite{Sha55,VD70}) to ``re-sum''
Goldstein's series expansion for $C_D$ (Eqn. \ref{GoldsteinCD}).
However, these results have little intrinsic (as opposed to
methodological) value, as Goldstein's result is only valid to
\Order[R]{}. If applied to more accurate approximations, such as Eqn.
\ref{illingworth}, these methods could be worthwhile. Alas, even
improved approximations lack a sufficient numbers of terms in the
expression for $C_D$ to make this practicable.

{\it h. Summary}

Simply put, Oseen's equations resolved the paradoxes of Stokes and
Whitehead, put Stokes' results on firm theoretical ground, and led to
the first solution for the drag on a cylinder. Although the Oseen
equations happen to provide a uniformly valid first approximation, it
is difficult to extend this work to higher order approximations.

\begin{figure}[tb]
\psfrag{Re}{\mbox{\Large $R$}}
\psfrag{CD}{\mbox{\Large $C_D R/6\pi - 1$}}
\psfrag{CD2 XXXXXXX}{\tiny{$C_D R/6\pi - 1$}}
\psfrag{Maxworthy XXXXXX}{Maxworthy}
\psfrag{Pruppacher}{Eqn. \ref{Pruppacher1}}
\psfrag{Le Clair}{Le Clair}
\psfrag{Dennis}{Dennis}
\psfrag{Stokes}{Stokes, Eqn. \ref{stokes:famoussol}}
\psfrag{Oseen}{Oseen, Eqn. \ref{Oseen:dragsphere}}
\psfrag{Goldstein}{Goldstein, Eqn. \ref{GoldsteinCD}}
\psfrag{Illingworth}{Illing., Eqn. \ref{illingworth}}
\begin{center}
\includegraphics[width=.8 \textwidth]{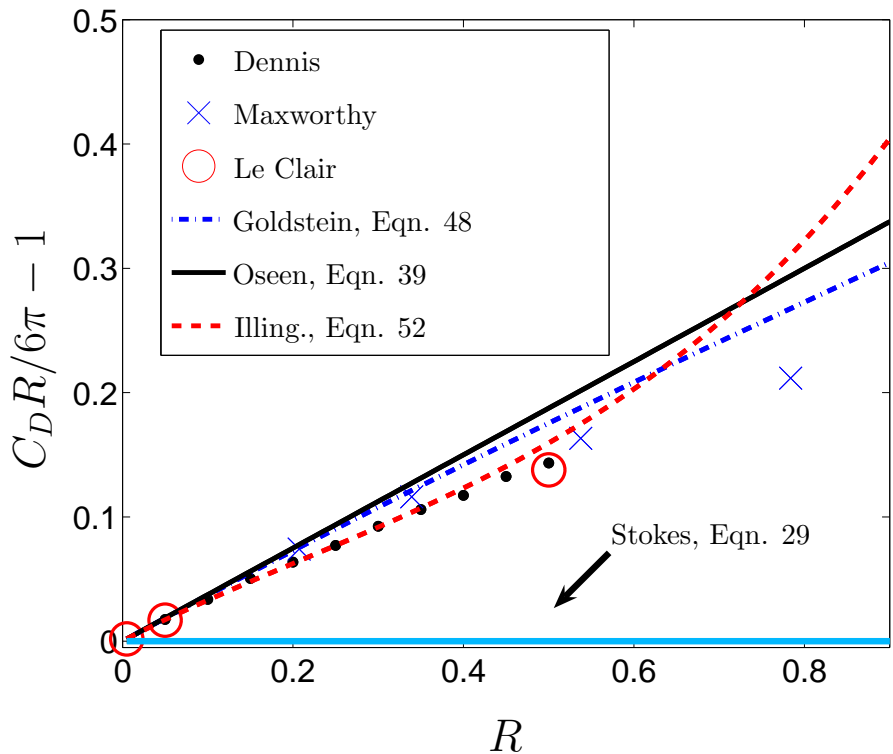}
\caption{(Color online) Drag on a sphere, experiment vs. Oseen theory
\cite{Maxworthy65,LC70,Den71}. The Stokes' solution (Eqn.
\ref{stokes:famoussol}) is shown at the bottom for reference. In these
coordinates, it is defined by the line $y=0$.} \label{OseenSpherePlot1}
\end{center}
\end{figure}

Figure \ref{OseenSpherePlot1} compares the ``predictions'' of Oseen
theory to experimental and numerical data for the drag on a sphere.
Again, Oseen's first order theory is, strictly speaking, not adequate
to make the predictions with which it is traditionally credited. The
theoretical drag coefficients are roughly valid for $R \lesssim 0.2$,
with Goldstein's solution (Eqn. \ref{GoldsteinCD}) being slightly
better than Oseen's prediction (Eqn. \ref{Oseen:dragsphere}). All are
clearly superior to Stokes' formula (Eqn. \ref{stokes:famoussol}).

Figure \ref{OseenSpherePlot1} also shows the prediction of
Illingworth's second order Oseen theory (Eqn. \ref{illingworth}). Not
surprisingly, it gives the best prediction of $C_D$, particularly when
compared to Dennis' numerical results.

%TODO: subfig? references

\begin{figure}[tb]
\psfrag{Re}{\mbox{\Large $R$}}
\psfrag{CD}{\mbox{\Large $C_D R/4\pi$}}
\psfrag{Jayaweera XXXXXX}{Jayaweera}
\psfrag{Tritton}{Tritton}
\psfrag{Imai}{Imai, Eqn. \ref{imaisol}}
\psfrag{Bairstow}{Bairstow, Eqn. \ref{Bairstowsol}}
\psfrag{Lamb}{Lamb, Eqn. \ref{Oseen:LambDrag}}
\begin{center}
\includegraphics[width=.8 \textwidth]{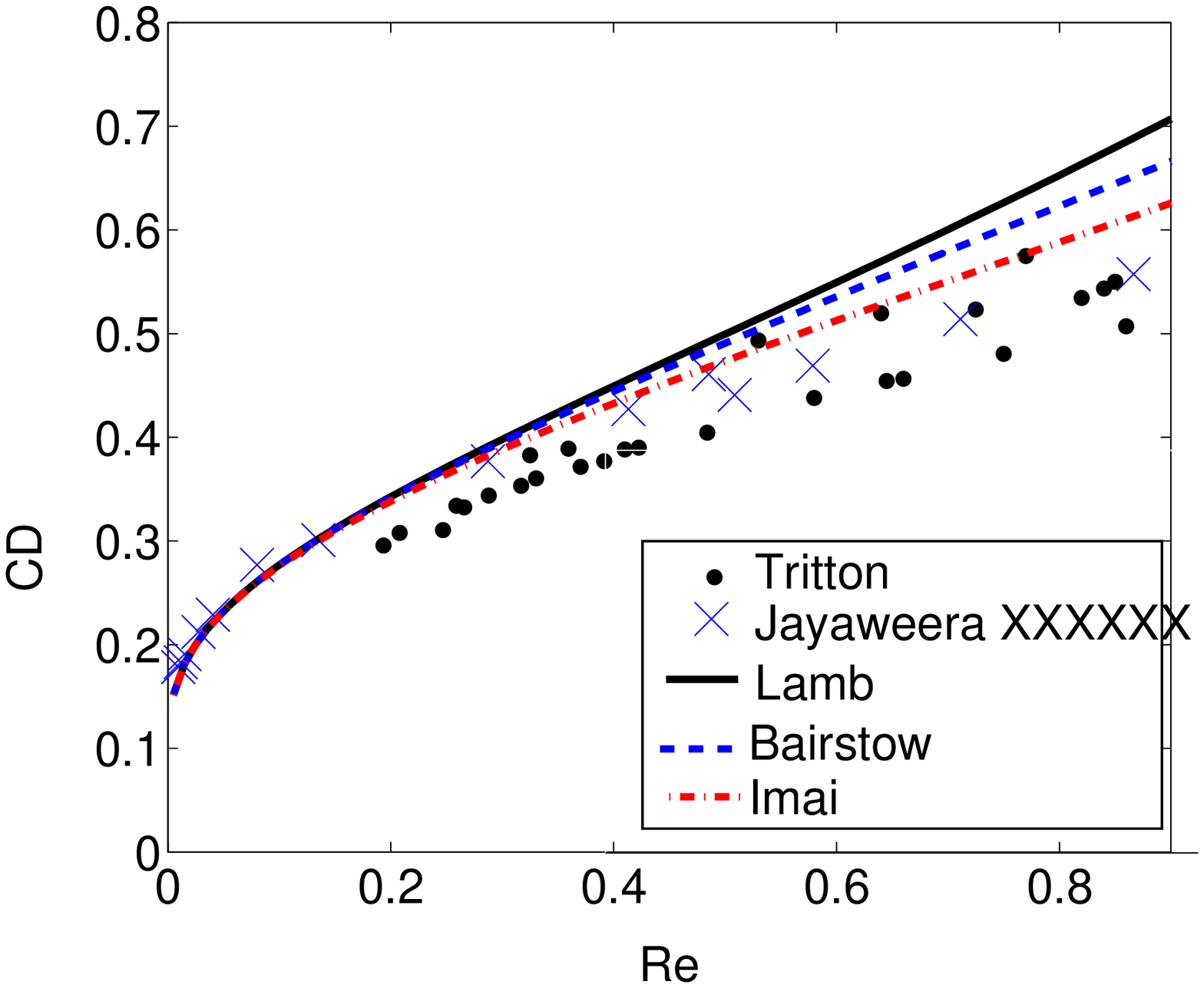}
\caption{(Color online) Drag on a cylinder, experiment vs. Oseen theory \cite{Jay65,Tritton59}.}
\label{OseenCylPlot1}
\end{center}
\end{figure}

Figure \ref{OseenCylPlot1} shows the important predictions of Oseen
theory for the drag on an infinite cylinder. As with the sphere, the
theory is only truly entitled to predict the lowest order term. Figure
\ref{OseenCylPlot1} shows decent agreement with the data. Although more
``exact'' solutions (such as Bairstow's and Imai's) do better than
Lamb's lowest order solution, this is purely coincidental. Tomotika's
solutions exhibit similar characteristics to these two solutions.

\subsubsection{Matched asymptotics}

Efforts to systematically improve Oseen's results led to the
development of \emph{matched asymptotic expansions}.\footnote{This
technique is also known as the method of \emph{inner and outer
expansions} or \emph{double asymptotic expansions}.} This branch of
applied mathematics was developed gradually, with systematic work
beginning with papers by Kaplun and Lagerstrom et al.
\cite{Lag55,Kap54}. Kaplun subsequently used these techniques to
calculate the drag on a cylinder, obtaining an entirely new result for
$C_D$ \cite{Kaplun67}. Proudman and Pearson subsequently applied
matched asymptotics to both the sphere and the cylinder, deriving a new
result for the drag on a sphere \cite{Proudman57}.
\begin{quote}
``The principle of asymptotic matching is simple. The interval on which
a boundary-value problem is posed is broken into a sequence of two or
more \emph{overlapping} subintervals. Then, on each subinterval
perturbation theory is used to obtain an asymptotic approximation to
the solution of the differential equation valid on that interval.
Finally, the matching is done by requiring that the asymptotic
approximations have the same functional form on the overlap of every
pair or intervals. This gives a sequence of asymptotic approximations
... the end result is an approximate solution to a boundary-value
problem valid over the entire interval'' \cite{BenderOrzag}.
\end{quote}

Both of the two low Reynolds number problems are attacked in similar
fashion. The problem is divided into only two regions. The first region
is near the surface of the solid body. In this region, inertial terms
are small, the approximation of Stokes ($R\approx0$) applies, and the
problem is solved perturbatively (in $R$). At each order in $R$, the
two no-slip boundary conditions at the surface are applied. One
undetermined constant remains (at each order in $R$). Loosely speaking,
it is determined by the boundary condition as $\vert \vec{r} \vert \to
\infty$. This expansion is referred to as the \emph{Stokes expansion}.

The second region is far from the sphere, where inertial terms are
important. In this region, $R\vert r\vert \sim \Order[1]{}$, and the
approximations which led to Oseen's governing equation apply. The Oseen
problem is then solved perturbatively, and the boundary condition as
$|\vec{r}| \to \infty$ is applied. There are two undetermined constants
remaining; they are related to the boundary conditions on the surface.
This perturbative expansion is referred to as the \emph{Oseen
expansion}.

The next part of this calculation is \emph{asymptotic matching}, which
determines the remaining coefficients.\footnote{At this point, there
are two unknown coefficients in the Oseen expansion, and one in the
Stokes expansion.} In this process, we expand the Oseen expansion for
small $ R |\vec{r}|$, and the Stokes expansion for large $|\vec{r}|$.
By choosing the three hitherto undetermined coefficients appropriately,
these two limiting forms are made to agree order by order in $R$. For
this to be possible, the two asymptotic functional forms must overlap.
With the coefficients determined, the two unique, locally valid
perturbative approximations are complete. If desired, they can be
combined to make a single uniformly valid approximation.

While straightforward in theory, asymptotic matching is difficult in
practice, particularly for an equation like the Navier-Stokes equation.
However, it is still far simpler than alternatives, such as iteratively
solving the Oseen equations. Van Dyke's book is an excellent
presentation of the many subtleties which arise in applying matched
asymptotics to problems in fluid mechanics \cite{VanDyke1975}. We now
present the matched asymptotic solutions for Eqns. \ref{CylinderEqn}
and \ref{SphereEqn}. These solutions result in the ``state of the art''
drag coefficients for both the sphere and the cylinder.

{\it a. Sphere}

Although Lagerstrom and Cole initially applied matched asymptotics to
the problem of the sphere, the seminal work came in an elegant 1957
paper by Proudman and Pearson \cite{Lag55, Proudman57}. Chester and
Breach extended this paper via a difficult calculation in 1969
\cite{Chester69}. We summarize the results of both papers here. These
workers used a perturbative solution in the Stokes regime of the form:
\begin{equation}
\psi(r,\mu) = \psi_0 + R \psi_1 + R^2 \log{R} \psi_{2L} + R^2 \psi_2 + R^3\log{R} + R^3\psi_3 + \Order[R]{3}
\label{PnPStokes1}
\end{equation}
This rather peculiar perturbative form cannot be determined a priori.
Rather, it arose in a fastidious incremental fashion, calculating one
term at a time. The procedure of asymptotic matching \emph{required}
including terms like $R^2\log{R}$ in the expansion; otherwise, no
matching is possible. Note that matched asymptotics gives no
explanation for the origin of these singular terms.

The first step to finding a perturbative solution in the Oseen region
is to define the \emph{Oseen variables}:
\begin{displaymath}
\rho = R r, \qquad \Psi(\rho,\mu) = R^2 \psi(r,\mu)
\end{displaymath}
Part of the reason for this transformation can be understood via the
derivation of the Oseen equation. The region where inertial effects
become important has been shown to be where $ R \vert r\vert \sim
\Order[1]{}$. Intuitively, the variable $\rho = R r$ is a natural
choice to analyze this regime, as it will be of \Order[1]{}. The
precise reasons for the selection of these variables is a technique
from boundary layer theory known as a \emph{dominant balance} argument,
which we will revisit later \cite{BenderOrzag}.

The perturbative expansion in the Oseen region takes the form:
\begin{equation}
\Psi(\rho,\mu) = \Psi_0 + R \Psi_1 + R^2 \Psi_2 + R^3\log{R}\Psi_{3L} + \Order[R]{3}
\label{PnPOseen1}
\end{equation}
Note that there is no $R^2\log{R}$ term in this expansion; none is
required to match with the Stokes expansion. As with the Stokes
expansion, this form cannot be determined a priori.

Proudman and Pearson completely solved for the Stokes' expansion
through $\mathcal{O}(R$ $\log{R})$, and partially solved for the
\Order[R]{2} term. They determined the Oseen expansion through
\Order[R]{}. Chester and Breach extended these results up to a partial
solution for \Order[R]{3} in the Stokes' expansion, and to
\Order[R^3\log{R}]{} in the Oseen expansion.

The exact form of these expansions is given in Chester and
Breach.\footnote{Note that the expression for $\psi_2$ in Proudman, is
incorrect \cite{Proudman57}. There is also a mistake in Chester and
Breach \cite{Chester69}, Eqn. 3.5; the coefficient of $c_8$ should be
$r^{-3}$ not $r^{-2}$.} Some aspects of these results have been seen
before: The leading order in the Stokes' expansion ($\psi_0$) is simply
the Stokes solution (Eqn. \ref{stokessol}). In the Oseen expansion,
$\Psi_0$ is simply the formula for the uniform stream expressed in
Oseen variables. The second term, $\Psi_1$, is the rotational part of
Oseen's solution (Eqn. \ref{Oseen:1sol}).

% The Oseen expansion has the following solutions:
% \begin{widetext}
% \begin{subequations}
% \begin{eqnarray}
% \Psi_0 &=& -\rho^2 Q_1(\mu) \\
% \Psi_1 &=& -\frac{3}{2}\left(1+\mu)\right)\left(1-e^{-\frac{1}{2}\rho\left(1-\mu\right)}\right) \\
% \Psi_2 &=& -\frac{9}{16}\left(1+\mu\right)\left(1-e^{-\frac{1}{2}\rho\left(1-\mu\right)}\right) +\rho \sqrt{1-\mu^2}A_2  \\
% &\approx&-\frac{9\left(\rho^2-4\rho\right)}{64}Q_1(\mu)+\frac{9\left(\rho^2-4\rho\right)}{64}Q_2(\mu)-\frac{\rho^2}{8}Q_1(\mu)\left(\frac{9}{5}\left(\log{\rho}+\gamma\right) + 3\log{2} - \frac{747}{200}-\frac{9}{8}\mu+\frac{129}{400}\left(5 \mu^2-1\right)\right) + \Order[\rho]{3} \\
% \Psi_{3L} &=& -\frac{27}{80}\left(1+\mu\right)\left(1-e^{-\frac{1}{2}\rho\left(1-\mu\right)}\right)
% \end{eqnarray}
% \end{subequations}
% \end{widetext}
%
% The first term, $\Psi_0$ is simply the a uniform stream, expressed in Oseen variables. The second term $\Psi_1$ is the rotational part of Oseen's solution (Eqn. \ref{Oseen:1sol}). The final term cannot be solved for in closed form; the constant $A_2$ is the magnitude of a horrendous integral (see Chester and Breach \cite{Chester69}). Only the approximate form, valid near the matching region (small $\rho$) is reproduced here.

Both sets of authors then used their result for the Stokes expansion to
calculate $C_D$, which is given in Eqn. \ref{matchedcd}.
\begin{eqnarray}
C_D &=& \frac{6 \pi}{R} \bigg(
\underbrace{1\vphantom{\bigg(}}_{\textrm{``Stokes''}}
\underbrace{\vphantom{\bigg(}+ \frac{3}{8}R}_{\textrm{``Oseen''}}
\underbrace{\vphantom{\bigg(}+ \frac{9}{40} R^2
\log{R}}_{\textrm{Proudman}} \nonumber \\ & &
\underbrace{\vphantom{\bigg(}+ \frac{9}{40}R^2\left( \gamma +
\frac{5}{3} \log{2} - \frac{323}{360} \right) + \frac{27}{80} R^3
\log{R}}_{\textrm{Chester and Breach}} + \Order[R]{3} \Bigg)
\label{matchedcd}
\end{eqnarray}
Here $\gamma$ is Euler's constant. This formula reproduces and extends
nearly all earlier work. Eqn. \ref{matchedcd} shows both the original
results of Proudman and Pearson and the higher order contributions of
Chester and Breach \cite{Chester69,Proudman57}. The ``Stokes'' term is
Stokes' original result (Eqn. \ref{stokes:famoussol}), which was
rigorously justified by Oseen. The ``Oseen'' term is generally credited
to Oseen (Eqn. \ref{Oseen:dragsphere}), although it is really beyond
the accuracy of his work, and is only justified by this
calculation.\footnote{Illingworth's unpublished result also justifies
this term.}

\begin{figure}
\psfrag{Re}{$R$}
\psfrag{CD}{$C_D R/6\pi - 1$}
\psfrag{CD2 XXXXXX}{\tiny{$C_D \frac{R}{6\pi} - 1$}}
\psfrag{Maxworthy XXXXXX}{Maxworthy \cite{Maxworthy65}}
\psfrag{Le Clair}{Le Clair \cite{LC70}}
\psfrag{Dennis}{Dennis \cite{Den71}}
\psfrag{Oseen}{Oseen (\ref{matchedcd})}
\psfrag{Proudman}{Proudman (\ref{matchedcd})}
\psfrag{Chester and Breach 2}{Chester (\ref{matchedcd})}
\begin{minipage}[t]{0.49\columnwidth}
\includegraphics[width=1 \textwidth]{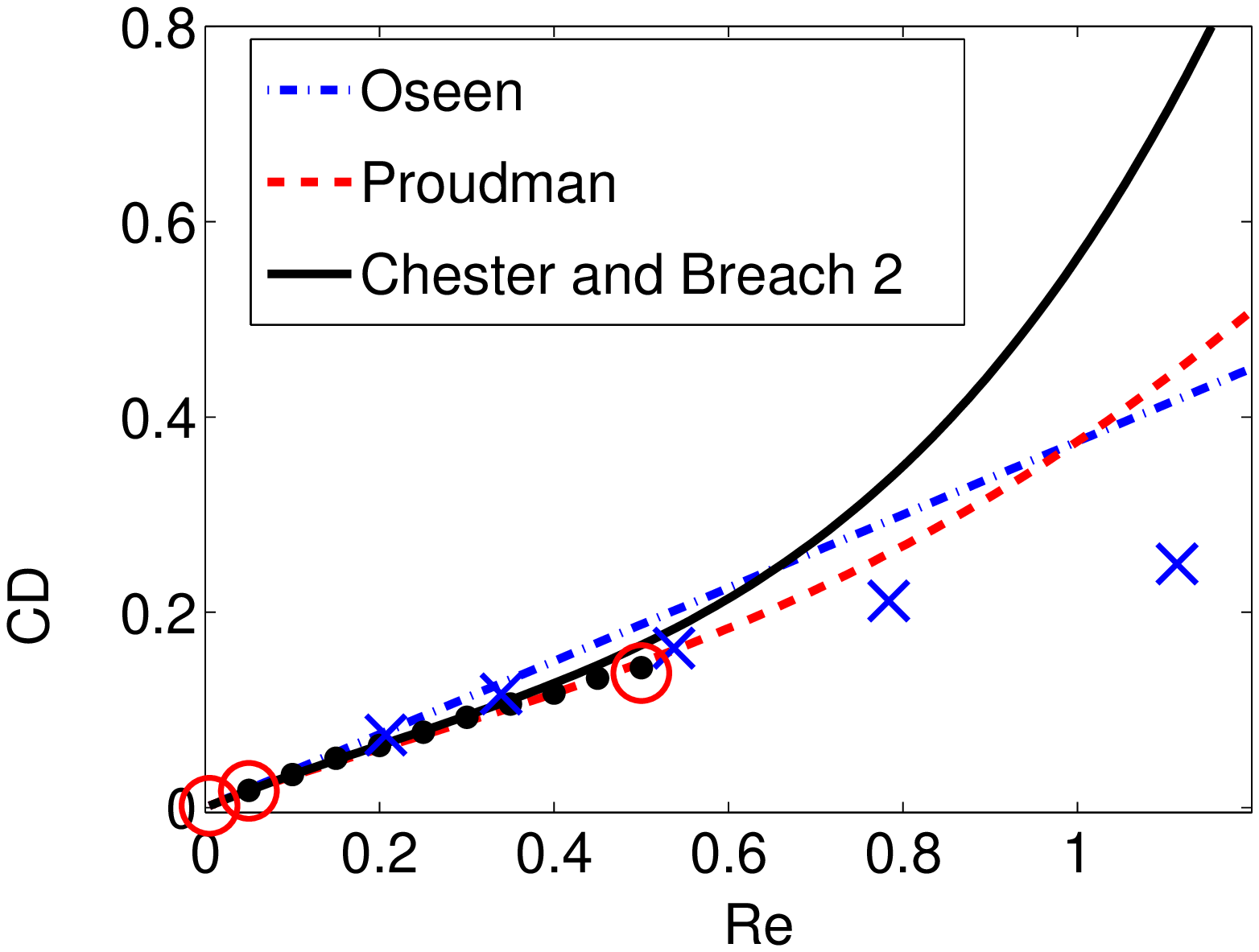}
\label{MatchedSphere1}
\end{minipage}
\begin{minipage}[t]{0.49\columnwidth}
\includegraphics[width=1 \textwidth]{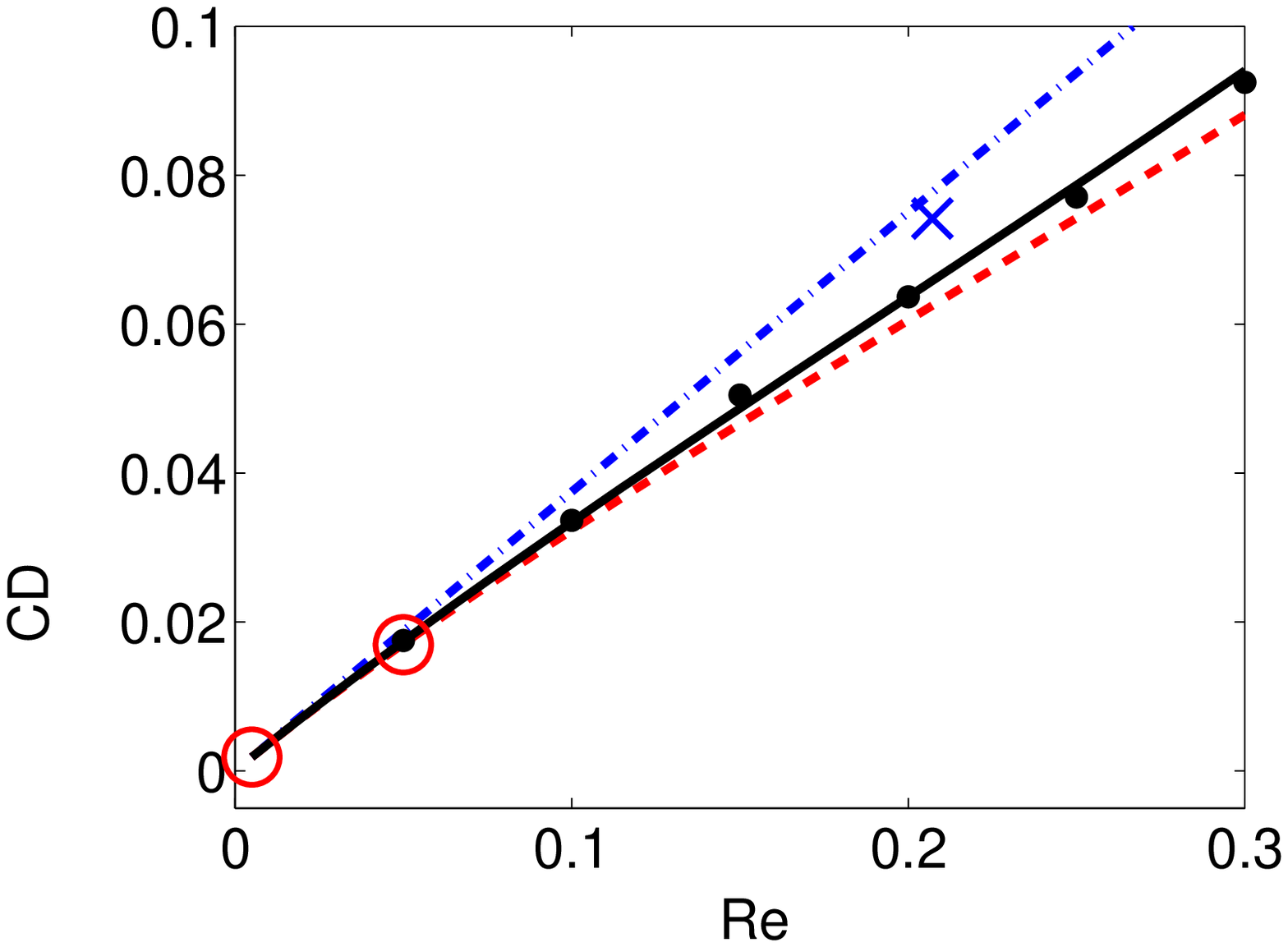}
\label{MatchedSphere2}
\end{minipage}
\caption{(Color online) Drag on a sphere, experiment vs. matched asymptotic theory.
Experimental and numerical results are plotted as in Figure
\ref{OseenSpherePlot1}.} \label{MatchedSphere0}
\end{figure}

Figure \ref{MatchedSphere0} compares the results of matched asymptotics
(Eqn. \ref{matchedcd}) with experimental data, numerical results, and
the basic prediction of Oseen's equation (Eqn. \ref{Oseen:dragsphere}).
This plot has been the source of some confusion. Maxworthy examined his
data and concluded that $C_D$ as computed by Oseen and Goldstein (Eqn.
\ref{GoldsteinCD}) were as good as any matched asymptotics predictions
\cite{Maxworthy65}. The calculations of Dennis and Le Clair, however,
refute that claim, and demonstrate the systematic improvement that
results from matched asymptotics.

Neither is it immediately clear that the extra terms in Eqn.
\ref{matchedcd} due to Chester and Breach are actually any improvement
on the work of Proudman and Pearson. Van Dyke notes, ``This result is
disappointing, because comparison with experiment suggests that the
range of applicability has scarcely been increased.''
\cite{VanDyke1975}, and Chester himself remarks that ``there is little
point in continuing the expansion further.'' At very low Reynolds
number, however, the results of Dennis ``indicate that the expression
of Chester and Breach gives a better approximation to the drag
coefficient than any other asymptotic solution until about [$R=0.3$].''
\cite{Den71} Figure \ref{MatchedSphere0} shows the excellent low $R$
agreement between Dennis' numerical results and Eqn. \ref{matchedcd}.

The prediction of matched asymptotics (Eqn. \ref{matchedcd}) is close
to Illingworth's second order Oseen theory (Eqn. \ref{illingworth}).
Close examination shows that the matched asymptotics results are
slightly closer to the Dennis' calculations in the limit of low
Reynolds number. Strictly speaking, these two theories should only be
compared as $r \rightarrow 0$, and in this regime matched asymptotics
is superior. This is not surprising, as the best matched asymptotic
calculation is a higher order approximation than that of Illingworth.

{\it b. Cylinder}
\label{matchedcylinder}

In 1957, Kaplun applied matched asymptotics to the problem of the
cylinder, and produced the first new result for $C_D$ \cite{Kaplun57c}.
Additional stream function calculations (but without a drag
coefficient) were done by Proudman and Pearson \cite{Proudman57}.
Kaplun's calculations were extended to higher order by Skinner, whose
work also explored the structure of the asymptotic expansions
\cite{Ski75}. We summarize results from all three papers here.

Near the surface of the cylinder, the Stokes' expansion applies, and
the perturbative solution takes the following form.
\begin{equation}
\psi(r,\theta)=\psi_0(r,\theta,\delta) + R \psi_2(r,\theta,\delta) + R^2 \psi_3(r,\theta,\delta) + \Order[R]{3}
\label{matchedcylstokes}
\end{equation}
Here, $\delta = \delta(R)$ is defined as in Eqn. \ref{Oseen:Lambcyl1}.
What is remarkable about the structure of this expansion is its
dependence on $\delta$. To be precise, each function $\psi_n$ is
actually another perturbative expansion, in $\delta$:
\begin{equation}
\psi_n(r,\theta,\delta) = \delta F_{n,1}(r,\theta) + \delta^2 F_{n,2}(r,\theta) + \Order[\delta]{3}
\label{form1}
\end{equation}
This formulation is equivalent to an asymptotic expansion in terms of
$\log{R}^{-1}$, which is used by Proudman and Pearson:
\begin{equation}
\psi_n(r,\theta,\log{R}) = \frac{\tilde{F}_{n,1}(r,\theta)}{(\log{R})^1} + \frac{\tilde{F}_{n,2}(r,\theta)}{(\log{R})^2} + \Order[\frac{1}{(\log{R})^3}]{}
\label{form2}
\end{equation}

This form is much less efficient than that given in Eqn. \ref{form1},
in the sense that more terms in the Stokes and Oseen expansions are
needed to obtain a given number of terms in $C_D$. For that reason,
expansions in $\delta$ are used here.

This curious asymptotic form is necessitated by matching requirements.
It is also the source of a number of bizarre complications. The first
implication is that \emph{all} terms in Eqn. \ref{matchedcylstokes} of
\Order[R]{} and higher will be transcendentally smaller than \emph{any}
of the terms in the expansion for $\psi_0$. This is true
asymptotically, as $R \to 0$. The reason for this is that inertial
terms \emph{never} enter into any of the governing equations for the
Stokes' expansion; they enter only through the matching process with
the Oseen expansion.

As with the sphere, the first step to finding a perturbative solution
in the Oseen region is to transform into the relevant Oseen variables.
In this case,
\begin{equation}
\rho = R r, \qquad \Psi(\rho,\mu) = R \psi(r,\mu)
\end{equation}
The perturbative expansion which can solve the problem in the Oseen
region has the same generic form as Eqn. \ref{matchedcylstokes}.
\begin{equation}
\Psi(\rho,\theta)=\Psi_0(\rho,\theta,\delta) + R \Psi_1(\rho,\theta,\delta) + \Order[R]{2}
\label{matchedcyloseen}
\end{equation}
The functions $\Psi_n(\rho,\theta,\delta)$ can also be expressed as a
series in $\delta(R)$. However, the formula cannot be written down as
conveniently as it could in Eqn. \ref{form1}. The first two terms take
the forms given in Eqn. \ref{eqn:form2}.
\begin{subequations}
\begin{eqnarray}
\Psi_0(\rho,\theta,\delta) &=& F_{0,0}(\rho,\theta) + \delta F_{0,1}(\rho,\theta) + \Order[\delta]{2} \\
\Psi_1(\rho,\theta,\delta) &=& \delta^{-1} F_{1,-1}(\rho,\theta) + F_{0,0}(\rho,\theta) + \Order[\delta]{}
\end{eqnarray}
\label{eqn:form2}
\end{subequations}

Kaplun and Proudman both considered only terms of \Order[R]{0} in the
Stokes' expansion. As $R \to 0$, this is an excellent approximation, as
all higher terms are transcendentally smaller. In this limit, the
Stokes expansion takes a particularly simple form:
\begin{equation}
\psi(r,\theta) = \psi_1(r,\theta,\delta) = \sum_{n=1}^\infty a_n \delta^n \left(2 r \log{r} -r + \frac{1}{r} \right)\sin{\theta}
\end{equation}
Kaplun obtained terms up to and including $n=3$. Proudman et al. also
obtained expressions for the Oseen expansion, albeit expressed as a
series in $\log{R}^{-1}$. Skinner extended Kaplun's Stokes expansion to
include terms up to \Order[\delta]{3}, \Order[R\delta]{}, and
\Order[R^2\delta]{} \cite{Ski75}. He obtained approximate solutions for
the Oseen expansion, including terms up to \Order[\delta]{} and
\Order[R]{}. The lowest order solutions in the Oseen expansion are
related to the expression for a uniform stream and the solution of Lamb
(Eqn. \ref{Oseen:Lambcyl1}).

Using his solution, Kaplun computed a new result for the drag
coefficient (Eqn. \ref{cylinder:KaplunCD}) which agrees with Lamb's
result (Eqn. \ref{Oseen:LambDrag}) at lowest order.
\begin{equation}
\label{cylinder:KaplunCD}
C_D =\frac{4 \pi}{R} \left( \delta - k \delta^3\right)
\end{equation}
Here, $k = \int_0^{\infty} K_0(x) K_1(x) \left( x^{-1} I_1(2 x) - 4
K_1(x)I_1(x)+1\right) \textrm{d}x \approx 0.87$. Skinner extended these
results, showed that terms of \Order[R]{0} do not contribute to the
drag, and calculated the first transcendentally smaller contribution,
which is of \Order[R]{1}. His result is given in Eqn.
\ref{cylinder:SkinnerCD}.
\begin{equation}
\label{cylinder:SkinnerCD} C_D=\frac{4 \pi}{R} \left( \delta - k
\delta^3 + \Order[\delta]{4} - \frac{R^2}{32} \left(1 -
\frac{\delta}{2} + \Order[\delta]{2} \right) + \Order[R]{4} \right)
\end{equation}
The value of these new terms is questionable, and Skinner himself noted
that they are likely negligible in comparison to the neglected terms of
\Order[\delta]{4}. Asymptotically this is unequivocally true.

\begin{figure}[tb]
\psfrag{Re}{\mbox{\large $R$}}
\psfrag{CD}{\mbox{\large $C_D R/4\pi$}}
\psfrag{Jayaweera XXXXXX}{Jayaweera}
\psfrag{Tritton}{Tritton}
\psfrag{Kaplun}{Kaplun, Eqn. \ref{cylinder:KaplunCD}}
\psfrag{Skinner}{Skinner, Eqn. \ref{cylinder:SkinnerCD}}
\psfrag{Lamb}{Lamb, Eqn. \ref{Oseen:LambDrag}}
\begin{center}
\includegraphics[width=.8 \textwidth]{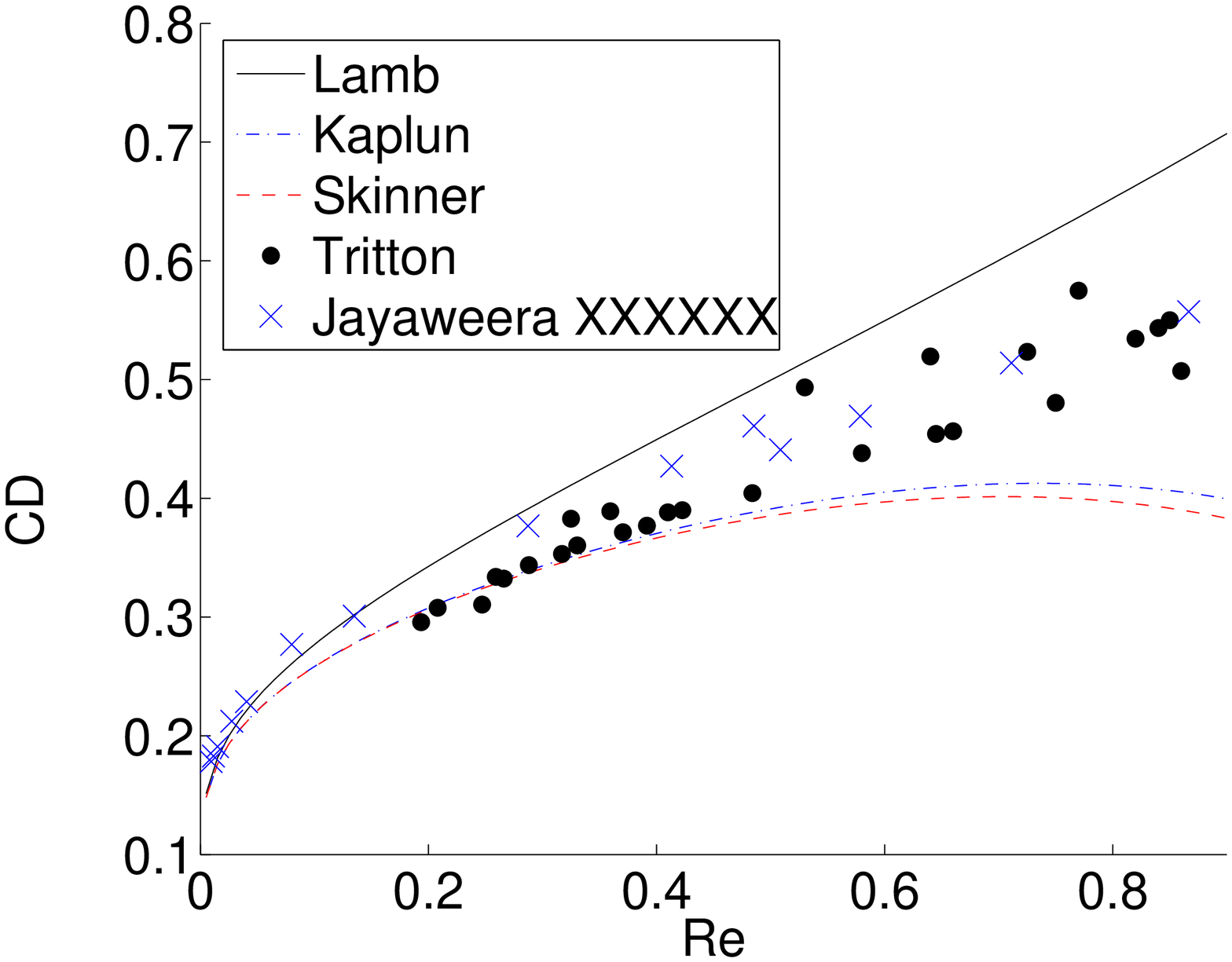}
\caption{(Color online) Drag on a cylinder, experiment vs. matched asymptotic theory \cite{Jay65, Tritton59}.}
\label{MatchedCylPlot1}
\end{center}
\end{figure}

Figure \ref{MatchedCylPlot1} compares the predictions of matched
asymptotic theory with Lamb's result (Eqn. \ref{Oseen:LambDrag}) based
on Oseen's equation. Although both theories agree as $r \rightarrow 0$,
matched asymptotic results seem no better than Lamb's solution. The
comparison is further complicated by the scatter in the different
experiments; matched asymptotics agree more closely with Tritton's
measurements, while Lamb's solution agrees better with Jayaweera's. We
draw two conclusions from Figure \ref{MatchedCylPlot1}: Both theories
break down for $R\gtrsim 0.1$ and neither theory is demonstrably more
accurate. Even more disappointingly, Skinner's result is nowhere better
than Kaplun's --- it is actually worse at higher Reynolds numbers.

Part of the problem with the matched asymptotics approach arises from
the need for two expansions, in $\delta$ and $R$. Because infinitely
many orders of $\delta$ are needed before \emph{any} higher orders in
$R$ are relevant means that infinitely many terms in the Oseen
expansion must be calculated before the second order term in the Stokes
expansion. This is  inefficient, and is the reason for Skinner's lack
of success.

A recent paper by Keller et al. solved this problem numerically
\cite{Kel96}. They developed a method to sum \emph{all} of the orders
of $\delta$ for the first two orders of $R$. Their ``beyond all
orders'' numerical results prove the importance of these higher order
terms. When such terms are accounted for, the resulting $C_D$ is vastly
improved from Kaplun's, and is superior to any of the analytic
solutions discussed here. Interestingly, it seems to agree very well
with the experiments of Tritton, although it is difficult to tell from
the plot in their paper, which does not remove the leading order
divergence.

\subsubsection{Other theories}

Amongst the community interested in settling velocities and
sedimentation, there are many theoretical models of the drag on a
sphere. These workers specify $C_D$ as a function of $R$ by means of a
``sphere drag correlation.'' An overview of these formula is given by
Brown \cite{Bro03}. These results are generally semi-empirical, relying
on a blend of theoretical calculations and phenomenologically fit
parameters to predict $C_D$ over a large range of Reynolds number.
While practically useful, these results are not specific to low
Reynolds numbers, and cannot be derived from the Navier-Stokes
equations. They address a different problem, and will not be further
considered here.

One other semi-empirical theory is due to Carrier \cite{Car53}. He
argued that the inertial corrections in the Oseen equation were
over-weighted, and multiplied them by a coefficient which he
constrained to be between $0$ and $1$. Consequently, his theory is in
some sense ``in between'' that of Stokes and that of Oseen. He
ultimately determined this coefficient empirically.

\subsubsection{Terminology}
\label{section:terminology}

Confusing terminology, particularly in the matched asymptotics
literature, riddles the history of these problems. We previously
detailed discrepancies in the definition of $C_D$. In this section we
explain the sometimes conflicting terms used in the matched asymptotics
literature, introduce a convention which eliminates confusion, and also
explain how some authors adopt different definitions of the Reynolds'
number.

Matched asymptotics literature discusses numerous perturbative
expansions, each of which are valid in a different regime, or ``domain
of validity.'' Different authors use different labels for these
expansions. Most workers define the ``inner'' expansion to be the
expansion which is valid inside the \emph{boundary layer}
\cite{BenderOrzag}. A boundary layer is a region of rapid variation in
the solution. The ``outer'' expansion is valid outside of the boundary
layer, where the solution is slowly varying \cite{BenderOrzag}.
Problems with multiple boundary layers require additional terminology.
The outer expansion is based ``upon the primary reference quantities in
the problem,'' and the inner expansion is usually obtained by
stretching the original variables by dimensionless functions of the
perturbation parameter \cite{VanDyke1975}. The appropriate stretching,
or scaling functions are obtained through a \emph{dominant balance}
analysis, which can be difficult. After this rescaling, the argument of
the inner expansion will be of $\mathcal{O}(1)$ inside the boundary
layer. Accompanying these inner and outer expansions are ``inner
variables'', ``outer variables'', ``inner limits'', and ``outer
limits''.

The low Reynolds number flow problems are complicated by the fact that
some authors, including Van Dyke, also define expansions on the basis
of their physical location \cite{VanDyke1975}. The ``outer'' limit is
valid far from the solid body ($|\vec{r}|$ is large), and the ``inner''
limit is valid near the surface of the body ($|\vec{r}| \approx 1$).

This is consistent with yet another definition, based on proximity to
the origin of the chosen coordinate system. In a review paper
Lagerstrom and Casten define the ``inner limit'' as being ``valid near
the origin,'' and the ``outer limit'' as being ``valid except near the
origin.'' \cite{Lag72} Part of their motivation for this new definition
was to distinguish between the \emph{domain of validity} of an
expansion, and the limit process by which it is obtained.

Finally, Kaplun refers to the inner and outer limits based on their
correspondence to high Reynolds number flow \cite{Kaplun57c}. He
identifies the Stokes' approximation as the ``inner'' limit, and
Oseen's equation as the ``outer'' limit.

Part of the confusion arises because of disagreements over the location
of the boundary layer. Van Dyke claims that ``it is the neighborhood of
the point at infinity'', while Kaplun argues that the boundary layer is
near the surface. Definitions referenced to the boundary layer disagree
when there are disagreements about its location.

To eliminate this confusion, a preferable alternative notation has
emerged from subsequent work\cite{Proudman57,Kaplun57a}. We follow this
notation, defining the``Oseen'' and ``Stokes'' expansions, which were used
in the previoussection. The Oseen expansion is valid far from the surface,
and is expressed in stretched coordinates. The Stokes limit is valid near
the surface of the sphere, where $r$ is small, and is expressed in the
original variables.\footnote{Van Dyke's book is not consistent in
relating ``inner'' and ``outer'' expansions to the Stokes and Oseen
expansions.}

Matched asymptotics workers also discuss \emph{uniform approximations},
\emph{intermediate} expansions, or \emph{composite} expansions
\cite{BenderOrzag, Kaplun67, VanDyke1975}. The basic idea is that the
Stokes and Oseen expansions can be blended together to form a single
expression which is valid everywhere. This result reduces to the two
original expansions when expanded asymptotically in the two limits. How
to calculate a uniform expansion is discused below.

There are also minor differences in the definition of the Reynolds
number, $R$. Some authors define $R$ based on the diameter of the
solid, while others base it on the radius. This factor of $2$ can be
difficult to track. We define the Reynolds number using the
\emph{radius} of the fixed body: $R  = |\vec{u}_{\infty}| a/\nu$. It is
worth noting that Kaplun \cite{Kaplun67}, Tomotika \cite{Tomotika50},
Goldstein \cite{Goldstein38}, Liebster \cite{Liebster26}, Thom
\cite{Tho33} and Tritton \cite{Tritton59} all use the \emph{diameter}.

%
% \section{Open Problems}
%
% \subsection{First Edy Problem}
%
% Damn well ought to be solved with a uniform expansion.

\subsection{Uniformly valid approximations}
\label{section:uniform}

As mentioned previously, the inner and outer expansions may be combined
into a single, \emph{uniformly valid} approximation, which is
applicable everywhere. For a function of one variable, the uniform
approximation is constructed as in Eqn. \ref{benderuniform}
\cite{BenderOrzag}.
\begin{equation}
\label{benderuniform}
y_{\textrm{uniform}}(x) = y_{\textrm{outer}}(x) + y_{\textrm{inner}}(x) -
y_{\textrm{overlap}}(x)
\end{equation}
$y_{\textrm{overlap}}(x)$ consists of the common ``matching'' terms
between the inner and outer expansions.

Kaplun demonstrates that $y_{\textrm{uniform}}(x) \to y(x)$ as the
expansion variable $R \to 0$, i.e. the uniform approximation tends to
the exact solution everywhere \cite{Kaplun67}. To be more precise, if
the matched asymptotics solution is constructed to \Order[R]{1}, then
\begin{equation*}
\lim_{R\rightarrow 0} y(x) - y_{\textrm{uniform}}(x) \sim \Order[R]{1}
\end{equation*}

As a matter of practice, calculating the uniform solution is
mechanistic. First, express the inner and outer expansions in the same
coordinates; in our case, express the Oseen expansion in Stokes
variables.\footnote{Note that this transformation affects both the
radial coordinates and the stream function, and that it differs for the
sphere and cylinder.} Alternatively, one can express the Stokes
expansion in Oseen variables. Next, express both solutions as a power
series in the expansion parameter, $R$. By construction the Stokes
expansion is already in this form  but the transformed Oseen expansion
is not, and must be expanded to the same power in $R$ as the Stokes
solution.

From these two power series we can identify the  ``overlap'' function,
$y_{\textrm{overlap}}$. This function consists of the terms which are
in common between the two expansions, and is usually obtained by
inspection. Of course,  $y_{\textrm{overlap}}$ is only valid to the
same order as the original matched asymptotics solution, and higher
order terms should be discarded. The uniformly valid approximation is
then obtained using $y_{\textrm{overlap}}$ and Eqn.
\ref{benderuniform}.

\subsubsection{The correct way to calculate $C_D$}

Proudman and Pearson argue that ``uniformly valid approximations
\emph{per se} are not usually of much physical interest ... In the
present problem, for instance, it is the Stokes expansion that gives
virtually all the physically interesting information.''
\cite{Proudman57} All matched asymptotics calculations are based solely
on the Stokes expansion, and are therefore influenced by the Oseen
expansion only via the boundary conditions. For instance, the drag
coefficient is calculated using only the Stokes' expansion. Other
properties of the stream function, such as the size of the dead water
wake directly behind the sphere or cylinder, are also calculated using
the Stokes' expansion.

In this section we argue that this approach is incorrect, and that
uniformly valid approximation should be used to calculate all
quantities of interest. By adopting this viewpoint, we obtain new
results for $C_D$, and demonstrate that these drag coefficients
systematically improve on previous matched asymptotics results.

Matched asymptotics workers argue that the drag coefficient is
calculated at the surface of the solid (Eqns. \ref{CDsphere},
\ref{CDcylinder}), where $r=1$. Since the Oseen solution applies for
large $r$, the Stokes solution applies for small $r$, and the Stokes
solution ought to be used to calculate $C_D$. In fact, by construction,
any uniformly valid approximation must reduce to the Stokes expansion
in the limit as $R r \to 0$.

Curiously, proponents of the Oseen equation argue conversely
\cite{Happel73,Faxen27}. They claim that because the Oseen expansion
\emph{happens} to apply everywhere, it should be used to calculate all
sorts of quantities of interest, including $C_D$. In fact, Hapel and
Brenner wrote a book essentially devoted to this premise
\cite{Happel73}. In fairness, it must be mentioned that all of these
authors were well aware of their choices, and motivated their approach
pragmatically: They obtained useful solutions to otherwise intractable
problems.

In reality, both approaches converge to the exact solution for suitably
small Reynolds' numbers. However, for small but non-infinitesimal $R$,
the best estimate of derivative quantities such as $C_D$ is obtained
not by using the Stokes expansion, but by using a uniformly valid
approximation calculated with both the Stokes and Oseen expansions.
Such a drag coefficient must agree with results derived from the Stokes
expansion as $R r \to 0$, and it can \emph{never} be inferior.
Moreover, this approach makes determination of the drag coefficient's
accuracy straightforward; it is determined solely by the accuracy of
the uniform expansion, without any need to be concerned about its
domain of applicability.

We now calculate the drag coefficients for both the sphere and the
cylinder using uniformly valid approximations, using previously
published inner and outer expansions. These corrections are small but
methodologically noteworthy, and are absent from the existing
literature.

{\it a. Cylinder}

Although the state-of-the-art matched asymptotics solutions are due to
Kaplun, it is more convenient to use stream functions \cite{Kaplun57c}.
Skinner conveniently combines previous work, providing a concise
summary of Stokes and Oseen stream functions \cite{Ski75}. We base our
derivation of a uniformly valid approximation on the results in his
paper. The Stokes expansion is given by Eqn. \ref{cylinder:stokes}
\cite{Ski75}.
\begin{equation}
\label{cylinder:stokes}
\psi(r,\theta) = \frac{1}{2}\left(\delta - k \delta^3 +
\Order[\delta]{4} \right) \left(2 r \log{r} - r + \frac{1}{r} \right) \sin{\theta}
+ \Order[R]{1}
\end{equation}
The Oseen Expansion is given by Eqn. \ref{cylinder:oseenU}.
\begin{equation}
\label{cylinder:oseenU}
\Psi(\rho,\theta) = \left( \rho \sin{\theta} - \delta \sum_{n=1}^\infty \phi_n \left(\frac{\rho}{2}\right) \frac{\rho}{n} \sin{n \theta} + \Order[\delta]{2} + \Order[R]{1} \right)
\end{equation}

% Where $K_i$, $I_i$ are the usual modified Bessel functions.
With these results, creating the uniform approximation and calculating
$C_D$ is straightforward. The only subtlety is the sine series in Eqn.
\ref{cylinder:oseenU}. However, Eqn. \ref{cylinder:convenientdrag}
tells us that, for the purposes of calculating the drag, only the
coefficient of $\sin{\theta}$ matters. We calculate the overlap between
the two functions by expanding Eqn. \ref{cylinder:oseenU} about
$\rho=0$. The result is given by Eqn. \ref{cylinder:overlap}.
\begin{equation}
\label{cylinder:overlap}
\psi_{\textrm{overlap}}(r,\theta) = \delta \frac{r}{2} \left(2 \log{r} - 1\right) \sin{\theta} +
  \Order[\delta]{2} + \Order[R]{1}
\end{equation}
Combining this with the Oseen and Stokes expansions, we obtain the
uniformly valid approximation given by Eqn. \ref{cylinder:uniform}.
\begin{eqnarray}
\label{cylinder:uniform}
\psi_{\textrm{uniform}}(r,\theta) &=& \left(r + \delta \left(\frac{1}{2r} - r \phi_1(\frac{r R}{2}) \right) + k \delta^3 \left( \frac{r}{2} - r \log{r} - \frac{1}{2r} \right) \right) \sin{\theta} - \nonumber \\
& &\delta \sum_{n=2}^\infty \phi_n \left(\frac{R r}{2}\right) \frac{r}{n} \sin{n \theta} +
  \Order[\delta]{2} + \Order[R]{1}
\end{eqnarray}
By substituting this result into Eqn. \ref{cylinder:convenientdrag}, we
obtain a new result for $C_D$:
%\begin{equation}
%\label{cylinder:uniformCD}
%C_D = \frac{\pi \delta}{32 R} \left(96 - 128 k \delta^2 + R \left( R
%\left( 2 R I_1\left(\frac{R}{2}\right) + 3 R I_3\left(\frac{R}{2}\right) - 2
%I_4\left(\frac{R}{2}\right) \right) K_0\left(\frac{R}{2}\right) -2 \left(24
%I_2\left(\frac{R}{2}\right) + R \left(6 I_3\left(\frac{R}{2}\right) + R
%I_4\left(\frac{R}{2}\right) \right)\right)
%K_1\left(\frac{R}{2}\right)\right)+R^2 \left(-5 R I_1\left(\frac{R}{2}\right)+16
%I_2\left(\frac{R}{2}\right) \right) K_2\left(\frac{R}{2}\right) + 2 R^2
%I_0\left(\frac{R}{2}\right) \left( 3 K_0\left(\frac{R}{2}\right)-6
%K_2\left(\frac{R}{2}\right) + R K_3\left(\frac{R}{2}\right) \right) \right)
%\end{equation}
%\end{widetext}
\begin{equation}
\label{cylinder:uniformCD}
C_D = \frac{\pi \delta \left( 24 - 32 k \delta^3 + 6 R^2
  \phi_1^{''}(R/2) + R^3 \phi_1^{'''}(R/2) \right)}{8 R}
\end{equation}

\begin{figure}[tb]
\psfrag{Re}{\mbox{\large $R$}}
\psfrag{CD}{\mbox{\large $C_D R/4\pi$}}
\psfrag{Jayaweera XXXXXX}{Jayaweera}
\psfrag{Tritton}{Tritton}
\psfrag{Kaplun}{Kaplun, Eqn. \ref{cylinder:KaplunCD}}
\psfrag{Uniform CD}{Eqn. \ref{cylinder:uniformCD}}
\begin{center}
\includegraphics[width=.8 \textwidth]{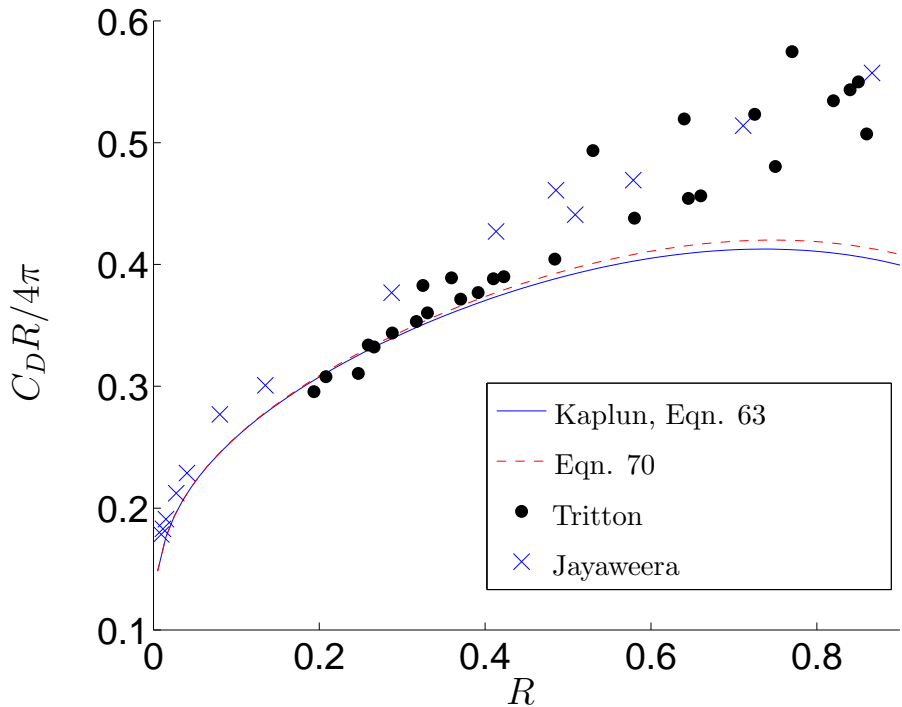}
\caption{(Color online) Drag on a cylinder, comparing a uniformly valid calculations
and matched asymptotics results\cite{Jay65,Tritton59}.}
\label{fig:cylinder:uniform}
\end{center}
\end{figure}

Fig. \ref{fig:cylinder:uniform} compares Eqn. \ref{cylinder:uniformCD}
with Kaplun's usual result (Eqn. \ref{cylinder:KaplunCD}). The new drag
coefficient (Eqn. \ref{cylinder:uniformCD}) is a small but
\emph{systematic} improvement over the results of Kaplun. Because they
are asymptotically identical up to \Order[\delta]{4} and \Order[R]{},
they agree as $R \rightarrow 0$. However, at small but
non-infinitesimal $R$, our new result is superior. Comparing Figures
\ref{fig:cylinder:uniform} and \ref{MatchedCylPlot1}, we can also see a
second surprise: The new result betters Skinner's $C_D$, even though
they were based on the same stream functions. If Skinner had used a
uniformly valid approximation, his result would not have misleadingly
appeared inferior to Kaplun's.

{\it b. Sphere}

As with the cylinder, calculating $C_D$ from a uniformly valid
expansion yields an improved result. However, there is a substantial
difference in this case. Although matched asymptotics calculations have
been done through \Order[R]{3} in Eqn. \ref{PnPStokes1} and
\Order[R^3\log{R}]{} in Eqn. \ref{PnPOseen1},  the higher order terms
in the Oseen expansion are impossible to express in a simple analytic
form. Asymptotic expressions exist (and have been used for matching),
but these cannot be used to construct a uniformly valid expansion.
Consequently, we can only compute the uniform expansion through
\Order[R]{}, and its predictions can only be meaningfully compared to
the first two terms in Eqn. \ref{matchedcd}.

The solutions for the Stokes and Oseen expansions are given in Chester
and Breach, and are quoted here \cite{Chester69}. The Stokes expansion:
\begin{eqnarray}
\psi(r,\mu) &=&-\frac{1}{2}\left(2r^2-3r+\frac{1}{r}\right)Q_1(\mu) - R \frac{3}{16} \Bigg( \left(2r^2-3r+\frac{1}{r} \right) Q_1(\mu) - \nonumber \\
 & &\left(2r^2-3r+1-\frac{1}{r}+\frac{1}{r^2}\right) Q_2(\mu) \Bigg) + \Order[R^2\log{R}]{}
\label{sphere:unif:stokes}
\end{eqnarray}
The Oseen expansion:
\begin{equation}
\Psi(\rho,\mu)=-\rho^2Q_1(\mu) - R \frac{3}{2}
\left(1+\mu\right)\left(1-e^{-\frac{1}{2}\rho\left(1-\mu\right)}
\right) + \Order[R]{2} \label{sphere:unif:oseen}
\end{equation}
By taking the $\rho \to 0$ limit of Eqn. \ref{sphere:unif:oseen}, we
can calculate the overlap between these two expansions. The result is
given in Eqn. \ref{sphere:unif:overlap}.
\begin{equation}
\label{sphere:unif:overlap} \psi_{\textrm{overlap}}(r,\mu) =
\frac{r}{8}\left(12 - 8 r \right)Q_1(\mu) + \frac{r R}{8}\left(3 r
Q_2(\mu) - 3 r Q_1(\mu) \right) + \Order[R]{2}
\end{equation}
Eqns. \ref{sphere:unif:overlap}, \ref{sphere:unif:oseen}, and
\ref{sphere:unif:stokes} can be combined to form a uniformly valid
approximation:
\begin{equation}
\psi_{\textrm{uniform}}(r,\mu)=\psi(r,\mu)-\psi_{\textrm{overlap}}(r,\mu)
+\frac{\Psi(r R,\mu)}{R^2} + \Order[R^2\log{R}]{}
\label{sphere:uniform}
\end{equation}

Due to the $e^{-\frac{1}{2}\rho\left(1-\mu\right)}$ term, we cannot use
the simple expression for $C_D$ (Eqn. \ref{sphere:simpledrag}).
Instead, we must use the full set of Eqns. \ref{spherestreamfunction},
\ref{CDsphere}, and \ref{sphpressure}. After completing this procedure,
we obtain a new result for $C_D$, given by Eqn. \ref{sphere:uniformcd}.
\begin{eqnarray}
\label{sphere:uniformcd} C_D&=&\frac{6 \pi}{R} \Bigg( \frac{e^{-2
R}}{320 R^3} \Big( 40 e^{R} \left(1728+1140R + 335R^2+56R^3+6R^4
\right) - 60 R \left(1+R\right) \nonumber \\ & & + e^{2 R} \big(-69120
+ 23580 R -2420 R^2 +20 (10 + \pi) R^3 + 10 (18-\pi)R^4 - 8 R^5
\nonumber \\ & & -3R^6 \big) \Big) - \frac{e^{-R/2} \pi I_1(R/2)}{4 R}
\Bigg) + \Order[R]{1}
\end{eqnarray}

\begin{figure}[tb]
\psfrag{Re}{\mbox{\large $R$}}
\psfrag{CD}{\mbox{\large $C_D R/6\pi - 1$}}
\psfrag{Maxworthy XXXXXX}{Maxworthy}
\psfrag{Le Clair}{Le Clair}
\psfrag{Dennis}{Dennis}
\psfrag{Oseen}{Eqn. \ref{matchedcd}, 2 terms}
\psfrag{Chester and Breach 2}{Eqn. \ref{matchedcd}, 5 terms}
\psfrag{Uniform}{Eqn. \ref{sphere:uniformcd}}
\begin{center}
\includegraphics[width=.8 \textwidth]{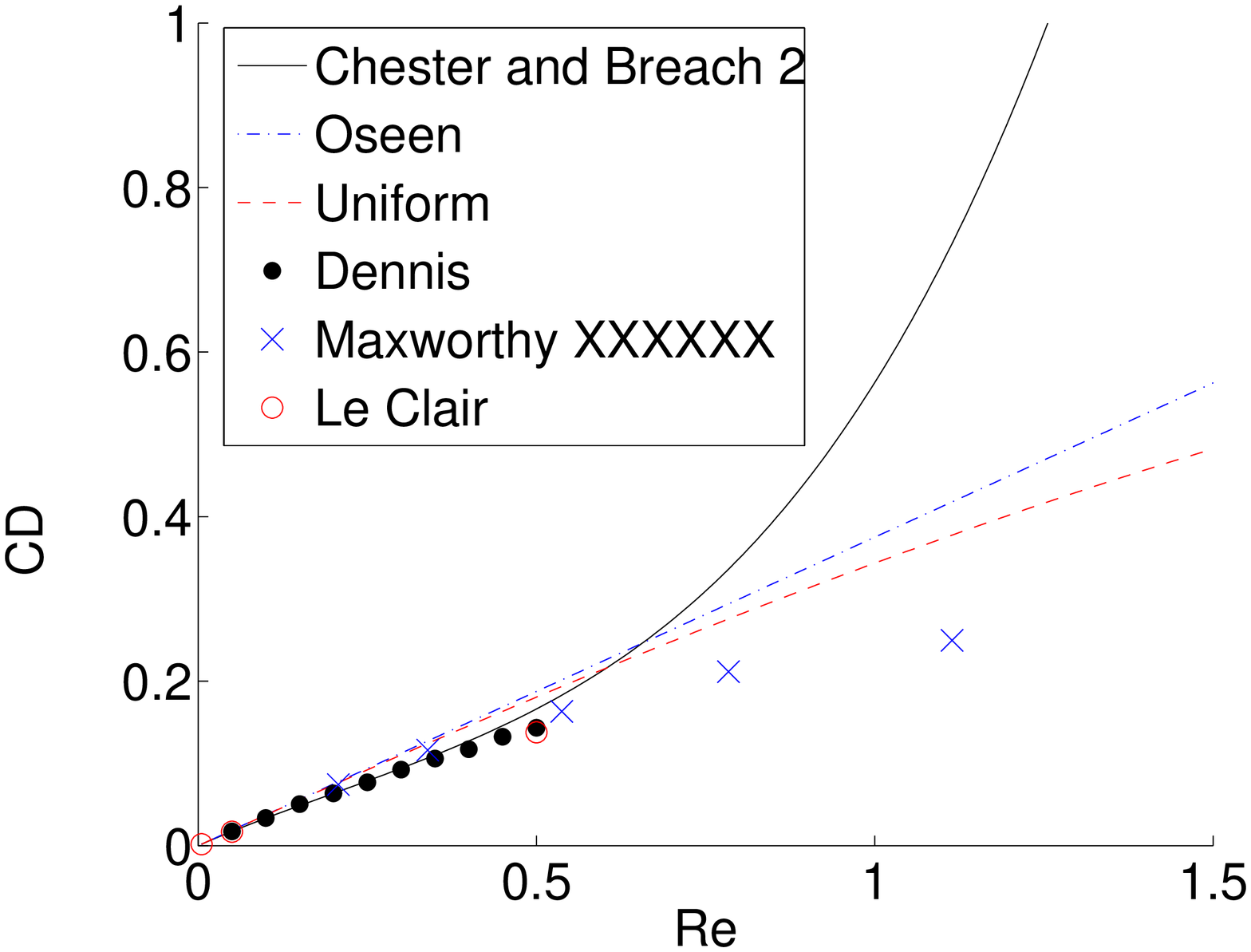}
\caption{(Color online) Drag on a sphere, experiment vs. theory \cite{Maxworthy65,LC70,Den71}.}
\label{UniformSphere1}
\end{center}
\end{figure}

This result is plotted in Figure \ref{UniformSphere1}. Asymptotically,
it agrees with the matched asymptotics predictions to \Order[1]{}, as
it must, and reproduces the $3/8R$ ``Oseen'' term. As $R$ increases,
however, the uniform calculation becomes superior to the first two
terms of the matched asymptotic $C_D$. Although it is a much higher
order solution than either of the other two results, we show the full
matched asymptotics prediction for comparison.

\section{THE RENORMALIZATION GROUP APPLIED TO LOW $R$ FLOW}
\label{chap:RG}

\subsection{Introduction to the renormalization group}

In 1961, Lagerstrom proposed the first of a number of ``model
problems'', ordinary differential equations which exhibited many of the
same asymptotic features as the low Reynolds number problems. They were
used to study and develop the theory of matched asymptotic expansions.
The mathematical solution of these problems is closely analogous to the
actual solutions of the Navier-Stokes equations.

A review of these equations, and of their matched asymptotic solutions,
is given in Lagerstrom \cite{Lag72}. The relevant models can be
summarized by the following equation:
\begin{equation}
\frac{\mathrm{d}^2 u}{\mathrm{d} x^2} + \frac{n-1}{x}\frac{\mathrm{d}
u}{\mathrm{d} x} + u \frac{\mathrm{d} u}{\mathrm{d} x} + \delta
\left(\frac{\mathrm{d} u}{\mathrm{d} x} \right)^2 = 0 \label{modeleqn1}
\end{equation}
This ODE is subject to the boundary conditions $u(\epsilon) = 1$,
$u(\infty)=0$. In this equation, $n$ corresponds to the number of
spatial dimensions ($n=2$ for the cylinder, $n=3$ for the sphere).
$\delta$ = 0 characterizes incompressible flow, and $\delta=1$
corresponds to compressible flow. This equation is similar to the
Navier-Stokes equations expressed in Oseen variables. There are
fundamental differences between the structure of the incompressible and
compressible flow equations.

These model problems are posed in Hinch, albeit in terms of ``Stokes''
(rather than ``Oseen'') variables \cite{Hinch91}. Hinch begins by
examining the model describing incompressible flow past a sphere. He
next examines incompressible flow past a cylinder, which he calls ``A
worse problem.'' Finally, he treats compressible flow past cylinder,
which he dubs ``A terrible problem.''

These problems, which have historically been the proving ground of
matched asymptotics, were recently solved using new
Renormalization Group (RG) techniques in two papers by Chen et al.
\cite{CGO96,CGO94}. These techniques afford both quantitative and
methodological advantages over traditional matched asymptotics. The RG
approach derives all of the subtle terms (e.g., $R^2 \log{R}$) which
arise during asymptotic matching, demonstrating that origin of these
terms lies in the need to correct flaws inherent in the underlying
expansions. Moreover, RG does not require multiple rescalings of
variables, and its results, while asymptotically equivalent to those of
matched asymptotics, apply over a much larger range (e.g., they extend
to higher $R$).

In particular, Chen et al. solved Hinch's first model, which describes
incompressible flow past a sphere ($n=3$, $\delta=0$), as well as the
model for both kinds of flow past a cylinder ($n=2$, $\delta=0,1$)
\cite{CGO96,CGO94}. In a notation consistent with Hinch, they termed
these models the ``Stokes-Oseen caricature'' and the ``terrible
problem.''

The dramatic success of the RG techniques in solving the model problems
inspired their application to the original low Reynolds number flow
problems. That is our primary purpose here, as the low Reynolds number
problems are the traditional proving ground for new methodologies. We
will show that the RG techniques perform well when applied to these
problems. RG produces results superior to and encompassing the predictions
of matched asymptotics. More importantly, the RG calculations are
considerably simpler than matched asymptotics, requiring half the work.

The utility of the RG approach is most easily seen through an example,
which will also provide a framework for understanding the analysis
presented in subsequent sections. Several pedagogical examples can also
be found in the references (e.g.,   \cite{CGO94,CGO96,Oon00,GoldenfeldBook}). We begin here with an analysis of the most complicated model problem, the ``terrible problem,'' which caricatures compressible flow past a cylinder.

\subsubsection{Detailed analysis of the ``terrible problem''}

Although the ``terrible problem,'' is solved in a paper by Chen et al.,
we re-examine it here in considerably more detail, as its solution is
closely analogous to those of the low Reynolds number flow problems.
This switchback problem is exceptionally delicate\footnote{Hinch notes,
``It is unusual to find such a difficult  problem ...''
\cite{Hinch91}.}, requiring the calculation of an infinite number of
terms for the leading order asymptotic matching.

There are pitfalls and ambiguities in applying  RG techniques, even to
the ``terrible problem,'' which while terrible, is considerably simpler
than the real low Reynolds number problems. Understanding these
subtleties in this simpler context provides essential guidance when
attacking the Navier-Stokes' equations.

We want to solve the ODE given in Eqn. \ref{terribleeqn}, subject to
the boundary conditions \ref{terriblebc}. This equation can be derived
from Eqn. \ref{modeleqn1} by setting $n=2$, $\delta = 1$, and
transforming to the ``Stokes'' variables, $r=x/\epsilon$. Unlike Eqn.
\ref{modeleqn1}, Eqn. \ref{terrible} is obviously a singular
perturbation in $\epsilon$, which has been removed from the boundary
conditions. The last term in the equation vanishes when $\epsilon=0$.
\begin{subequations}
\begin{eqnarray}
\frac{d^2 u(r)}{d r^2} + \frac{1}{r}\frac{d u(r)}{d r} + \left(\frac{d
    u(r)}{d r}\right)^2 + \epsilon u(r) \frac{d u(r)}{d r}  =0 \quad \label{terribleeqn}\\
u(1)=0, \quad u(r=\infty)=1 \quad \label{terriblebc}
\end{eqnarray}
\label{terrible}
\end{subequations}
This problem cannot be solved exactly, although numerical solution is
straightforward. Trouble arises due to the \emph{boundary
layer}\footnote{A boundary layer is a region of rapid variation in the
solution, $y(t)$.} located near $r = \infty$. RG analysis requires that
we work in the ``inner'' variable for our approximation to capture the
correct behavior near the boundary layer\footnote{Here we use ``inner'' in
the usual sense \cite{BenderOrzag}. For further discussion, see Section
\ref{section:terminology}}. This requirement may also be
qualitatively motivated by arguing that one must choose coordinates to
``stretch out'' the boundary layer so that it can be well characterized
by our approximate solution.

To determine the appropriate change of variables, we need to analyze
Eqn. (\ref{terrible}) using a \emph{dominant balance} argument
\cite{BenderOrzag}. As it stands, the first three terms of Eqn.
(\ref{terribleeqn}) will dominate, since $\epsilon$ is small. The
rescaling $x=\epsilon r$ yields inner Eqn. (\ref{terrible2}). This, of
course, is the same equation originally given by Lagerstrom (Eqn.
\ref{modeleqn1}).
\begin{subequations}
\begin{eqnarray}
\frac{d^2 u(x)}{d x^2} + \frac{1}{x}\frac{d u(x)}{d x} + \left(\frac{d
    u(x)}{d x}\right)^2 + u(x) \frac{d u(x)}{d x}  =0 \quad \label{terribleeqn2}\\
u(\epsilon)=0, \quad u(x=\infty)=1 \quad \label{terriblebc2}
\end{eqnarray}
\label{terrible2}
\end{subequations}

The next step in the RG solution is to begin with the ansatz that the
solution to Eqn. (\ref{terrible2}) can be obtained from a perturbation
expansion (Eqn. \ref{naive terrible}).  We fully expect this ansatz to
fail, since we have a singular perturbation in our ODE. We therefore
refer to this starting point as the \emph{na\"\i ve} perturbation
expansion.
\begin{equation}
\label{naive terrible}
u(x) = u_0(x) + \epsilon u_1(x) + \epsilon^2 u_2(x) +
\mathcal{O}(\epsilon^3)
 \end{equation}
Collecting powers of $\epsilon$, we obtain differential equations for
$u_0(x)$, $u_1(x)$, etc:
\begin{eqnarray}
\label{terriblegovern1}
\Order{0} &:& \frac{u_0^{'}(x)}{x} + u_0(x)u_0^{'}(x)+u_0^{'}(x)^2+u0^{''}(x)=0 \\
\label{terriblegovern2} \Order{1} &:&
u_1u_0^{'}+\frac{u_1^{'}}{x}+u_0u_1^{'}+2u_0^{'}u_1^{'}+u_1^{''}=0
\\
\label{terriblegovern3} \Order{2} &:& u_2 u_0^{'} + u_1^{'} u_1 + u_0 u_2^{'}
+ (u_1^{'})^2 + 2 u_0^{'}u_2^{'} +\frac{u_2^{'}}{x} + u_2^{''} = 0
\end{eqnarray}

{\it a. $\mathcal{O}(\epsilon^0)$ solution}

The first complication of the terrible problem arises when we attempt to
solve Eqn. \ref{terriblegovern1}, a nonlinear ODE. Although one solution --- $u_0(x)=A_0$
--- is seen by inspection, an additional integration constant is not
forthcoming, and our solution to the $\mathcal{O}(\epsilon^0)$ problem
cannot satisfy both of the boundary conditions (Eqn.
\ref{terriblebc2}). The resolution to this quandary is simple: Ignore
the problem and it will go away; continue constructing the na\"\i ve
solution as if $u_0(x)=A_0$ were wholly satisfactory. The qualitative
idea is that the $\mathcal{O}(\epsilon^0)$ solution is the uniform
field which we have far from any disturbance source. Why is this
acceptable?

The RG method is robust against shortcomings in the na\"\i ve
expansion. We know that singular perturbation problems cannot be solved
by a single perturbation expansion. We therefore expect problems, such
as secular behavior, to arise in our solution for the na\"\i ve
expansion. RG techniques can be used to remove these flaws from the
perturbative solution, turning it into a uniformly valid approximation
\cite{CGO96}. It does  not matter whether these defects arise from an
incomplete solution for $u_0(x)$, the intrinsic structure of the
equation, or a combination of the two. To solve the terrible problem
(and later the low Reynolds number problems), we must exploit this
flexibility.

For subsequent calculations, there are two ways to proceed. First, we
may retain $A_0$ as an arbitrary constant, one which will ultimately be
renormalized in the process of calculating a uniformly valid
approximation. Alternatively, we may set $A_0=1$, satisfying the
boundary condition at $x=\infty$.\footnote{Meeting the boundary
condition at $x=\epsilon$ results only in the trivial solution
$u_0(x)=0$.} This unconventional approach to the RG calculation
effectively shifts the freedom that usually comes with the
$\mathcal{O}(\epsilon^0)$ constants of integration into the
$\mathcal{O}(\epsilon^1)$ solution. This artifice greatly simplifies
subsequent calculations, and is invaluable in treating the
Navier-Stokes equations. Moreover, these two approaches are equivalent,
as we now show.

{\it b. $\mathcal{O}(\epsilon^1)$ solution}

If $u_0(x)=A_0$, Eqn. \ref{terriblegovern2} simplifies to Eqn. \ref{terrible1a}.
\begin{equation}
\label{terrible1a}
\frac{d^2 u_1}{d x^2} + \left( \frac{1}{x}+A_0\right) \frac{d u_1}{d
  x}=0
\end{equation}
The solution is: $u_1(x)=B_0 + B_1 e_1(A_0 x)$, where $e_n(x)\equiv
\int_x^\infty e^{-t} t^{-n} \text{d} t$. Notice that the first term is
linearly dependant on the $u_0(x)$ solution. There are many opinions
regarding how to utilize this degree of freedom
\cite{Kun95,Woodruff95}. In our approach, one is free to choose the
homogeneous solutions of $u_0, u_1, \textrm{etc.}$ for convenience. The
only constraint\footnote{Of course the solution must also satisfy the
governing equation.} is that the ``na\"\i ve'' solution (Eqn.
\ref{naive terrible}) must have a sufficient number of integration
constants to meet the boundary conditions. In this example, that means
two constants of integration.

Different choices of particular solutions will ultimately result in
different approximate solutions to the ODE. However, all of these
solutions will agree within the accuracy limitations of the original
approximation (in this case the na\"\i ve expansion). This can be shown
explicitly. In this example, as in the low Reynolds number problems, we
choose a particular solution which simplifies subsequent
calculations. Setting $B_0 = 0$ (note that this is \emph{not} the same as
a redefinition of the constant $A_0$), we obtain the solution:
\begin{equation}
\label{Oeterriblesol}
u(x)=A_0+ \underbrace{\epsilon  B_1 e_1(A_0 x)}_{\textrm{divergent as } x
  \rightarrow 0} +
\mathcal{O}(\epsilon^2)
\end{equation}

The second term in Eqn. \ref{Oeterriblesol} diverges logarithmically as
$x\rightarrow 0$. One may argue that this divergence is irrelevant,
since the range of the original variable is $r \in [1,\infty)$, and
numerical solutions demonstrate that the solutions to Eqn.
\ref{terrible} in $[1,\infty)$ diverge when extended to $r < 1$. But
the argument that the divergence in Eqn. \ref{Oeterriblesol} is an
intrinsic part of the solution (and therefore should not be considered
problematic) is incorrect. Although the original variable, $r$, is
limited to $r \in [1,\infty)$, the transformed variable, $x = \epsilon
r$, has the range $x \in [0,\infty)$. This occurs because there are no
restrictions on the lower limit of $\epsilon$. The divergence exhibited
by the second term of Eqn. \ref{Oeterriblesol} must be removed via
renormalization in order to turn the flawed na\"\i ve solution into a
uniformly valid approximation.

This divergence arises for two reasons. First, we are perturbing about
an $\mathcal{O}(\epsilon^0)$ solution which is deficient; it is missing
the second integration constant (and concomitant fundamental
solution). More fundamentally, Eqn \ref{naive terrible} attempts to
solve a singular perturbation problem with a regular expansion, an
approach which must fail. The RG technique solves these problems by
restructuring the na\"ive expansion and eliminating the flaws in
$u_0(x)$.

Although $A_0$ is simply a constant of integration when $\epsilon = 0$,
it must be modified when $\epsilon \neq 0$. We will absorb the
divergences into a modification, or renormalization, of the constant of
integration $A_0$. Formally, one begins by ``splitting'' the secular
terms, replacing  $e_1(A_0 x)$ by $e_1(A_0 x) - e_1(A_0 \tau) + e_1(A_0
\tau)$, where $\tau$ is an arbitrary position. This results in Eqn.
\ref{Oeterriblesol2}:
\begin{equation}
\label{Oeterriblesol2}
u(x)=A_0+ \epsilon  B_1 (e_1(A_0 x) - e_1(A_0 \tau)+ e_1(A_0 \tau)) + \mathcal{O}(\epsilon^2)
\end{equation}
Since $\tau$ is arbitrary, it can be chosen such that $e_1(A_0 x) - e_1(A_0 \tau)$ is
non-secular (for a given $x$). The divergence is now contained in the
last term of Eqn. (\ref{Oeterriblesol2}), and is exhibited as a function
of $\tau$.

It is dealt with by introducing a multiplicative renormalization
constant, $Z_1 = 1 + \sum_{i=1}^\infty a_i(\tau) \epsilon^i$,  and then
renormalizing $A_0$ as $ A_0 = Z_1 A_0(\tau)$.\footnote{$A_0$ is the
only constant which can be renormalized to remove the divergences, as
$B_1$ is proportional to the secular terms.} The coefficients
$a_i(\tau)$ can then be chosen.\footnote{Note that the coefficients
\emph{must} also be independent of $x$.} order by order so as to
eliminate the secular term in Eqn. (\ref{Oeterriblesol2}). Substituting,
and choosing $a_1$ to eliminate the final term of Eqn.
\ref{Oeterriblesol2}, we obtain
\begin{equation}
\label{terribleRG1}
u(x)=A_0(\tau)+ \epsilon  B_1 (e_1(A_0(\tau) x) - e_1(A_0(\tau) \tau)) + \mathcal{O}(\epsilon^2)
\end{equation}
Where $a_1$ satisfies
\begin{equation}
a_1(\tau)= \frac{-B_1 e_1(\tau A_0(\tau)(1+\sum_{i=1}^\infty a_i(\tau)
  \epsilon^i))}{A_0(\tau)}
\end{equation}

Note that to obtain Eqn. \ref{terribleRG1} we needed to expand $e_1$
about $\epsilon=0$. Unusually in this equation, the renormalized
constant ($A_0(\tau)$) appears in the argument of the exponential
integral; this complicates the calculation. We will later show how to
avoid this problem by restructuring our calculations.

Qualitatively, the idea underlying Eqn. \ref{terribleRG1} is that
boundary conditions far away (from $x = \epsilon$) are unknown to our
solution at $x \gg \epsilon$, so that $A_0$ is undetermined at $x =
\tau$. RG determines $A_0$ in this regime through the renormalization
constant $Z_1$ (which depends on $\tau$). Afterward there will be new
constants which can be used to meet the boundary conditions.

The RG condition states that the solution $u(x)$ cannot depend on the
arbitrary position $\tau$. This requirement can be implemented in one
of two ways. First, since $\partial_\tau u(x) = 0$, apply
$\partial_\tau$ to the RHS of Eqn. \ref{terribleRG1} and set the result
equal to zero:
\begin{equation}
\label{terribleRGans1}
A_0^{'}(\tau) + \epsilon B_1 \left( \frac{e^{-A_0(\tau)
      \tau}}{\tau} + \frac{A_0^{'}(\tau)}{A_0(\tau)}\left(
      e^{-A_0(\tau) \tau} - e^{-A_0(\tau) x} \right) \right) +
      \mathcal{O}(\epsilon^2) = 0
\end{equation}
The next step in RG is to realize Eqn. \ref{terribleRGans1} implies
that $A_0^{'}(\tau) \sim \mathcal{O}(\epsilon)$. Retaining only terms of
$\mathcal{O}(\epsilon)$, we obtain:
\begin{equation}
\label{terribleRGans2}
\frac{d A_0(\tau)}{d \tau} + \epsilon B_1 \left( \frac{e^{-A_0(\tau)
      \tau}}{\tau} \right) + \mathcal{O}(\epsilon^2) = 0
\end{equation}

In principle, we simply solve Eqn. \ref{terribleRGans2} for
$A_0(\tau)$. Unfortunately, that is not possible, due to the presence
of $A_0(\tau)$ in the exponential.  This complication also occurs in
other switchback problems, as well as in the low Reynolds number
problems. Eqn. \ref{terribleRGans2} can be solved by an iterative
approach: Initially set $\epsilon=0$, and solve for
$A_0(\tau)=\alpha_0$, a constant. Next substitute this result into the
$\mathcal{O}(\epsilon)$ term in Eqn. \ref{terribleRGans2}, solving for
$A_0(\tau)$ again:
\begin{equation}
\label{A0ans}
A_0(\tau)= \alpha_0 + \epsilon B_1 e_1(\alpha_0 \tau)
\end{equation}
In this solution, we have a new integration constant, $\alpha_0$.
Having obtained this result, we again must exploit the arbitrary nature
of $\tau$. Setting $\tau = x$, and substituting into Eqn.
\ref{terribleRG1}, we obtain:
\begin{equation}
\label{terribleRG1b}
u(x)=\alpha_0 + \epsilon  B_1 e_1(\alpha_0 x) + \mathcal{O}(\epsilon^2)
\end{equation}

But this is identical to the original solution (Eqn. \ref{terrible1a})!
What have we accomplished? This renormalized result is guaranteed to be
a uniformly valid result, for $\forall x$. The renormalization
procedure ensures that the logarithmic divergence in Eqn.
\ref{terribleRG1b} is required by the solution, and is \emph{not} an
artifact of our approximations. Obtaining the same answer is a
consequence of solving Eqn. \ref{terribleRGans1} iteratively. Had we
been able to solve that equation exactly, this disconcerting
coincidence would have been avoided.

We obtain the final solution to Eqn. \ref{terribleeqn} by applying the
boundary conditions (Eqn. \ref{terriblebc2}) to Eqn.
\ref{terribleRG1b}: $\alpha_0 = 1$, $B_1 = -1/(\epsilon
e_1(\epsilon))$. Lastly, we undo the initial change of variables
($r=x/\epsilon$), yielding the result given in Eqn.
\ref{terribleO1sol}. As shown in Chen et al., this is an excellent
approximate solution \cite{CGO96}.
\begin{equation}
\label{terribleO1sol}
u(r) = 1 - \frac{e_1(r \epsilon)}{e_1(\epsilon)} + \Order{2}
\end{equation}
Furthermore, if we expand the coefficient $B_1=-1/(\epsilon
e_1(\epsilon))$ for $\epsilon \rightarrow 0^+$, $B_1(\epsilon)
/\epsilon \sim -1/\ln{(1/\epsilon)} - \gamma / \ln^2{(1/\epsilon)}$.
These logarithmic functions of $\epsilon$ are exactly those which are
required by asymptotic matching! These ``unexpected'' orders in
$\epsilon$ make the solution of this problem via asymptotic matching
very difficult. They must be deduced and introduced order by order, so
as to make matching possible. In the RG solution, they are seen to
arise naturally as a consequence of the term $1/e_1(\epsilon)$.

There are several other equivalent ways to structure this calculation.
It is worthwhile to examine these (and to demonstrate their
equivalence), in order to streamline our approach for the low Reynolds
number problems.

The first variation occurs in how we apply the RG condition. Rather than
applying $\partial_\tau$ to Eqn. \ref{terribleRG1}, we
may also realize that the original constants of integration, $ A_0
= Z_1(\tau) A_0(\tau)$, must be independent of $\tau$. Hence the
``alternative'' RG equation:
\begin{displaymath}
\frac{\partial A_0}{\partial \tau} = \frac{\partial (Z_1(\tau)
  A_0(\tau))}{\partial \tau} = 0
\end{displaymath}
Substituting $Z_1 = 1 + \epsilon \left( -B_1 e_1(\tau A_0(\tau)(1+\sum_{i=1}^\infty a_i(\tau)
  \epsilon^i))\right)/A_0(\tau) + \Order{2}$, one obtains:
\begin{equation}
\label{terribleRGans2b}
A_0^{'}(\tau) + \epsilon B_1 \left(\frac{e^{-A_0(\tau)
      \tau}}{\tau} + \frac{A_0^{'}(\tau)}{A_0(\tau)}
       e^{-A_0(\tau) \tau} \right) + \Order{2} = 0
\end{equation}
Because this implies $A_0^{'}(\tau) \sim \Order{1}$,
% Do we want to be concerned with the difference between self-consistent
% solutions and other solutions?
Eqn. \ref{terribleRGans2b} simplifies to
Eqn. \ref{terribleRGans2} (to within \Order{2}), and these two
methods of implementing the RG condition are equivalent.

In addition to this dichotomous implementation of the RG condition,
there is yet another way to structure the analysis from the outset: We
set $A_0=1$ in the zeroth order solution, and rely on the robustness of
the RG approach to variations in our perturbative solution. With this
$u_0(x)$ solution, there is no longer any freedom in our choice of
$u_1(x)$ integration constants --- both are needed to meet boundary
conditions. In this approach, our na\"\i ve perturbative solution is:
\begin{equation}
\label{Oeterriblesol2b}
u(x)=1 + \epsilon ( B_0 +  \underbrace{B_1 e_1(x)}_{\textrm{divergent}}) + \Order{2}
\end{equation}
Proceeding as before, replace  $e_1(x)$ by $e_1(x) - e_1(\tau) + e_1(\tau)$:
\begin{displaymath}
u(x)=1 + \epsilon \left(B_0 + B_1 \left(e_1(x) - e_1(\tau) + e_1(\tau)\right)\right) + \Order{2}
\end{displaymath}
Again introduce renormalization constants ($Z_1 = 1 + \sum_{i=1}^\infty
a_i(\tau) \epsilon^i$,  $Z_2 = 1 + \sum_{i=1}^\infty b_i(\tau)
\epsilon^i$), and renormalize $B_0$, $B_1$ as $B_0 = Z_1 B_0(\tau)$ and
$B_1 = Z_2 B_1(\tau)$. In fact, only $B_0$ needs to be renormalized, as
the $B_1$ term multiplies the secular term and consequently cannot
absorb that divergence. This can be seen systematically by attempting
to renormalize both variables.
With an appropriate choice of coefficients, $a_1 = -B_1(\tau)
e_1(\tau)$ and $b_1 = 0$, the final term in the last equation is
eliminated. $b_1=0$ demonstrates that $B_1$ does not need to be
renormalized at \Order{1}. The resulting equation is given in Eqn.
\ref{terribleRGb2}.
\begin{equation}
\label{terribleRGb2}
u(x)=1 + \epsilon \left(B_0(\tau) + B_1(\tau) \left(e_1(x) - e_1(\tau)\right)\right) + \Order{2}
\end{equation}
We did not actually need to determine $a_1$ or $b_1$ in order to write
the above equation; it could have been done by inspection.
Determination of these quantities is useful for two reasons. First, it
helps us see which secular terms are being renormalized by which
integration constants. Secondly, it allows the second implementation of
the RG condition which was described above. This can sometimes simplify
calculations.

Using the first implementation (requiring  $\partial_\tau u(x) = 0$), and using
Eqn. \ref{terribleRGb2}, we obtain:
\begin{equation}
\label{terribleRGb1}
B_0^{'}(\tau) + B_1^{'}(\tau)\left( e_1(x) - e_1(\tau) \right) +  B_1(\tau)
\frac{e^{-\tau}}{\tau} = \Order{1}
\end{equation}
This can only be true $\forall x$ if $B_1^{'}(\tau) = 0$, or $B_1(\tau) =
\beta_2$, a constant (as expected). Knowing this, we solve for $B_0(\tau) =
\beta_1 + \beta_2 e_1(\tau)$. Substituting this result into
Eqn. \ref{terribleRGb2}, and setting $\tau = x$, we obtain the
renormalized solution:
\begin{equation}
\label{terribleRGsolv2}
u(x) = 1 + \epsilon \left( \beta_1 + \beta_2 e_1(x) \right)
\end{equation}
The boundary conditions in Eqn. \ref{terriblebc2} are satisfied if
$\beta_1 = 0$ and $\beta_2 = -1/(\epsilon
  e_1(\epsilon))$. Returning to the original variable
($r=x/\epsilon$), we obtain:
\begin{equation}
\label{terribleO1solb}
u(r) = 1 - \frac{e_1(r \epsilon)}{e_1(\epsilon)} + \Order{2}
\end{equation}
This is identical to Eqn. \ref{terribleO1sol}, demonstrating the
equivalence of these calculations. The latter method is preferable, as
it avoids the nonlinear RG equation (Eqn. \ref{terribleRGans2}). We
will use this second approach for analyzing the low Reynolds number
problems.

The RG analysis has shown us that the logarithmic divergences present
in Eqn. \ref{Oeterriblesol} are an essential component of the solution,
Eqn. \ref{terribleO1solb}. However, we must work to \Order{2} in order
to see the true utility of RG and to understand all of the nuances of
its application.

{\it c. \Order{2} solution}

We base our treatment of the \Order{2} on the second analysis
presented above. Through \Order{1}, the na\" \i ve solution is:  $u_0(x)
= 1$, $u_1(x)= B_0 +  B_1 e_1(x)$. Substituting into
Eqn. \ref{terriblegovern3}, we obtain the governing equation for $u_2(x)$:
\begin{equation}
\label{terriblegovern3b}
u_2^{''} + \left( 1 + \frac{1}{x} \right)u_2 = \frac{B_0 B_1
e^{-x}}{x} - \frac{B_1^2 e^{-2x}}{x^2} + \frac{B_1^2 e^{-x} e_1(x)}{x}
\end{equation}
This has the same homogeneous solution as $u_1(x)$, $u_2^{(h)}(x) = C_0 +
C_1 e_1(x)$. A particular solution is:
\begin{displaymath}
u_2^{(p)}(x) = -B_1 B_0 e^{-x} + 2 B_1^2 e_1(2x) - \frac{1}{2}B_1^2
e_1^2(x) - B_1^2 e^{-x}e_1(x)
\end{displaymath}

As discussed previously, we are free to choose $C_0$, $C_1$ to simplify
subsequent calculations. The constants $B_0$, $B_1$ are able to meet
the boundary conditions, so there is no need to retain the \Order{2}
constants: We choose $C_0 = 0$, $C_1 = 0$. In this case, the differing
differing choices of ${C_0, C_1}$ correspond to a redefinition of
${B_0, B_1}$ plus a change of \Order{3}, i.e. $\tilde B_0 = B_0 +
\epsilon C_0$.\footnote{This was not true at the previous order.} Our
na\" \i ve solution through \Order{2} is thus:
\begin{eqnarray}
\label{o2terriblenaive}
 u(x) &=& 1 + \epsilon \left( B_0 + \underline{B_1
   e_1(x)}\right) + \\
   & & \epsilon^2 \left( -B_1 B_0 e^{-x} + \underline{2 B_1^2 e_1(2x)} - \underline{\underline{\frac{1}{2}B_1^2 e_1^2(x)}} - \underline{B_1^2 e^{-x}e_1(x)} \right) + \Order{3} \nonumber
\end{eqnarray}

The underlined terms in this expression are divergent as $x \rightarrow
0$; the doubly underlined term is the most singular ($ \sim \ln (x)^2$).
RG can be used to address the divergences in Eqn.
\ref{o2terriblenaive}. However, there is a great deal of flexibility in
its implementation; while most tactics yield equivalent approximations,
there are significant differences in complexity. We now explore all of
the organizational possibilities in the terrible problem, an exercise
which will subsequently guide us through the low Reynolds number
calculations.

The first possibility is to treat only the most secular term at
\Order{2}. The doubly underlined term dominates the divergent behavior,
and contains the most important information needed for RG to construct
a uniformly valid approximation. The approximation reached by this
approach is necessarily inferior to those obtained utilizing additional
terms. However it is nonetheless valid and useful, and eliminating most
of the \Order{2} terms simplifies our calculations.

Discarding all \Order{2} terms except the doubly underlined term, we
begin the calculation in the usual manner, but come immediately to the
next question: Ought we replace $e_1^2(x)$ by $e_1^2(x) - e_1^2(\tau) +
e_1^2(\tau)$ or by $\left(e_1(x)-e_1(\tau)\right)^2 + 2 e_1(x) e_1(\tau)
- e_1^2(\tau)$? Each option eliminates the divergence in $x$, replacing
it with a divergence in $\tau$. Both merit consideration. Beginning with the
latter, the renormalized perturbative solution is:
\begin{eqnarray}
\label{o2terrible1}
 u(x) &=& 1 + \epsilon \left( B_0(\tau) + B_1(\tau) \left(e_1(x)-e_1(\tau)
 \right) \right) - \epsilon^2 \left(\frac{1}{2}B_1(\tau)^2 \left(e_1(x)-e_1(\tau)
 \right)^2 \right)  \nonumber \\
 & & + \epsilon^2 \left(\textrm{less divergent terms} \right) + \Order{3}
\end{eqnarray}

Applying the RG condition ($\partial_\tau u(x) = 0$) results in a
lengthy differential equation in $\tau$. Because we want our solution
to be independent of $x$, we group terms according to their $x$
dependence. Recognizing that $B^{'}_1(\tau) \sim \Order{1}$,
$B^{'}_0(\tau) \sim \Order{1}$, and working to \Order{3}, we obtain two
equations which must be simultaneously satisfied:
\begin{subequations}
\label{o2terriblesol1}
\begin{eqnarray}
B_1^{'}(\tau) - \frac{\epsilon e^{-\tau} B_1^2(\tau)}{\tau} &=& \Order{3}
\label{b1sol} \\
\frac{e^{-\tau} \left( B_1(\tau) +
  \epsilon B_1^2(\tau)e_1(\tau)\right) }{\tau}-e_1(\tau)B_1^{'}(\tau)+B_0^{'}(\tau) &=&
  \Order{3} \label{b2sol}
\end{eqnarray}
\end{subequations}
Eqn. \ref{b1sol} has the solution
\begin{displaymath}
B_1(\tau) = \frac{1}{\beta_1 + \epsilon e_1(\tau)} + \Order{2}
\end{displaymath}
Substituting this result into Eqn. \ref{b2sol}, and solving, we obtain the result
\begin{displaymath}
B_0(\tau)= \beta_0 + \frac{\ln{(\beta_1 + \epsilon
    e_1(\tau)})}{\epsilon} + \Order{2}
\end{displaymath}
Both $\beta_0$ and $\beta_1$ are constants of integration which can be
later used to meet the boundary conditions. Substituting these solutions into
Eqn. \ref{o2terrible1}, setting $\tau=x$, disregarding terms of
\Order{2} and higher we obtain the renormalized solution:
\begin{equation}
\label{o2terriblesol0}
u(x)=1 + \epsilon \left( \beta_0 + \frac{\ln{\left(\beta_1 + \epsilon
      e_1(x)\right)}}{\epsilon}\right) + \Order{2}
\end{equation}
Choosing $\beta_0$ and $\beta_1$ to satisfy Eqn. \ref{terriblebc2}, results in Eqn. \ref{o2terrible1sol}.
\begin{equation}
\label{o2terrible1sol}
u(x) = \ln \left(e + \frac{\left(1-e\right)
    e_1(x)}{e_1(\epsilon)}\right) + \Order{2}
\end{equation}
Expressing this in the original variable ($r=x/\epsilon$), results in the
final answer (Eqn. \ref{finalo2terrible}).
\begin{equation}
u(r) = \ln \left(e + \frac{\left(1-e\right)
    e_1(\epsilon r)}{e_1(\epsilon)}\right) + \Order{2}
\label{finalo2terrible}
\end{equation}

This is the solution previously obtained by Chen et al., albeit with a
typographical error corrected \cite{CGO96}. We will now  revisit this
analysis, using the alternative ``splitting'' of the most secular term
in  Eqn. \ref{o2terriblenaive}, but not yet considering less secular
(or
 non-secular) terms of \Order{2}.

If we replace replace $e_1^2(x)$ in Eqn. \ref{o2terriblenaive} by
$e_1^2(x) - e_1^2(\tau) + e_1^2(\tau)$, we obtain the new na\"ive
expansion given by Eqn. \ref{o2terrible2}.
\begin{eqnarray}
\label{o2terrible2}
 u(x) &=& 1 + \epsilon \left( B_0(\tau) + B_1(\tau) \left(e_1(x)-e_1(\tau)
 \right) \right) - \epsilon^2 \left(\frac{1}{2}B_1(\tau)^2 \left(e_1^2(x)-e_1^2(\tau)
 \right) \right) \nonumber \\
 & & + \epsilon^2 \left(\textrm{less divergent terms}  \right) + \Order{3}
\end{eqnarray}
We now repeat the same calculations:
\begin{enumerate}
\item Apply the RG condition ($\partial_\tau u(x) = 0$).
\item Group the resulting equation according to $x$ dependence. This
  will result in two equations which must be satisfied independently.
\item Discard terms of \Order{3}, observing that $B_0^{'}(\tau),
  B_1^{'}(\tau)$ must be of $\Order{1}$.
\item Solve these differential equations simultaneously for
  $B_0(\tau),B_1(\tau)$.
\item Substitute these solutions into the original equation
  (i.e. Eqn. \ref{o2terrible2}), and set $\tau = x$.
\item Choose the integration constants in this result to satisfy Eqn. \ref{terriblebc2}.
\item Obtain the final solution by returning to the original variable, $r=x/\epsilon$.
\end{enumerate}

For Eqn. \ref{o2terrible2}, steps 1 - 4 result in the following
solutions for our renormalized constants: $B_1(\tau) = \beta_1 +
\Order{2}$, $B_0(\tau) = \beta_0 + \beta_1 e_1(\tau) - \epsilon
\beta_1^2 e_1^2(\tau)/2 + \Order{2}$. Completing step 5, we obtain
the renormalized result:
\begin{equation}
\label{o2terrible2sol}
u(x) = 1 + \epsilon \left( \beta_0 + \beta_1 e_1(x) \right) - \epsilon^2
\frac{\beta_1 ^2 e_1^2(x)}{2} + \Order{2}
\end{equation}

This is identical to our starting point, Eqn. \ref{o2terriblenaive}
(retaining only the most secular terms). This should no longer be
surprising, as we observed the same phenomena in the \Order{1} analysis
(Eqn. \ref{terribleRG1b}). However, it is worth noticing that we
obtained two different results (Eqns. \ref{o2terrible1sol},
\ref{o2terrible2sol}) depending on how we structured our RG
calculation. This apparent difficulty is illusory, and the results are
equivalent: Expanding Eqn. \ref{o2terriblesol0} for small $\epsilon$
reproduces Eqn. \ref{o2terrible2sol}. Here, as in previous cases, we
are free to structure the RG calculation for convenience. This easiest
calculation is the second approach --- in which only one constant of
integration is actually renormalized --- and our renormalized result is
the same as our na\" \i ve starting point.

This simplified analysis (considering only the most secular terms)
illustrates some of the pitfalls which can arise in applying RG to
switchback problems. However, we must finish the \Order{2} analysis by
considering all terms in Eqn. \ref{o2terriblenaive} to understand the
final nuances of this problem.  There is a new complication when we
attempt to renormalize all terms of Eqn. \ref{o2terriblenaive}: The
final term, $-B_1^2 e^{-x}e_1(x)$, has the same kind of ``splitting''
ambiguity which we encountered in dealing with the doubly underlined
term.

We introduce our arbitrary position variable, $\tau$, which we want to
choose so as to eliminate the secular term in $x$ by replacing it with
a divergence in $\tau$. In many cases, it is clear how to deal with the
secular term. For example, a linear divergence --- $x$ --- can be
replaced with $ x - \tau + \tau$. The final $\tau$ will be absorbed
into the renormalized constants of integration, and the $x - \tau$ term
(which is now considered non-secular), will ultimately disappear after
renormalization. However the term $-B_1^2e^{-x}e_1(x)$ is confusing. As
seen above, there are two ways to ``split'' the $B_1^2 e_1^2(x)/2$
term. There are \emph{four} different ways to split $e^{-x}e_1(x)$. It
may be replaced by any of the following:
\begin{enumerate}
\item $\left(e^{-x}-e^{-\tau}\right)e_1(x)+e^{-\tau}e_1(x)$
\item $e^{-x}e_1(x)-e^{-\tau}e_1(\tau)+e^{-\tau}e_1(\tau)$
\item $\left(e^{-x}-e^{-\tau}\right)\left(e_1(x)-e_1(\tau)\right)+e^{-\tau}e_1(x)+e^{-x}e_1(x)-e^{-\tau}e_1(\tau)$
\item $e^{-x}\left(e_1(x)-e_1(\tau)\right)+e^{-x}e_1(\tau)$
\end{enumerate}

All four of these options ``cure'' the divergent term (i.e. the secular
term will vanish when we subsequently set $\tau = x$), and are equal to
$e^{-x}e_1(x)$. If handled properly, any of these options can lead to a
valid renormalized solution. However, we will show that the fourth and
final option is most natural, and results in the simplest algebra.

How do we choose? The first consideration is subtle: \emph{The overall
renormalized perturbative result must satisfy the governing equation
(Eqn. \ref{terriblegovern1}) independently for each order in
$\epsilon$}. How we renormalize the \Order{1} divergences (Eqn.
\ref{terribleRGb2}) has implications for \Order{2} calculations. For
example, in \Order{1} renormalization, there is an important difference
between Eqn. \ref{terribleRGb2} and Eqn. \ref{Oeterriblesol2}. The
former has the additional term $- \epsilon B_1(\tau) e_1(\tau)$. This
term requires the presence of an additional \Order{2} term:
$\epsilon^2e^{-x}B_1^2(\tau)e_1(\tau)$. Without this term the \Order{2}
renormalized solution will not satisfy Eqn. \ref{terriblegovern3}, and
the renormalization procedure will yield an incorrect solution. We were
able to gloss over this before because we were considering only the
most secular term at \Order{2}.

Inspecting the four possible splittings enumerated above, we see that
only the last two options provide the necessary
$\epsilon^2e^{-x}B_1^2(\tau)e_1(\tau)$ term, and can satisfy Eqn.
\ref{terriblegovern3} without contrivances.\footnote{The first two
options \emph{can} satisfy the governing equation \emph{if} we
carefully choose a different homogeneous solution at \Order{2}. With
the proper non-zero choice of $C_0$ and $C_1$ we can use the first two
splittings enumerated, and they will result in an equivalent RG
solution.} In examining both of these options, we split the $e_1^2(x)$
term for simplicity, as in the derivation of Eqn.
\ref{o2terrible2sol}.\footnote{In principle, each of the possible
\Order{1} splittings could be paired with all possibilities at
\Order{2}, resulting in eight total possibilities.} Considering the
third option first, our renormalized perturbation solution becomes:
\begin{eqnarray}
\label{terrible:I}
 u(x) &=& 1 + \epsilon \left( B_0(\tau) + B_1(\tau) \left(e_1(x)-e_1(\tau)
 \right) \right) + \epsilon^2 \Big(-B_1(\tau)B_0(\tau)e^{-x} - \nonumber \\
 & & B_1^2(\tau)\left(e^{-x}-e^{-\tau}\right)\left(e_1(x)-e_1(\tau)\right)
-\frac{1}{2}B_1(\tau)^2 \left(e_1^2(x)-e_1^2(\tau)\right) + \nonumber \\
& & 2 B_1^2(\tau)\left(e_1(2x)-e_1(2 \tau)\right)\Big) + \Order{3}
\end{eqnarray}

As it must, this result satisfies Eqn. \ref{terriblegovern3} to
\Order{2}. By applying the RG condition ($\partial_\tau u(x) = 0$) to
this equation, and grouping the resulting equation according to $x$
dependence, we obtain a lengthy equation which can only be satisfied to
\Order{3} $\forall x$ if:
\begin{eqnarray}
\label{terrible:Iconditions}
B_1^{'}(\tau) e^{\tau} &=& \epsilon B_1^2(\tau) \\ e^{2 \tau} \tau B_0^{'}(\tau) &=&
e^{2 \tau} \tau e_1(\tau) B_1^{'}(\tau)-e^{\tau}B_1(\tau)-3\epsilon B_1^2(\tau)+e^{\tau}\epsilon B_1^2(\tau) e_1(\tau) - e^{\tau} \tau \epsilon B_1^2(\tau) e_1(\tau)  \nonumber \\ 0 &=& \epsilon \left( \epsilon B_1(\tau) + e^{\tau} \tau \epsilon  B_0^{'}(\tau) \right) \nonumber
\end{eqnarray}
Generally, no solution will exist, as we have two unknown functions and
three differential equations. In this case, however, the first equation requires that:
\begin{equation}
\label{terrible:almost1}
B_1(\tau) = \frac{e^{\tau}}{-\epsilon + e^\tau \beta_1}
\end{equation}
For this $B_1(\tau)$ solution, it is actually possible to satisfy the latter equations simultaneously to \Order{3}: This occurs because the last equation is simply the lowest order of
the second one.\footnote{This can be seen  explicitly by substituting Eqn. \ref{terrible:almost1}.}
There is another noteworthy point regarding the second part of Eqn. \ref{terrible:Iconditions}. In all previous calculations, we discarded terms like $\epsilon^2 B_0^{'}(\tau)$, since $B_0^{'}(\tau)$ and $B_1^{'}(\tau)$ had to be of $\Order{1}$. To solve these equations, however, $B_0^{'}(\tau)$ can \emph{not} be $\Order{1}$ (although $B_1^{'}(\tau)$ is). Solving for $B_0$,
\begin{equation}
\label{terrible:almost}
 B_0(\tau)= \beta_0 - \int_\epsilon^\tau \frac{2 \epsilon +
 e^{\sigma}\beta_1 + e^{\sigma}\left( 2 \sigma - 1 \right) \epsilon
 e_1(\sigma)}{\sigma \left( \epsilon - e^{\sigma}\beta_1 \right)} \textrm{d}\sigma
\end{equation}

This solution, while valid, is cumbersome. Consider instead the fourth
possible ``split'' enumerated above. Eqn. \ref{terrible:II} gives our
renormalized perturbation solution, which satisfies Eqn.
\ref{terriblegovern3}.
\begin{eqnarray}
\label{terrible:II}
 u(x) &=& 1 + \epsilon \left( B_0(\tau) + B_1(\tau) \left(e_1(x)-e_1(\tau)
 \right) \right) + \epsilon^2 \Big(-B_1(\tau)B_0(\tau)e^{-x}- \nonumber \\ & & B_1^2(\tau)e^{-x}\left(e_1(x)-e_1(\tau)\right)
-\frac{1}{2}B_1(\tau)^2 \left(e_1^2(x)-e_1^2(\tau)\right) + \nonumber \\ & & 2
 B_1^2(\tau)\left(e_1(2x)-e_1(2 \tau)\right)\Big) + \Order{3}
\end{eqnarray}

Applying the  RG condition ($\partial_\tau u(x) = 0$),
and requiring that it be satisfied $\forall x$, we obtain the following
solutions for $B_0(\tau)$, and $B_1(\tau)$:
\begin{subequations}
\begin{eqnarray}
\label{terrible:O2sol}
B_1(\tau)&=&\beta_1 +\Order{3}\\
B_0(\tau)&=&\beta_0 + \beta_1 e_1(\tau) + \epsilon \left(
 - \frac{\beta_1^2 e_1^2(\tau)}{2}+2\beta_1^2 e_1(2 \tau) \right) + \Order{3}
\end{eqnarray}
\end{subequations}
Substituting these results into Eqn. \ref{terrible:II} and setting
$\tau = x$, we obtain the final RG result, given by Eqn. \ref{terrible:final}.
\begin{eqnarray}
\label{terrible:final}
 u(x) &=& 1 + \epsilon \left( \beta_0 + \beta_1
   e_1(x)\right) + \\ & & \epsilon^2 \Bigg( -\beta_1 \beta_0  e^{-x} + 2 \beta_1^2 e_1(2x) - \frac{1}{2}\beta_1^2 e_1^2(x) - \beta_1^2 e^{-x}e_1(x) \Bigg) + \Order{3} \nonumber
\end{eqnarray}
This is, of course, identical to our na\"\i ve staring point, a
happenstance we have seen several times previously. It is worth noting
that the renormalized solutions obtained using Eqns.
\ref{terrible:almost1} and \ref{terrible:almost} are asymptotically
equivalent to Eqn. \ref{terrible:final}.

It may seem that we have needlessly digressed into the ``terrible''
problem. However, a clear-cut ``best'' strategy has emerged from our
detailed exploration. Furthermore, we have identified --- and resolved
--- a number of subtleties in the application of RG. Before applying
these lessons to the problem of low Reynolds number flow past a
cylinder, we summarize our conclusions.

The ``best'' strategy is the one used to derive Eqn.
\ref{terrible:final}, a result which is identical to our na\" \i ve
solution (Eqn. \ref{o2terriblenaive}). First, transform to the inner
equation. Solve the \Order{0} equation incompletely (obtaining just one
constant of integration), which can then be set to satisfy the boundary
condition at $\infty$. This ``trick'' necessitates retention of
integration constants at \Order{1}, but results in computational
simplifications (a non-linear RG equation) which are \emph{essential}
in dealing with the Navier-Stokes equations.

At \Order{2}, the homogeneous solution are identical to
those at \Order{1}. Consequently, the \Order{2} integration constants
need not be retained, as we can meet the boundary conditions with the
\Order{1} constants. We just pick a convenient particular solution.

To apply RG to the terrible problem, we first ``split'' the secular
terms. There are several ways to do this, even after requiring that the
renormalized perturbation expansions satisfy the governing equations at
each order. We can again choose for simplicity, bearing in mind that
\Order{1} renormalization can impact \Order{2} calculations. It is
easiest to apply the RG condition to the renormalized perturbation
expansion, rather than applying it to the integration constants
directly. In solving the resulting equation, we want solutions which
are valid $\forall x$. To solve the RG equation, care must be taken to
satisfy several conditions simultaneously, and it cannot be assumed
that our renormalized constants have a derivative of \Order{1}.

Although there is quite a bit of flexibility in implementing the RG
technique, our results are robust: Regardless of how we structure the
calculation, our solutions agree to within an accuracy limited by the
original na\" \i ve perturbative solution; they are asymptotically
equivalent. It is this robustness which makes RG a useful tool for the
low Reynolds number problems, where the complexity of the Navier-Stokes
equations will constrain our choices.

\subsection{Flow past a cylinder}

\subsubsection{Rescaling}
To solve Eqn. \ref{CylinderEqn} using RG techniques, we begin by
transforming the problem to the Oseen variables. As in the terrible
problem, to find a solution which is valid for all $\vec{r}$, we need
to analyze Eqn. \ref{CylinderEqn} using a \emph{dominant balance}
argument. As it stands, different terms of Eqn. \ref{CylinderEqn} will
dominate in different regimes.\footnote{i.e. the LHS, which is
comprised of \emph{inertial terms} dominates for small $|\vec{r}|$
whereas at large $|\vec{r}|$ the \emph{viscous} terms which comprise
the RHS are of equal or greater importance.} Looking for a rescaling of
$\psi$ and $r$ which makes all terms of the same magnitude (more
precisely, of the same order in $R$), yields the rescaling given in
Eqn. \ref{Oseendef} \cite{Proudman57}.
\begin{equation}
\label{Oseendef}
\rho = R r,  \qquad \Psi = R \psi
\end{equation}
Transforming to these variables, Eqn. \ref{CylinderEqn} becomes:
\begin{equation}
\label{OseenCylinder}
\nabla_\rho^4 \Psi(\rho,\theta) = - \frac{1}{\rho} \frac{\partial (\Psi,
\nabla_\rho^2)}{\partial(\rho,\theta)}
\end{equation}
The boundary conditions (Eqn. \ref{Cylinder BC}) become:
\begin{equation}
\label{Oseen Cylinder BC}
\Psi(\rho = R, \theta) = 0, \qquad \frac{\partial \Psi(\rho,\theta)}{\partial
\rho} \bigg|_{\rho=R} = 0, \qquad
\lim_{\rho \to \infty} \frac{\Psi(\rho,\theta)}{\rho} = sin(\theta)
\end{equation}

\subsubsection{Na\"\i ve perturbation analysis}

The next step in obtaining the RG solution is to begin with the ansatz
that the solution can be obtained from a perturbation expansion (Eqn.
\ref{cylinder:naive}).
\begin{equation}
\label{cylinder:naive}
\Psi(\rho,\theta) = \Psi_0(\rho,\theta) + R \Psi_1(\rho,\theta) + R^2
\Psi_2(\rho,\theta)+ \Order[R]{2}
\end{equation}
Substituting Eqn. \ref{cylinder:naive} into Eqn. \ref{OseenCylinder}, and
collecting powers of $R$ yields a series of equations which must be satisfied:
% \begin{widetext}
% \begin{subequations}
% \setlength\arraycolsep{1pt}
% \begin{eqnarray}
% O(R^0):
% \nabla_\rho^4 \Psi_0(\rho,\theta)  &=& \frac{1}{\rho} \left(
% \frac{\partial \Psi_0}{\partial \theta}  \frac{\partial}{\partial \rho} -
% \frac{\partial \Psi_0}{\partial \rho}  \frac{\partial}{\partial \theta}\right)
% \nabla_\rho^2 \Psi_0
% \label{cylinder:0}
% \\
% O(R^1):
% \nabla_\rho^4 \Psi_1(\rho,\theta)   &=& \frac{1}{\rho} \Bigg( \left(
% \frac{\partial \Psi_1}{\partial \theta}  \frac{\partial}{\partial \rho} -
% \frac{\partial \Psi_1}{\partial \rho}  \frac{\partial}{\partial \theta}\right)
% \nabla_\rho^2 \Psi_0 + {}
% \left(
% \frac{\partial \Psi_0}{\partial \theta} \frac{\partial}{\partial \rho} -
% \frac{\partial \Psi_0}{\partial \rho}  \frac{\partial}{\partial \theta}\right)
% \nabla_\rho^2 \Psi_1 \Bigg)
% \label{cylinder:1}
% \\
% O(R^2):
% \nabla_\rho^4 \Psi_2(\rho,\theta)   &=& \frac{1}{\rho} \Bigg( \left(
% \frac{\partial \Psi_2}{\partial \theta}  \frac{\partial}{\partial \rho} -
% \frac{\partial \Psi_2}{\partial \rho}  \frac{\partial}{\partial \theta}\right)
% \nabla_\rho^2 \Psi_0 + {} \nonumber \\
% & & \left(
% \frac{\partial \Psi_0}{\partial \theta} \frac{\partial}{\partial \rho} -
% \frac{\partial \Psi_0}{\partial \rho}  \frac{\partial}{\partial \theta}\right)
% \nabla_\rho^2 \Psi_2 +
% \left( \frac{\partial \Psi_1}{\partial \theta}  \frac{\partial}{\partial \rho} -
% \frac{\partial \Psi_1}{\partial \rho}  \frac{\partial}{\partial \theta}\right)
% \nabla_\rho^2 \Psi_1  \Bigg)
% \label{cylinder:2}
%  \end{eqnarray}
% \label{cylinder:gov}
% \end{subequations}
% \end{widetext}
\setlength\arraycolsep{1pt}
\begin{eqnarray}
\Order[R]{0}:
\nabla_\rho^4 \Psi_0(\rho,\theta)  &=& \frac{1}{\rho} \left(
\frac{\partial \Psi_0}{\partial \theta}  \frac{\partial}{\partial \rho} -
\frac{\partial \Psi_0}{\partial \rho}  \frac{\partial}{\partial \theta}\right)
\nabla_\rho^2 \Psi_0
\\
\Order[R]{1}:
\nabla_\rho^4 \Psi_1(\rho,\theta)   &=& \frac{1}{\rho} \Bigg(\! \left(
\frac{\partial \Psi_1}{\partial \theta}  \frac{\partial}{\partial \rho} -
\frac{\partial \Psi_1}{\partial \rho}  \frac{\partial}{\partial \theta}\right)
\nabla_\rho^2 \Psi_0 + {}
\left(
\frac{\partial \Psi_0}{\partial \theta} \frac{\partial}{\partial \rho} -
\frac{\partial \Psi_0}{\partial \rho}  \frac{\partial}{\partial \theta}\right)
\nabla_\rho^2 \Psi_1 \Bigg)
\nonumber \\
\Order[R]{2}:
\nabla_\rho^4 \Psi_2(\rho,\theta)   &=& \frac{1}{\rho} \Bigg(\! \left(
\frac{\partial \Psi_2}{\partial \theta}  \frac{\partial}{\partial \rho} -
\frac{\partial \Psi_2}{\partial \rho}  \frac{\partial}{\partial \theta}\right)
\nabla_\rho^2 \Psi_0 + \left(
\frac{\partial \Psi_0}{\partial \theta} \frac{\partial}{\partial \rho} -
\frac{\partial \Psi_0}{\partial \rho}  \frac{\partial}{\partial \theta}\right)
\nabla_\rho^2 \Psi_2 \nonumber \\ & & + \left( \frac{\partial \Psi_1}{\partial \theta}  \frac{\partial}{\partial \rho} - \frac{\partial \Psi_1}{\partial \rho}  \frac{\partial}{\partial \theta}\right) \nabla_\rho^2 \Psi_1  \!\Bigg) \nonumber
\label{cylinder:gov}
 \end{eqnarray}

\subsubsection{\Order[R]{0} solution}

The zeroth order part of Eqn. \ref{cylinder:gov} is the same as Eqn.
\ref{OseenCylinder}, and is equally hard to solve. But RG does not need
a complete solution; we just need a starting point. We will begin with
the equation which describes a uniform stream. This is analogous to
the constant \Order{0} solution in the ``terrible'' problem.

A first integral to the \Order[R]{0} equation can be obtained by noting
that any solutions of $\nabla_\rho^2 \Psi_0(\rho,\theta) = 0$ are also
solutions of Eqn. \ref{cylinder:gov}. This is Laplace's equation in
cylindrical coordinates, and has the usual solution (assuming the
potential is single-valued):
\begin{equation}
\label{laplacesoln}
\Psi_0(\rho,\theta) = A_0 + B_0 \ln{\rho} + \sum_{n=1}^\infty \left( \left(A_n
  \rho^n + B_n \rho^{-n} \right) \sin{n \theta} + \left(C_n \rho^n + D_n \rho^{-n} \right)\cos{n \theta} \right)
\end{equation}
We are only interested in solutions with the symmetry imposed by the
uniform flow (Eqn. \ref{Cylinder BC}). Hence $A_0=B_0=C_n=D_n=0$.
Furthermore, the boundary conditions at infinity require that $A_n = 0$
for $n>1$. For simplicity at higher orders, we set $C_n=0$; this is not
required, but these terms will simply re-appear at \Order[R]{1}.
Finally set $A_1=1$ to satisfy the boundary condition at $\infty$ (Eqn.
\ref{Oseen Cylinder BC}). As in the ``terrible'' problem, this is done
for technical convenience, but will not change our results. We are left
with the potential describing the uniform flow:
\begin{equation}
\label{cylinder:0sol}
\Psi_0(\rho,\theta) = \rho \sin(\theta)
\end{equation}

\subsubsection{\Order[R]{1} solution}

By substituting Eqn. \ref{cylinder:0sol} into the \Order[R]{1} governing equation, we
obtain Eqn. \ref{cylinder:oseen}.
\begin{equation}
\label{cylinder:oseen}
\nabla_\rho^4 \Psi_1(\rho,\theta) = \left(\cos(\theta)
\frac{\partial}{\partial \rho} -
\frac{\sin (\theta)}{\rho}\frac{\partial}{\partial \theta} \right) \nabla_\rho^2
\Psi_1
\end{equation}
This equation is formally identical to Oseen's equation (Eqn.
\ref{Oseen:CylinderEqn}), albeit derived through a different argument. This is
fortuitous, as its solutions are known \cite{Tomotika50}. Unfortunately, when
working with stream functions, the solution can only be expressed as an infinite
sum involving combinations of modified Bessel functions, $K_n$, $I_n$.

The general solution can be obtained either by following Tomotika or by
using variation of parameters \cite{Proudman57}. It is comprised of two
parts, the first being a solution of Laplace's equation (as at
\Order[R]{0}). The same considerations of symmetry and boundary
conditions limit our solution: In Eqn. \ref{laplacesoln},
$A_0=B_0=C_n=D_n=0$; $A_n = 0$, if $n>1$. Here, however, we retain the
constants $B_n$, and do not fix $A_1$. This is analogous to what was
done with the homogeneous terms at \Order{1} in the ``terrible''
problem. The second part of the general solution is analogous to a
particular solution in the ``terrible'' problem, and can be obtained
from Tomotika's solution (Eqn. \ref{cylinder:1sol}). These two results
are combined in Eqn. \ref{cylinder:1solb}, which will be the basis for
our RG analysis.
\begin{equation}
\label{cylinder:1solb}
\Psi_1(\rho,\theta) = A_1 \rho \sin{\theta} + \sum_{n=1}^\infty \left( B_n \rho^{-n} + \sum_{m=0}^\infty X_m \rho \Phi_{m,n}(\rho/2) \right) \sin{n \theta}
\end{equation}

Before discussing the application of RG to Eqn. \ref{cylinder:1solb}, it
is worthwhile to discuss Eqn. \ref{cylinder:oseen} in general
terms. Eqn. \ref{cylinder:oseen} may be re-written as:
\begin{equation}
\label{cylinder:oseen2}
\mathcal{L}\Psi_1 \equiv \left(\nabla_\rho^2  - \cos(\theta) \frac{\partial}{\partial \rho} +
\frac{\sin (\theta)}{\rho}\frac{\partial}{\partial \theta} \right) \nabla_\rho^2
\Psi_1 = 0
\end{equation}
We see explicitly that this equation is a linear operator
($\mathcal{L}$) acting on $\Psi_1$, and that the RHS is zero. This is
the \emph{homogeneous Oseen equation}. It is only because of our
judicious choice of $\Psi_0$ that we do not need to deal with the
inhomogeneous counterpart, i.e. with a non-zero RHS. However, the
inhomogeneous Oseen equation governs $\Psi_n$ at all higher orders.
This can be seen for \Order[R]{2} from Eqn. \ref{cylinder:gov}.

In general, the solutions to the inhomogeneous Oseen equation are found
using the method of variation of parameters. It is worth exploring
these solutions, as they provide some insight into the structure of
Eqn. \ref{cylinder:1sol}. We now solve Eqn. \ref{cylinder:oseen2} for a
particular kind of inhomogeneity, one which can be written as a Fourier
sine series.\footnote{The symmetry of the problem precludes the
possibility of cosine terms in the governing equations for $\Psi_n$,
$\forall n > 1$.} We want to solve:
\begin{equation}
\label{cylinder:oseen3}
\mathcal{L} \Psi_1 = \sum_{n=1}^\infty \tilde F_n(\rho) \sin{n \theta}
\end{equation}

The substitution $\nabla^2\Psi_1 = e^{\rho \cos{\theta}/2}
\Pi(\rho,\theta)$\footnote{ $\nabla^2 \Psi_1(\rho,\theta)$ is the
\emph{vorticity}.}, allows us to obtain the first integral of Eqn.
\ref{cylinder:oseen3}. This result is given by Eqn.
\ref{cylinder:oseen4} \cite{Proudman57}.
\begin{equation}
\label{cylinder:oseen4}
\left( \nabla^2 - \frac{1}{4}\right)\Pi(\rho,\theta)  = \sum_{n=1}^\infty F_n(\rho) \sin{n \theta}
\end{equation}
Here $F_n(\rho) = e^{-\rho\cos{\theta}/2} \tilde F_n(\rho)$. To solve for
    $\Pi(\rho,\theta)$, begin by noting that the symmetry of the
    inhomogeneous terms implies that $\Pi(\rho,\theta)$ can be written
    as a sine series. Consequently, substitute $\Pi(\rho,\theta) = \sum_{n=1}^\infty
    g_n(\rho)\sin{n\theta}$ into Eqn. \ref{cylinder:oseen4} to obtain:
\begin{equation}
\label{oseen:1stint}
g_n^{''}(\rho) + \frac{1}{\rho}g_n^{'}(\rho) - \left( \frac{1}{4} +
\frac{1}{\rho^2} \right) g_n(\rho) = F_n(\rho)
\end{equation}
The fundamental solutions of this equation are $K_n(\rho/2)$,
$I_n(\rho/2)$. Using variation of parameters, the
general solution of Eqn. \ref{oseen:1stint} may be written:
\begin{equation}
\label{oseen:1stintsol}
g_n(\rho)=-I_n\left(\frac{\rho}{2}\right)\left(\alpha_n +
\mathcal{J}_1^{(n)}(\rho) \right) +
K_n\left(\frac{\rho}{2}\right)\left(\beta_n+\mathcal{J}_2^{(n)}
(\rho)\right)
\end{equation}
Here, $\mathcal{J}_1^{(n)}(\rho)=\int\textrm{d}\rho \rho F_n(\rho)
K_n(\rho/2)$, $\mathcal{J}_2^{(n)}(\rho) = \int\textrm{d}\rho \rho
F_n(\rho)  I_n(\rho/2)$, and $\alpha_n$, $\beta_n$ are constants. The
next step is to undo our original transformation, and to solve the
resulting equation:
\begin{eqnarray}
\nabla^2 \Psi_1(\rho,\theta) &=& e^{\frac{\rho
    \cos{\theta}}{2}} \sum_{n=1}^\infty g_n(\rho) \sin{n \theta} \\
&=& \sum_{n=1}^\infty b_n(\rho)\sin{n\theta} \nonumber
\label{oseen:2ndint}
\end{eqnarray}
In this equation, $b_n(\rho)=\sum_{m=1}^\infty
g_m(\rho)\left(I_{n-m}\left(\rho/2\right)-I_{n+m}\left(\rho/2\right)\right)$.
We have the unfortunate happenstance that each $b_n$ depends on the
\emph{all} of the harmonics of the first integral. This is the origin
of the nested sum (over $m$) in Tomotika's solution (Eqn. \ref{cylinder:1sol}).

As before, symmetry will require that $\Psi_1(\rho,\theta)$ be representable as a
sine series: $\Psi_1(\rho,\theta) = \sum_{m=1}^\infty
X_m(\rho)\sin{m\theta}$. With this substitution we obtain (for each
$m$), the radial component of Poisson's equation in cylindrical coordinates:
\begin{equation}
\label{possion}
X_m^{''}(\rho) + \frac{1}{\rho} X_m^{'}(\rho) -
\frac{m^2}{r^2}X_m(\rho)=b_m(\rho)
\end{equation}
The fundamental solutions were discussed before in the context of
Laplace's equation: $\rho^m$, $\rho^{-m}$. As before, a particular
integral is obtained through variation of parameters, and the general
solution may be written:
\begin{equation}
\label{cylinder:sol2}
X_n(\rho) = -\rho^n \left(A_n + \mathcal{I}_1^{(n)}(\rho)\right) +
\frac{1}{\rho^n}\left(B_n+\mathcal{I}_2^{(n)}(\rho)\right)
\end{equation}
Here $\mathcal{I}_1^{(n)}(\rho)=\int\textrm{d}\rho -\rho b_n(\rho)
  /(2n\rho^n)$, $\mathcal{I}_2^{(n)}(\rho)=\int\textrm{d}\rho -\rho b_n(\rho)
  \rho^n/(2n)$, and $A_n$, $B_n$ are integration constants.

It is useful to relate Eqn. \ref{cylinder:sol2} to Tomotika's solution
(Eqn. \ref{cylinder:1sol}). There are four integration constants for
each angular harmonic. Two are obvious: $A_n$,$B_n$. The other two
arise in the first integral (the vorticity solution), Eqn.
\ref{oseen:1stintsol}. However, every vorticity integration constant
appears in each harmonic of Eqn. \ref{cylinder:sol2}. For example, one
cannot uniquely assign $\alpha_1$ and $\beta_1$ to the $\sin{\theta}$
harmonic of Eqn. \ref{cylinder:sol2}. However, if one considers $n$
terms from Eqn. \ref{oseen:1stintsol} and $n$ terms from Eqn.
\ref{cylinder:sol2}, there will be $4n$ integration constants --- four
per retained harmonic of Eqn. \ref{cylinder:sol2}.
% TODO
In passing we note that matched asymptotics workers avoid this problem
by using the vorticity directly, and thereby simplify their treatment
of boundary conditions. This approach does not work in conjunction with
RG.

It is mildly disconcerting to have four integration constants, as there
are only three boundary conditions for each harmonic (Eqn. \ref{Oseen
Cylinder BC}). However, two of the constants --- $A_n$ and $\alpha_n$
--- will be determined by the boundary conditions at infinity.  This
claim is not obvious, particularly since terms which are divergent
prior to renormalization might not be present after the renormalization
procedure. We outline here an argument which can be made rigorous.
There are two kinds of divergences in Eqn. \ref{cylinder:sol2}: Terms
which are secular as $\rho \rightarrow 0$, and terms which diverge too
quickly as $\rho \rightarrow \infty$.\footnote{To be precise, terms
which diverge faster than $\rho$ as $\rho \rightarrow \infty$ are
problematic, and prevent satisfying the boundary conditions (Eqn.
\ref{Oseen Cylinder BC}).}

After renormalization, we will try to need to meet the boundary
conditions (Eqn. \ref{Oseen Cylinder BC}). As in the case of the
``terrible'' problem, it will turn out that the simplest approach to
renormalization yields a renormalized perturbation solution which is
the same as the na\" \i ve series. Consider Eqn. \ref{cylinder:sol2}.
The terms which are secular as $\rho \rightarrow 0$ will not preclude
satisfying the boundary conditions. Those which diverge too quickly as
$\rho \rightarrow \infty$, however, will conflict with Eqn. \ref{Oseen
Cylinder BC}.

These terms must be eliminated by a suitable choice of integration
constants. It turns out not to matter whether we do this before or
after the renormalization procedure. For simplicity, we will do it before
renormalizing. First, the coefficient of $\rho^n$ must vanish for all $n >
1$. This can happen, with an appropriate choice of $A_n$, if
\begin{displaymath}
\lim_{\rho \to \infty} \mathcal{I}_1^{(n)}(\rho) \sim \Order[1]{}
\end{displaymath}
For this requirement to be met, the coefficient of $I_n(\rho/2)$ in
Eqn. \ref{oseen:1stintsol} must vanish (e.g., $\alpha_n = \lim_{\rho
\rightarrow \infty} \mathcal{J}_1^{n}(\rho)$). It is always possible to
choose $\alpha_n$ appropriately, because the following condition is
satisfied for all $n$:
\begin{displaymath}
\lim_{\rho \to \infty} \mathcal{J}_1^{(n)}(\rho) \sim \Order[1]{}
\end{displaymath}
In our problem this is true because $F_n(\rho)$ is based on solutions
to the lower order governing equations. By construction, these are
well-behaved as $\rho \to \infty$. Therefore, for the inhomogeneous
Oseen equation under consideration (Eqn. \ref{cylinder:oseen4}), we see
that two of the four integration constants --- $A_n$, $\alpha_n$ ---
are needed to satisfy the boundary conditions at infinity.

More specifically, the immediate problem requires us to consider the
homogeneous Oseen's equation (Eqn. \ref{cylinder:oseen2}), and
Tomotika's solution (Eqn. \ref{cylinder:1sol}). For this problem,
$F_n(\rho) = 0$, and the coefficient of $I_n(\rho/2)$ in Eqn.
\ref{oseen:1stintsol} has no $\rho$ dependence. So we simply choose
$\alpha_n$ such that this coefficient vanishes. Simplifying Eqn.
\ref{oseen:1stintsol}, we then have the following solution for the
vorticity:
\begin{equation}
\nabla^2 \Psi_1(\rho,\theta) = e^{\frac{\rho
    \cos{\theta}}{2}} \sum_{n=1}^\infty
    K_n\left(\frac{\rho}{2}\right)\left(\beta_n \right) \sin{n \theta}
\end{equation}
Since this solution for the vorticity is well-behaved as $\rho \to
\infty$, it follows that we can choose $A_n$ ($n > 1$) in
Eqn. \ref{cylinder:sol2} so that the coefficient of $\rho^n$
vanishes as $\rho \to \infty$. We are left with the solution
\begin{equation}
\label{xneqn}
X_n(\rho) = A_n \rho \delta_{n,1} + \rho^n
 \left(\mathcal{I}_1^{(n)}(\rho)-\mathcal{I}_1^{(n)}(\infty)\right) +
 \rho^{-n} \left(B_n + \mathcal{I}_2^{(n)}(\rho) \right)
\end{equation}
For the homogeneous Oseen's equation, $\mathcal{I}_1^{(n)}(\rho)$ and
$\mathcal{I}_2^{(n)}(\rho)$ simplify to:
\begin{eqnarray}
\label{someintegrals}
\mathcal{I}_1^{(n)}(\rho) &=&
 \int \textrm{d}\rho \frac{-\rho}{2 n} \rho^{-n}
 \left(\sum_{m=1}^{\infty} \beta_m K_m\left(\frac{\rho}{2}\right) \left(
 I_{n-m}\left(\frac{\rho}{2}\right) - I_{n+m}\left(\frac{\rho}{2}\right)
\right) \right) \\
\mathcal{I}_2^{(n)}(\rho) &=&
 \int \textrm{d}\rho \frac{-\rho}{2 n} \rho^n
 \left(\sum_{m=1}^{\infty} \beta_m K_m\left(\frac{\rho}{2}\right) \left(
 I_{n-m}\left(\frac{\rho}{2}\right) - I_{n+m}\left(\frac{\rho}{2}\right)
\right) \right)
\end{eqnarray}

This result is fundamentally the same as Tomotika's (Eqn.
\ref{cylinder:1sol}). However, his solution is more useful, as he
accomplished the integrals in Eqn. \ref{someintegrals}. What is the
point of all this work? Firstly, the approach based on the variation of
parameters may be applied to the inhomogeneous Oseen's equation, which
must be solved for orders higher than \Order[R]{1}. Secondly, we see
explicitly what happens to the two sets of integration constants
$\alpha_n$ and $A_n$. Tomotika's solution has but two integration
constants\footnote{There is also $A_1$, but that is a special case.} --- $B_n$ and $\beta_n$. The other constants have already been chosen so as to satisfy the boundary conditions at $\infty$. We have shown explicitly how they must be determined, and
stated without proof that this may be done prior to renormalization. In
short, we have explained why Eqn. \ref{cylinder:1solb} is the
appropriately general \Order[R]{1} solution for our na\" \i ve
perturbation analysis.

In addition to explaining why Tomotika's solution is a suitable
starting point for RG, our analysis also connects with the \Order[R]{1}
solution of Proudman and Pearson \cite{Proudman57}. We have shown that the
vorticity must be well-behaved at $\rho=\infty$ if the overall
solution is to satisfy the boundary conditions.

{\it a. Secular behavior}

Combining Eqns. \ref{cylinder:0sol}, \ref{cylinder:1solb}, we begin the
following na\" \i ve solution:
\begin{equation}
\label{cylinder:naivesol}
\Psi_(\rho,\theta) = \rho \sin(\theta) + R \left( A_1 \rho
\sin{\theta} + \sum_{n=1}^\infty \left( B_n \rho^{-n} +
\sum_{m=0}^\infty X_m \rho \Phi_{m,n}\left(\frac{\rho}{2}\right) \right)
\sin{n \theta} \right) +\Order[R]{2}
\end{equation}
Although intimidating, this is conceptually equivalent to
Eqn. \ref{Oeterriblesol2} (in the terrible problem). The first step in
our analysis is identifying which terms are divergent. As explained
above, Eqn. \ref{cylinder:naivesol} is specifically
constructed to be of \Order[\rho]{1} as $\rho \to \infty$. In
fact, only the \Order[R]{0} and $A_1$ terms matter at large
$\rho$. As $\rho \to 0$, however, many other terms in
Eqn. \ref{cylinder:naivesol} diverge. All of the $B_n$ terms
diverge. Most of the $\Phi_{m,n}(\rho)$ terms are also secular.

Rather than enumerating and sorting through the different divergences,
we simply treat the problem abstractly. Eqn. \ref{cylinder:naivesol}
can be rewritten as:
\begin{equation}
\label{cylinder:beginRG}
\Psi_(\rho,\theta) = \rho \sin(\theta) + R \left( A_1 \rho
\sin{\theta} + \mathcal{R}(\rho,\theta; \{B_i\}; \{X_j\}) +
\mathcal{S}(\rho,\theta;  \{B_m\}; \{X_n\}) \right)
\end{equation}
Here, $\mathcal{S}$ includes the terms which are secular as
$\rho \to 0$, and $\mathcal{R}$ includes regular terms.

\subsubsection{Renormalization}

Equation \ref{cylinder:beginRG} is renormalized just like the
terrible problem. We begin with the renormalized perturbation
expansion given in Eqn. \ref{cylinder:RG2}. Note that we are not
specifying the details of which terms are secular, or how we are
``splitting'' these terms. The only term we are explicitly considering
is $A_1$. This is a trick built on consideration of the terrible
problem. Our ``best'' solution (Eqn. \ref{terrible:final}) to that
problem was built on the renormalization of just \emph{one} constant,
$B_0$ in Eqn. \ref{terrible:O2sol}. Essentially, we will repeat that
procedure here, using $A_1$ as that constant.
\begin{eqnarray}
\label{cylinder:RG2}
\Psi(\rho,\theta) &=& \rho \sin(\theta) + R \Big( A_1(\tau) \rho
\sin{\theta} + \mathcal{R}(\rho,\theta; \{B_i(\tau)\}; \{X_j(\tau)\}) + \\ \nonumber & &
\mathcal{S}(\rho,\theta;  \{B_m(\tau)\}; \{X_n(\tau)\}) - \mathcal{S}(\tau,\theta; \{B_m(\tau)\}; \{X_n(\tau)\}) + \Order[R]{2}
\Big)
\end{eqnarray}

We will now apply the RG condition --- $\partial_\tau \Psi(\rho,\theta) = 0$ --- to this
  equation. Accomplishing this in complete generality is
  difficult. However, using our experience from the terrible problem,
  we can see that this is not necessary. The RG condition may be
  satisfied as follows: First, suppose that $X_n^{'}(\tau) = \Order[R]{2} \quad \forall
  n$, $B_m^{'}(\tau) = \Order[R]{2} \quad \forall m$. These equations are
  satisfied by $X_n(\tau) = \chi_n$, $B_m(\tau)=\beta_m$. Substituting
  these results into Eqn. \ref{cylinder:RG2}, and applying the RG condition
  results in:
\begin{equation}
0 = R \left( A_1^{'}(\tau) \rho \sin{\theta} -
\mathcal{S}^{'}(\tau,\theta; \{\beta_m\}; \{\chi_n\})\right)
\end{equation}
This is easily solved for $A_1(\tau)$.
\begin{equation}
\label{cylinder:RGsol1}
A_1(\tau) = \frac{\mathcal{S}(\tau,\theta; \{\beta_m\}; \{\chi_n\})}{
  \rho \sin{\theta}} + \alpha_1
\end{equation}

We have explicitly validated our supposition that $\{X_n(\tau)\}$ and
$\{B_m(\tau)\}$ can be constants. With this supposition, we have shown
that the RG condition applied to Eqn. \ref{cylinder:RG2} can be
satisfied with an appropriate choice of $A_1(\tau)$. We have satisfied
the RG condition through clever tricks derived from our experience
with the terrible problem. However, this solution is entirely
valid, and our experience with the terrible problem has shown us that
more complicated solutions are asymptotically equivalent.

Substituting Eqn. \ref{cylinder:RGsol1} into Eqn. \ref{cylinder:RG2},
and setting $\tau = \rho$, we obtain our renormalized solution:
\begin{equation}
\label{cylinder:endRG}
\Psi_(\rho,\theta) = \rho \sin(\theta) + R \left( \alpha_1 \rho
\sin{\theta} + \mathcal{R}(\rho,\theta; \{\beta_i\}; \{\chi_j\}) +
\mathcal{S}(\rho,\theta;  \{\beta_m\}; \{\chi_n\}) \right)
\end{equation}
By now it should not be surprising that this is the same equation as
our naive perturbation solution (Eqn. \ref{cylinder:beginRG}), and by
extension the same solution obtained by Tomotika \cite{Tomotika50}. As in the case
of the terrible problem, however, we now know that this
is a uniformly valid approximation. We now may choose the integration
constants to satisfy the boundary conditions, and then calculate the
drag coefficient.

{\it a. Truncation}

Unfortunately, there are infinitely many integration constants, and it
is impossible to apply the boundary conditions to our renormalized
solution (or Eqn. \ref{cylinder:naivesol}). To progress further, we
must make the same sort of uncontrolled approximations made by previous
workers \cite{Proudman57,Tomotika50}.\footnote{Kaplun was able to avoid
this difficulty by using the velocity field instead of stream
functions, although his approach brings other problems: the solution
cannot be expressed in closed form, and must be approximated to apply
the boundary conditions (see section \ref{matchedcylinder}).}

Our approximation consists of a careful truncation, in both $m$ and
$n$, of the series in Eqn. \ref{cylinder:naivesol}. There are two
important points to consider. First is the $\sin{\theta}$ symmetry of
the overall problem: terms proportional to $\sin{\theta}$ reflect the
symmetries exhibited by the uniform flow which are imposed on our
solution via the boundary conditions at infinity. The importance of
this harmonic is further seen in Eqn. \ref{cylinder:convenientdrag}:
Only the coefficient of $\sin{\theta}$ will be needed for the
computation of $C_D$.

Secondly we recall that the remaining boundary conditions are imposed
at the surface of the sphere, at $\rho = R$ in Oseen coordinates. When
applying the boundary conditions, terms which are secular as $\rho \to
0$ will therefore be most important. Specifically, we cannot truncate
any terms which are divergent, although we are at liberty to set their
coefficients equal to zero.

These considerations allow exactly one solution. First, set all $B_n =
0 \quad n > 1$. Secondly, set all $X_m = 0 \quad m > 0$. We retain
three coefficients: $A_1, B_1, X_0$, which will permit the boundary
conditions to be satisfied for the $\sin{\theta}$ harmonic. What about
the higher harmonics? These terms are truncated in an
\emph{uncontrolled} approximation. However, as we will show, the
discarded terms are \Order[R^3\log{R}]{} or higher at the surface of
the sphere. They are regular terms, and thus negligible in comparison
to the secular terms retained (which are \Order[R]{-1}).

Now, suppose we follow Tomotika, and try to extend this approach, by
retaining a few more terms. The next step would be to retain the $B_2,
X_1$ terms, and to try to satisfy the boundary conditions for the
$\sin{2 \theta}$ harmonic. As before, all the higher $B_n,X_m$ are set
to zero. Why not include the next harmonic or two?

The answer lies in the terms we discard. If we satisfy the boundary
conditions at $\rho = R$ for the first $n$ harmonics, we must retain
the coefficients $X_0, ... ,X_n-1$. To minimize the amount of
truncation we do, first set $X_m =0$ for $\forall m > n-1$ and $B_k=0$
for $\forall k > n$. What, then, is the form of the terms which are
discarded from our solution?
\begin{equation}
\label{cylinder:discards}
\Psi_{\textrm{discard}}^{(n)}(\rho,\theta) = R \left(\sum_{k=n+1}^\infty
\sum_{m=0}^{n-1} X_m \Phi_{m,k}(\rho/2) \rho \sin{k \theta} \right)
\end{equation}
$\Psi_{\textrm{discard}}^{(n)}(\rho,\theta)$ is largest as $\rho \to
0$, and will be most important at $\rho = R$, on the surface of the
cylinder. If we retain only the $n=1$ harmonic,
$\Psi_{\textrm{discard}}^{(1)}(\rho,\theta) \sim \Order[R^3
\log{R}]{}$. Since we are only working to \Order[R]{1}, this is fine.
We must also consider the derivative, since we want to satisfy all of
the boundary conditions (Eqn. \ref{Cylinder BC}) to the same order.
$\Psi_{\textrm{discard}}^{'(1)}(\rho,\theta) \sim \Order[R^2\log{R}]{}$
Therefore, in the case where we retain only the $\sin{\theta}$
harmonic, the discarded terms are negligible, as we are working to
\Order[R]{1}.\footnote{This argument is somewhat simplistic: The
neglected terms also contribute, when meeting the boundary conditions,
to the values of the retained coefficients. i.e. All non-zero $X_m$
affect $X_0$. But these are lower order effects.} When we retain higher
harmonics, everything changes. Table \ref{tab:discards} shows the
magnitude of the discarded terms at $\rho = R$ for the first four
harmonics.

\begin{table}
\begin{center}
  \begin{tabular}{|c|cccc|} \hline
   n = & 1 & 2 & 3 & 4 \\
   \hline
   $\Psi_{\textrm{discard}}^{(n)}(\rho,\theta)$ & \Order[R^3
  \log{R}]{} & \Order[R]{2} & \Order[R]{1} & \Order[R]{0} \\
   $\Psi_{\textrm{discard}}^{'(n)}(\rho,\theta)$ &
   \Order[R^2\log{R}]{} & \Order[R]{1} & \Order[R]{1} & \Order[R]{-1} \\ \hline
  \end{tabular}
 \caption{Relative importance of discarded terms at $\rho=R$.}
  \label{tab:discards}
\end{center}
\end{table}

From Table \ref{tab:discards}, we see immediately that to retain
$\sin{2 \theta}$ harmonics, we must have an error in our derivative
boundary condition of \Order[R]{1} --- the order to which we are trying
to work. If we retain higher harmonics, this situation gets worse.
First we have an \Order[R]{1} error in the stream function itself, and
then we begin to have errors which are \emph{divergent} in $R$! For
$n>4$, both $\Psi_{\textrm{discard}}^{(n)}(\rho,\theta)$ and
$\Psi_{\textrm{discard}}^{'(n)}(\rho,\theta)$ are increasingly
divergent functions of $R$.

Since it is in practice impossible to fit the boundary conditions to
Eqn. \ref{cylinder:naivesol}, we must truncate the series expansion. We
have shown that there is only one truncation consistent with both the
symmetry requirements of the problem and the demand that we satisfy the
boundary conditions to \Order[R]{1}:
\begin{equation}
\label{cylinder:truncsol}
\Psi_(\rho,\theta) = \rho \sin(\theta) + R \left( A_1 \rho + B_1
\rho^{-1} + X_0 \rho \Phi_{0,1}\left(\frac{\rho}{2}\right) \right)
\sin{\theta}
\end{equation}

% Get syntax consistent with greek letters for Renormalized result.

This result is identical to Proudman's \Order[R]{1} result for the
Oseen stream function \cite{Proudman57}. However, he arrives at this
result by considering matching requirements with the \Order[R]{0}
Stokes expansion and by imposing $\sin{\theta}$ symmetry on the first
integral (Eqn. \ref{oseen:2ndint}). Our approach arrives at the same
conclusion, but without the need for asymptotic matching or the two
expansions it requires. Moreover, we did not need the expertise and
finesse which matched asymptotics workers needed to deduce the unusual
form of their expansions (e.g., the $1/\log{R}$ term in Eqn.
\ref{form2}). Finally, we note that Tomotika's numerical results
support our truncation \cite{Tomotika50}.

{\it b. Meeting boundary conditions}

It is straightforward to apply the boundary conditions
(Eqn. \ref{Cylinder BC}) to Eqn. \ref{cylinder:truncsol}. To satisfy
the condition at infinity, $A_1 = 0$. The other two requirements are
met by the following choice of coefficients:
\begin{eqnarray}
\label{cylinder:BC}
B_1 &=& \frac{-R^2 \Phi_{0,1}^{'}(R/2)}{4 \Phi_{0,1}(R/2) + R
  \Phi_{0,1}^{'}(R/2)}\\
X_0 &=& \frac{-4}{R \left( 4 \Phi_{0,1}(R/2)+R
  \Phi_{0,1}^{'}(R/2)\right)}
\end{eqnarray}
Notice that we are using the Oseen stream function. The Stokes'
function is related by: $\psi(r,\theta)=\Psi(rR,\theta)/R$. Putting
everything together, we have the new result given by Eqn.
\ref{cylinder:finalsol}.
\begin{eqnarray}
\label{cylinder:finalsol}
\Psi_(\rho,\theta) &=& \rho \sin(\theta) + R \Bigg( \frac{-R^2 \Phi_{0,1}^{'}(R/2)}{4 \Phi_{0,1}(R/2) + R
  \Phi_{0,1}^{'}(R/2)} \rho^{-1} + \\ & & \frac{-4}{R \left( 4 \Phi_{0,1}(R/2)+R
  \Phi_{0,1}^{'}(R/2)\right)} \rho \Phi_{0,1}(\rho/2) \Bigg) \sin{\theta} \nonumber
\end{eqnarray}
Remember that although our truncated solution satisfies the boundary
conditions \emph{exactly}, it only satisfies the governing equations
\emph{approximately}.

\subsubsection{Calculating the drag coefficient}

\begin{figure}[tb]
\psfrag{Re}{\mbox{\large $R$}}
\psfrag{CD}{\mbox{\large $C_D R/4\pi$}}
\psfrag{Kaplun}{Kaplun, Eqn. \ref{cylinder:KaplunCD}}
\psfrag{Imai}{Imai, Eqn. \ref{imaisol}}
\psfrag{Bairstow XXXXXXX}{Bairstow, Eqn. \ref{Bairstowsol}}
\psfrag{Lamb}{Lamb, Eqn. \ref{Oseen:LambDrag}}
\psfrag{RG}{RG, Eqn. \ref{cylinder:O1CD}}
\begin{center}
\includegraphics[width=.9 \textwidth]{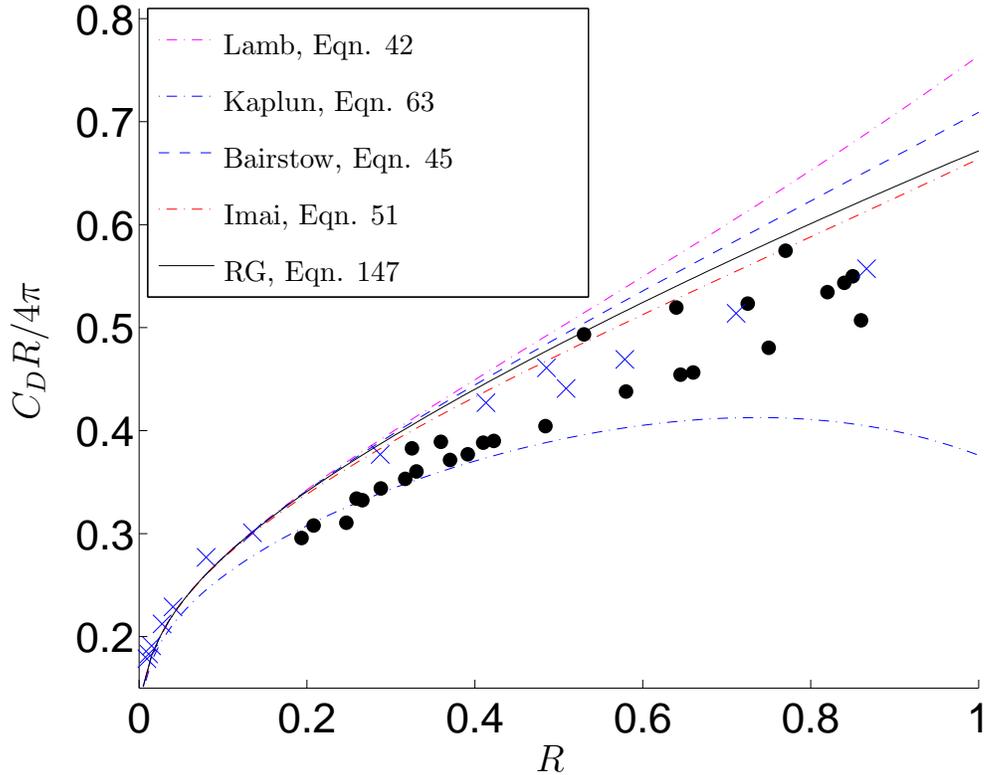}
\caption{(Color online) Drag on cylinder, comparing RG predictions to other theories at low $R$.}
\label{fig:cylinder:RG}
\end{center}
\end{figure}

We now transform Eqn. \ref{cylinder:finalsol} into Stokes'
coordinates, and substitute the result into
Eqn. \ref{cylinder:convenientdrag}.\footnote{Or, alternatively, into
  Eqns. \ref{cylinderstreamfunction}, \ref{CDcylinder},
and \ref{cylpressure}.} We thereby obtain a new result for $C_D$, given by Eqn. \ref{cylinder:O1CD}.
\begin{equation}
\label{cylinder:O1CD}
C_D = \frac{\pi \left( -12 \Phi_{0,1}^{'}(R/2) + R \left( 6
  \Phi_{0,1}^{''}(R/2) + R \Phi_{0,1}^{'''}(R/2) \right) \right)}{8
  \Phi_{0,1}(R/2) + 2 R \Phi_{0,1}^{'}(R/2)}
\end{equation}
This result is plotted in Figure \ref{fig:cylinder:RG}, where it is
compared against the principal results of Oseen theory, matched
asymptotic theory, and experiments. When compared asymptotically, all
of these theoretical predictions agree. At small but not infinitesimal
Reynolds number, the largest difference is seen between Kaplun's second
order result and the first order predictions, including Eqn.
\ref{cylinder:O1CD}. As explained previously, current experimental data
cannot determine whether Kaplun's second order matched asymptotics
solution is actually superior.

The RG result lies among the first order predictions. Fundamentally,
the RG calculation begins with an equation similar to Oseen's, so this
is not too surprising. Within this group Eqn. \ref{cylinder:O1CD}
performs very well, and is only slightly bettered by Imai's prediction
(Eqn. \ref{imaisol}). These two results are very close over the range
$0 < R < 1$.

\begin{figure}[tb]
\psfrag{Re}{\mbox{\large $R$}}
\psfrag{CD}{\mbox{\large $C_D R/4\pi$}}
\psfrag{Kaplun}{Kaplun, Eqn. \ref{cylinder:KaplunCD}}
\psfrag{Imai}{Imai, Eqn. \ref{imaisol}}
\psfrag{Bairstow XXXXXXX}{Bairstow, Eqn. \ref{Bairstowsol}}
\psfrag{Lamb}{Lamb, Eqn. \ref{Oseen:LambDrag}}
\psfrag{RG}{RG, Eqn. \ref{cylinder:O1CD}}
\begin{center}
\includegraphics[width=.9 \textwidth]{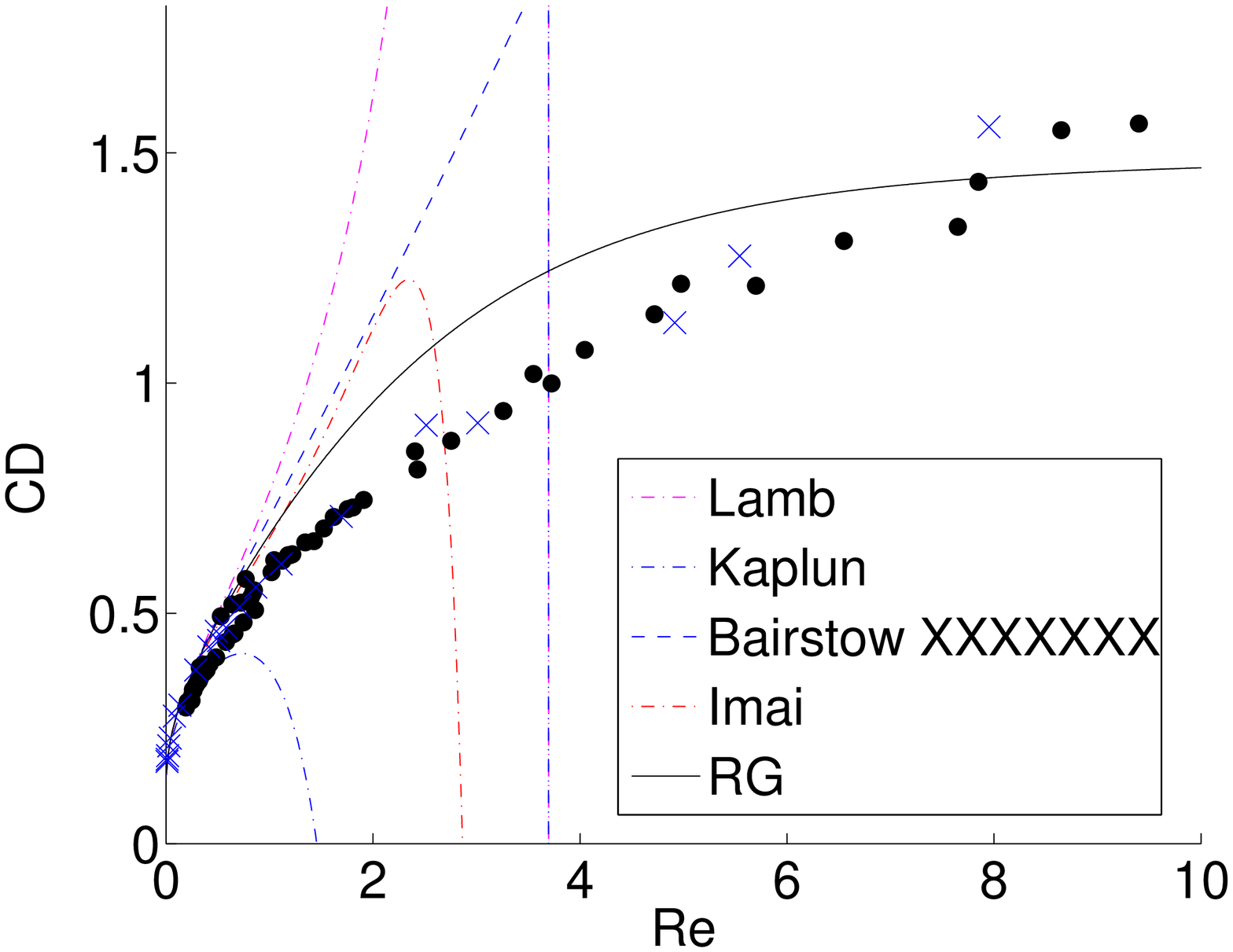}
\caption{(Color online) Drag on a Cylinder, comparing RG predictions to other theories at higher $R$.}
\label{fig:cylinder:RG2}
\end{center}
\end{figure}

The real strength of Eqn. \ref{cylinder:O1CD} can be seen in in Figure
\ref{fig:cylinder:RG2}. As the Reynolds number increases beyond $R=1$,
all other theories begin to behave pathologically. They diverge from
experimental measurements and behave non-physically (e.g., a negative
drag coefficient). The RG prediction suffers from none of these
problems; it is well behaved for all $R$. As it is still based on a
perturbative solution, it does become less accurate as $R$ increases.
%
% \subsection{$\mathcal{O}(\epsilon^2)$ Solution}
% Pain. Is this even possible with considerations of truncation error?
% Maybe can use the new terms at next Order to correct? What about the
% solution driven by just B1 at O1?

\subsection{Flow past a sphere}

\subsubsection{Rescaling}
Our analysis of low Reynolds number flow past a sphere closely follows
both the cylinder problem and the terrible problem. We omit redundant
explanations. As before, the first step is a rescaling of both $r$ and
$\psi$ --- the transformation into Oseen coordinates. A dominant
balance analysis identifies the rescaling given in Eqn.
\ref{sphere:rescale}.
\begin{equation}
\label{sphere:rescale}
\rho = Rr, \quad \Psi = R^2 \psi
\end{equation}
In Oseen variables, the governing equation (Eqn. \ref{SphereEqn}) becomes:
\begin{equation}
\label{sphere:oseeneqn}
D_\rho^4 \Psi(\rho,\mu) = \frac{1}{\rho^2}\left(\frac{\partial(\Psi(\rho,\mu), D_\rho^2 \Psi(\rho,\mu))}{\partial(\rho,\mu)} +
2 D_\rho^2 \Psi(\rho,\mu) L_\rho \Psi(\rho,\mu) \right)
\end{equation}
where
\begin{equation}
 \mu \equiv \cos{\theta}, \qquad D_\rho^2 \equiv \frac{\partial^2}{\partial \rho^2} + \frac{1-\mu^2}{\rho^2} \frac{\partial^2}{\partial \mu^2}, \qquad
L_\rho \equiv \frac{\mu}{1-\mu^2}\frac{\partial}{\partial \rho} + \frac{1}{\rho} \frac{\partial}{\partial \mu}
\end{equation}
The boundary conditions (Eqn. \ref{Sphere BC}) transform into:
\begin{equation}
\label{sphere:oseenbc}
\Psi(\rho = R, \mu) = 0, \qquad \frac{\partial \Psi(\rho,\mu)}{\partial \rho} \bigg|_{\rho=R} = 0, \qquad \lim_{\rho \to \infty} \frac{\Psi(\rho,\mu)}{\rho^2} = \frac{1}{2} \left( 1 - \mu^2 \right)
\end{equation}

\subsubsection{Na\"\i ve perturbation analysis}

We continue by substituting our na\" \i ve perturbation assumption
(Eqn. \ref{cylinder:naive}) into Eqn. \ref{sphere:oseeneqn}, and then
collecting powers of $R$.
\begin{subequations}
\setlength\arraycolsep{1pt}
\begin{eqnarray}
\Order[R]{0}:
D_\rho^4 \Psi_0(\rho,\mu) &=&\frac{1}{\rho^2}\left(\frac{\partial(\Psi_0(\rho,\mu), D_\rho^2 \Psi_0(\rho,\mu))}{\partial(\rho,\mu)} + 2 D_\rho^2 \Psi_0(\rho,\mu) L_\rho \Psi_0(\rho,\mu) \right) \qquad \qquad \label{sphere:0}
\\
\Order[R]{1}:
D_\rho^4 \Psi_1(\rho,\mu)
&=& \frac{1}{\rho^2}\Bigg(\frac{\partial(\Psi_0, D_\rho^2
  \Psi_1)}{\partial(\rho,\mu)} + \frac{\partial(\Psi_1, D_\rho^2
  \Psi_0)}{\partial(\rho,\mu)} + \nonumber \\ & & 2 \left(D_\rho^2 \Psi_0 L_\rho \Psi_1 + D_\rho^2 \Psi_1 L_\rho \Psi_0 \right)\Bigg) \label{sphere:1}
\\
\Order[R]{2}:
D_\rho^4 \Psi_2(\rho,\mu)
&=& \frac{1}{\rho^2}\Bigg(\frac{\partial(\Psi_0, D_\rho^2
  \Psi_2)}{\partial(\rho,\mu)} + \frac{\partial(\Psi_1, D_\rho^2
  \Psi_1)}{\partial(\rho,\mu)} + \frac{\partial(\Psi_2, D_\rho^2
  \Psi_0)}{\partial(\rho,\mu)} + \nonumber \\ & &
2 \left(D_\rho^2 \Psi_0 L_\rho \Psi_2 + D_\rho^2 \Psi_1 L_\rho \Psi_1
+ D_\rho^2 \Psi_2 L_\rho \Psi_0 \right) \Bigg)
\label{sphere:2}
\end{eqnarray}
\label{sphere:gov}
\end{subequations}

\subsubsection{\Order[R]{0} solution}
As seen with both the cylinder problem and the terrible problem, Eqn.
\ref{sphere:0} is the same as the original governing  equation (Eqn.
\ref{sphere:oseeneqn}). As before, we proceed using an incomplete
solution for $\Psi_0$: the uniform stream which describes flow far from
any disturbances. Analogously to the cylinder, we notice that Eqn.
\ref{sphere:0} is satisfied if $\Psi_0(\rho,\mu)$ obeys  $D_\rho^2
\Psi_0(\rho,\mu) = 0$. The general solution of this equation which also
satisfies the appropriate symmetry requirement ($\Psi_0(\rho,\mu = \pm
1) = 0$) is given by Eqn. \ref{sphere:laplace}.
\begin{equation}
\label{sphere:laplace}
\Psi_0(\rho,\mu)= \sum_{n=0}^\infty \left(A_n \rho^{n+1} + B_n
\rho^{-n} \right) Q_n(\mu)
\end{equation}
Here $Q_n(\mu)$ is defined as in Eqn. \ref{Oseen:Goldstein}. Following
the analysis used for the cylinder, we set all of the coefficients to
zero, excepting $A_1 = -1/2$. This choice of $A_1$ satisfies the
uniform stream boundary condition (Eqn. \ref{sphere:oseenbc}) at $\rho
= \infty$. We thereby obtain:
\begin{equation}
\label{sphere:0sol}
\Psi_0(\rho,\mu) = - \rho^2 Q_1(\mu)
\end{equation}

\subsubsection{\Order[R]{1} solution}
Substituting Eqn. \ref{sphere:0sol} into Eqn. \ref{sphere:1}, we
obtain Eqn. \ref{sphere:1eqn}:
\begin{equation}
\label{sphere:1eqn}
D_\rho^4 \Psi_1(\rho,\mu) = \left(\frac{1-\mu^2}{\rho}
  \frac{\partial}{\partial \mu} + \mu \frac{\partial}{\partial \rho}
  \right) D_\rho^2 \Psi_1(\rho,\mu)
\end{equation}
This result is also derived in matched asymptotic analysis, and is
formally identical to the Oseen equation for a sphere (Eqn. \ref
{Oseen:Eqn}). Structurally, this problem is similar to what we have
seen previously, and is solved in two steps \cite{Goldstein29}. First
use the transformation $D_\rho^2 \Psi_1 = e^{\rho\mu/2} \Phi(\rho,\mu)$
to obtain Eqn. \ref{sphere:1stintgov}.\footnote{$D_\rho^2
\Psi_1(\rho,\mu)$ is the vorticity.}
\begin{equation}
\label{sphere:1stintgov}
\left(D_\rho^2 - \frac{1}{4} \right) \Phi(\rho,\mu)=0
\end{equation}
This may be solved to obtain the first integral:
\begin{equation}
\label{oseen:1stintb}
D_\rho^2 \Psi_1(\rho,\mu)= e^{\frac{1}{2}\rho \mu} \sum_{n=1}^\infty \left(
\mathcal{A}_n \left(\frac{\rho}{2}\right)^{\frac{1}{2}}
K_{n+\frac{1}{2}}\left(\frac{\rho}{2}\right)+B_n
\left(\frac{\rho}{2}\right)^{\frac{1}{2}}
I_{n+\frac{1}{2}}\left(\frac{\rho}{2}\right) \right) Q_n(\mu)
\end{equation}

As in the case of the cylinder, the inhomogeneous term on the RHS of
Eqn. \ref{oseen:1stintb} consists of integration constants which
multiply the two modified Bessel functions. We are beset by the same
considerations, which (properly speaking) must be resolved by applying
boundary conditions (Eqn. \ref{sphere:oseenbc}) to the renormalized
solution. Following the same arguments given for the cylinder, we set
the coefficients $B_n=0$, which will later make it possible to satisfy
the boundary conditions at infinity.

Completing the second integration is difficult, but was accomplished by
Goldstein \cite{Goldstein29}. The requisite solution is essentially the second term in
Eqn. \ref{Oseen:Goldstein}:
\begin{equation}
\label{sphere:psi1gold}
\Psi_1^{\textrm{(a)}}(\rho,\theta)= A_1 \rho^2 Q_1(\mu) + \sum_{n=1}^\infty \left(B_n
\rho^{-n} + \sum_{m=0}^\infty X_m \rho^2 \Phi_{m,n}(\rho /2) \right) Q_n(\mu)
\end{equation}
Note that we have omitted the terms $A_n r^n Q_n(\mu)$ which diverge too quickly at
infinity (this was also done for the cylinder).

Alternatively, one may simplify the series in Eqn. \ref{oseen:1stintb},
by retaining only the $n=1$ term (setting all other $A_n = 0$). It is then
possible to complete the second integration with a closed form solution:
\begin{equation}
\label{sphere:oseen1sol}
\Psi_1^{\textrm{(b)}}(\rho,\theta)= A_1 \rho^2 Q_1(\mu) +
\mathcal{A}_1 \left( 1+ \mu \right)\left(1 - e^{-\frac{1}{2}\rho
  \left(1 - \mu\right)}\right) + \sum_{n=1}^\infty B_n\rho^{-n} Q_n(\mu)
\end{equation}
As before, we neglect the $A_n r^n Q_n(\mu)$ solutions. This is
essentially Oseen's solution (Eqn. \ref{Oseen:1sol}), expressed in the
appropriate variables and with undetermined coefficients.

We therefore have two solutions (Eqns. \ref{sphere:psi1gold},
\ref{sphere:oseen1sol}) which can be used for $\Psi_1$. For the moment,
we will consider both. We will later demonstrate that the former is the
preferred choice by considering boundary conditions.

\subsubsection{Secular behavior}

We consider our \Order[R]{1} na\" \i ve solution abstractly:
\begin{equation}
\label{sphere:sphereRsol}
\Psi(\rho,\mu) = -\rho^2 Q_1(\mu) + R \left(A_1 \rho^2
Q_1(\mu) + \sum_{n=1}^\infty B_n\rho^{-n} Q_n(\mu) + \cdots \right) +
\Order[R]{2}
\end{equation}
This generic form encompasses both Eqn. \ref{sphere:oseen1sol} and Eqn.
\ref{sphere:psi1gold}. It also possesses two key similarities with both
the terrible and the cylinder problems. First, there is a term at
\Order[R]{1} which is a multiple of the \Order[R]{0} solution ($A_1
\rho^2 Q_1(\mu)$). Secondly, the secular behavior in our na\" \i ve
solution occurs \emph{at the same order} as the integration constants
which we hope to renormalize.\footnote{These secular terms are not
written explicitly in Eqn. \ref{sphere:sphereRsol}. They can be found
in Eqns. \ref{sphere:oseen1sol} and \ref{sphere:psi1gold}.} This fact
is in essence related to equations like Eqn. \ref{terribleRGans2},
which must be solved iteratively. We avoided that kind of RG equation
by introducing the constant which could have been associated with the
\Order[R]{0} solution at \Order[R]{1}. But renormalizing divergences
into integration constants \emph{at the same order} limits the ability
of RG to ``re-sum'' our na\" \i ve series. In all of these cases, the
real power of RG techniques could be seen by extending our analysis to
\Order[R]{2}.

Because of the similarities between Eqn. \ref{sphere:sphereRsol} and
Eqn. \ref{cylinder:naivesol}, we can tackle this problem in a manner
formally the same as the cylinder. By construction, Eqn.
\ref{sphere:sphereRsol} is \Order[\rho]{2} as $\rho \to \infty$. Hence
the only terms with problematic secular behavior occurs in the limit
$\rho \to 0$. As before, these divergences need not even be explicitly
identified. We write:
\begin{equation}
\label{sphere:beginRG}
\Psi_(\rho,\mu) = -\rho^2 Q_1(\mu) + R \left( A_1 \rho^2
Q_1(\mu) +  \mathcal{R}(\rho,\mu; \{B_i\}; \{X_j\}) +
\mathcal{S}(\rho,\mu;  \{B_m\}; \{X_n\}) \right)
\end{equation}
Here, $\mathcal{S}$ includes the terms which are secular as
$\rho \to 0$, and $\mathcal{R}$ includes regular terms.

\subsubsection{Renormalization}

Eqn. \ref{sphere:beginRG} is only cosmetically different from
Eqn. \ref{cylinder:beginRG}. Renormalizing the two equations can
proceed in \emph{exactly} the same fashion. Therefore, we may
immediately write the renormalized solution:
\begin{equation}
\label{sphere:endRG}
\Psi_(\rho,\mu) = -\rho^2 Q_1(\mu) + R \left( \alpha_1 \rho^2
Q_1(\mu) +  \mathcal{R}(\rho,\theta; \{\beta_i\}; \{\chi_j\}) +
\mathcal{S}(\rho,\theta;  \{\beta_m\}; \{\chi_n\}) \right)
\end{equation}
This is, of course, the same solution from which we began. As in the
previous two problems, we now know that it is a uniformly valid
solution, and turn to the application of the boundary conditions.

\subsubsection{Meeting the boundary conditions}

We have two possible solutions for $\Psi_1(\rho,\mu)$. Considering the
boundary conditions on the surface of the sphere (Eqn.
\ref{sphere:oseenbc}) will demonstrate why Eqn. \ref{sphere:psi1gold}
is preferential. Eqn. \ref{sphere:oseen1sol} can never satisfy the two
requirements for all of the angular harmonics. Expanding the
exponential term, we see that although it has but one integration
constant, it contributes to \emph{all} of the powers of $\mu$.  The
second solution, Eqn. \ref{sphere:psi1gold}, can meet both of the
boundary conditions --- in principle. However, as in the case of the
cylinder, this is practically impossible, and we must consider
truncating our solution.

It is clear that we will need to approximate our solutions in order to
apply the boundary conditions. Our procedure is governed by the
following considerations. First, we demand that our approximate
solution satisfy the boundary conditions as accurately as possible.
This requirement is necessary because our goal is to calculate the drag
coefficient, $C_D$, a calculation which is done by evaluating
quantities derived from the stream function \emph{on the surface of the
sphere}. Hence it is necessary that the stream function be as accurate
as possible in that regime. Secondly, we want the difference between
our modified solution and the exact solution (one which satisfies the
governing equations) to be as small as possible.

{\it a. Oseen's solution}

First, consider trying to satisfy these requirements starting from Eqn.
\ref{sphere:oseen1sol}. Although this is the less general solution to
Oseen's equation, we consider Oseen's solution because of (1) its
historical importance, including widespread use as a starting point for
matched asymptotics work and (2) the appealling simplicity of a
closed-form solution.

We combine Eqns. \ref{sphere:oseen1sol} and \ref{sphere:0sol} to begin
from the solution: $\Psi(\rho,\mu) = \Psi_0(\rho,\mu) + R
\Psi_1^{\textrm{(b)}}(\rho,\mu)$. Since we are interested in the
solution near the surface of the sphere ($\rho = R$), and because there
is no other way to determine the integration constants, we expand the
exponential in that vicinity. Retaining terms up to $\Order[R\rho]{1}
\sim \Order[\rho]{2}$, we obtain:
\begin{equation}
\label{sphere:approxossensol}
\Psi(\rho,\mu) =  \left( - \rho^2 + R \left( A_1 \rho^2 -
 \mathcal{A}_1 \rho \right) \right) Q_1(\mu) + R \sum_{n=1}^\infty B_n\rho^{-n} Q_n(\mu)
\end{equation}
The boundary conditions are satisfied if $B_n = 0$ $\forall n > 1$,
$A_1=0$, $\mathcal{A}_1 = -3/2$, and $B_1 = -R^2/2$. In passing, we
note that substituting these values into Eqn. \ref{sphere:oseen1sol}
reproduces Oseen's original solution \cite{Oseen10}. Continuing, we
substitute these values into Eqn. \ref{sphere:approxossensol},
obtaining:
\begin{equation}
\label{sphere:truncoseen}
\Psi(\rho,\mu) =  \left( - \rho^2 + \frac{3 R \rho}{2} -
\frac{R^3}{2 \rho} \right) Q_1(\mu)
\end{equation}

This is nothing more than Stokes' solution (Eqn. \ref{stokessol}), albeit expressed in Oseen
variables. Consequently, when substituted into Eqns. \ref{spherestreamfunction}, \ref{CDsphere}, and
\ref{sphpressure} Eqn. \ref{sphere:truncoseen} reproduces $C_D = 6 \pi/R$.

How accurate is our approximate solution? The difference between
Eqn. \ref{sphere:truncoseen} and Eqn. \ref{sphere:oseen1sol} is given
by:
\begin{equation}
\Delta \Psi = -\frac{3}{4} R \left(1+\mu\right) \left( -2 + 2
 e^{-\frac{1}{2}\rho(1-\mu)} + \rho \left(1-\mu\right) \right)
\end{equation}
At the surface of the sphere ($\rho = R$), this equates to an
\Order[R]{3} error in the stream function, and an \Order[R]{2} error in
the derivative. That is entirely acceptable. However, at large $\rho$,
$\Delta \Psi$ grows unbounded, being of \Order[\rho]{1}. This is the
fundamental problem with the solution given by Eqn.
\ref{sphere:truncoseen}. By beginning from Eqn. \ref{sphere:psi1gold},
we can avoid this difficulty.

It is at first a little disconcerting that Oseen used his solution to
obtain the next approximation to $C_D$ (Eqn. \ref{Oseen:dragsphere})
\cite{Ose13}. How can our results be worse? As noted previously,
 ``Strictly, Oseen's method gives only the leading term ... and is scarcely
to be counted as superior to Stokes' method for the purpose of obtaining
the drag.''\cite{Proudman57}

{\it b. Goldstein's solution}

We now apply the boundary conditions to
Eqn. \ref{sphere:psi1gold}. By starting from the more general solution to
Oseen's equation, we can remedy the difficulties encountered above. This
analysis will be very similar to the truncation performed on Tomotika's
solution for the cylinder problem.

We combine Eqns. \ref{sphere:psi1gold} and \ref{sphere:0sol} to begin
from the solution: $\Psi(\rho,\mu) = \Psi_0(\rho,\mu) + R
\Psi_1^{\textrm{(a)}}(\rho,\mu)$. As with the cylinder, we will
approximate the full solution by truncating the series in both $m$ and
$n$. Our first consideration is again symmetry: The uniform flow
imposes a $\sin{\theta}$, or $Q_1(\mu)$ symmetry on the problem. Hence
we must retain the $n=1$ term in Eqn. \ref{sphere:psi1gold}. The
importance of this term is clearly seen from
Eqn. \ref{sphere:simpledrag}: \emph{Only} the coefficient of
$Q_1(\mu)$ is needed to calculate the drag if the stream function
satisfies the boundary conditions.

As in the case of the cylinder, if we retain $n$ harmonics, we must
retain $m=n-1$ terms in the second sum (the sum over $m$) in order to
meet both boundary conditions. To minimize the error introduced by our
approximations we set all \emph{other} $B_n$, $X_m$ equal to zero. The
remaining terms, those which would violate the boundary conditions and
must be truncated, are then given by Eqn. \ref{sphere:discards}.
\begin{equation}
\label{sphere:discards}
\Psi_{\textrm{discard}}^{(n)}(\rho,\mu) = R \left(\sum_{k=n+1}^\infty
\sum_{m=0}^{n-1} X_m \Phi_{m,k}(\rho/2) \rho^2 Q_k(\mu) \right)
\end{equation}

We want to estimate the magnitude of the error in our approximation,
both overall and at the surface (the error in the boundary conditions).
The error is given by Eqn. \ref{sphere:discards}. First, we calculate
the magnitude of both $\Psi_{\textrm{discard}}^{(n)}(\rho,\mu)$ and its
derivative at the surface ($\rho = R$) with $n$ retained harmonics. The
results are given in Table \ref{tab:spherediscards}.
\begin{table}
\begin{center}
    \begin{tabular}{|c|cccc|} \hline
   n = & 1 & 2 & 3 & 4 \\
   \hline
   $\Psi_{\textrm{discard}}^{(n)}(\rho,\mu)$ & \Order[R]{3} & \Order[R]{2} & \Order[R]{1} & \Order[R]{0} \\
   $\Psi_{\textrm{discard}}^{'(n)}(\rho,\mu)$ &
   \Order[R]{2} & \Order[R]{1} & \Order[R]{1} & \Order[R]{-1} \\ \hline
  \end{tabular}
\caption{Importance of discarded terms at $\rho = R$.}
  \label{tab:spherediscards}
\end{center}
\end{table}

From Table \ref{tab:spherediscards}, we see that to retain the $Q_2(\mu)$
harmonics, we must have an error in our derivative boundary condition
of \Order[R]{1} --- the order to which we are trying to work. If we
retain higher harmonics, this situation gets worse.

Since it is in practice impossible to fit the boundary conditions to
all harmonics, we must truncate the series
expansion. We see that there is only one truncation consistent
with both the symmetry requirements of the problem and the demand that
we satisfy the boundary conditions to \Order[R]{1}:
\begin{equation}
\label{sphere:truncsol}
\Psi(\rho,\mu) = -\rho^2 Q_1(\mu) + R \left( A_1 \rho^2 + B_1
\rho^{-1} + X_0 \Phi_{0,1}(\rho/2) \rho^2 \right) Q_1(\mu) + \Order[R]{2}
\end{equation}
We also must consider the overall error, e.g., how big can
$\Psi_{\textrm{discard}}^{(1)}(\rho,\mu)$ get? Although, at the
surface of the sphere, Eqn. \ref{sphere:truncsol} is no better than
Eqn. \ref{sphere:truncoseen}, it is superior for $\rho \ne R$. The
magnitude of the error is maximized as $\rho \to \infty$. It can be
shown by Taylor expansion (separately accounting for $m=0,m=n$, etc.) that
$\Phi_{m,n}(x \to \infty) \sim x^{-2}$. Therefore,
\begin{displaymath}
\lim_{\rho \to \infty} \Psi_{\textrm{discard}}^{(1)}(\rho,\mu) =
\Order[R]{1}
\end{displaymath}
Although this is somewhat unsatisfactory, this solution does not suffer from the
same shortcomings as Eqn. \ref{sphere:approxossensol}. The error
remains bounded.

Eqn. \ref{sphere:truncsol} will satisfy the boundary conditions
(Eqn. \ref{sphere:oseenbc}) if $A_1 = 0$ and
\begin{eqnarray}
X_0 &=& \frac{6}{6 R \Phi_{0,1}(R/2) + R^2 \Phi_{0,1}^{'}(R/2)}\\
B_1 &=& \frac{R^3 \Phi_{0,1}^{'}(R/2)}{6 \Phi_{0,1}(R/2) + R \Phi_{0,1}^{'}(R/2)}
\end{eqnarray}
As in the case of the cylinder, the resulting stream function
satisfies the boundary conditions exactly, and the governing equations
approximately. Our final solution is:
\begin{eqnarray}
\label{sphere:goldfinal}
\Psi(\rho,\mu) &=& -\rho^2 Q_1(\mu) + R \Bigg( \frac{R^3 \Phi_{0,1}^{'}(R/2)}{6 \Phi_{0,1}(R/2) + R \Phi_{0,1}^{'}(R/2)}
\rho^{-1} + \\ & & \frac{R^3 \Phi_{0,1}^{'}(R/2)}{6 \Phi_{0,1}(R/2) + R \Phi_{0,1}^{'}(R/2)} \Phi_{0,1}(\rho/2) \rho^2 \Bigg) Q_1(\mu) + \Order[R]{2} \nonumber
\end{eqnarray}
For reference,
\begin{displaymath}
\Phi_{0,1}(x) = -\frac{3 \pi}{4 x^2} \left(2 -
\frac{2}{x}+\frac{1}{x^2}-\frac{e^{-2x}}{x^2}\right)
\end{displaymath}

\subsubsection{Calculating the drag coefficient}

We calculated the drag coefficient by substituting Eqn.
\ref{sphere:goldfinal} into Eqn. \ref{sphere:simpledrag}, giving this
new result:
\begin{equation}
\label{sphere:cdphi}
C_D = \frac{ \pi \left( -16 \Phi_{0,1}^{'}(R/2) + R \left( 8
  \Phi_{0,1}^{''}(R/2) + R \Phi_{0,1}^{'''}(R/2) \right) \right)}{2
  \left(6 \Phi_{0,1}(R/2) + R \Phi_{0,1}^{'}(R/2) \right)}
\end{equation}
This can be expressed in terms of more conventional functions by
substituting for $\Phi_{0,1}(x)$, resulting in the drag coefficient
given by Eqn. \ref{sphere:cdresult}.
\begin{equation}
\label{sphere:cdresult}
C_D = \frac{4 \pi \left(24 + 24 R + 8 R^2 + R^3 + 4 e^R \left(R^2 - 6
  \right)\right)}{R \left( 2 \left(R + 1\right) + e^R \left(R^2 -2
  \right)\right)}
\end{equation}

\begin{figure}[tb]
\psfrag{Re}{\mbox{\Large $R$}}
\psfrag{CD}{\mbox{\Large $C_D R/6\pi - 1$}}
\psfrag{CD2 XXXXXX}{\tiny{$C_D \frac{R}{6\pi} - 1$}}
\psfrag{Maxworthy XXXXXX}{Maxworthy}
\psfrag{Le Clair}{Le Clair}
\psfrag{Dennis}{Dennis}
\psfrag{Oseen}{Oseen, Eqn. \ref{matchedcd}}
\psfrag{Proudman}{Proudman, Eqn. \ref{matchedcd}}
\psfrag{Chester and Breach 2}{Chester, Eqn. \ref{matchedcd}}
\psfrag{Goldstein}{Goldstein, Eqn. \ref{GoldsteinCD}}
\psfrag{RG}{RG, Eqn. \ref{sphere:cdresult}}
\begin{center}
\includegraphics[width=.9 \textwidth]{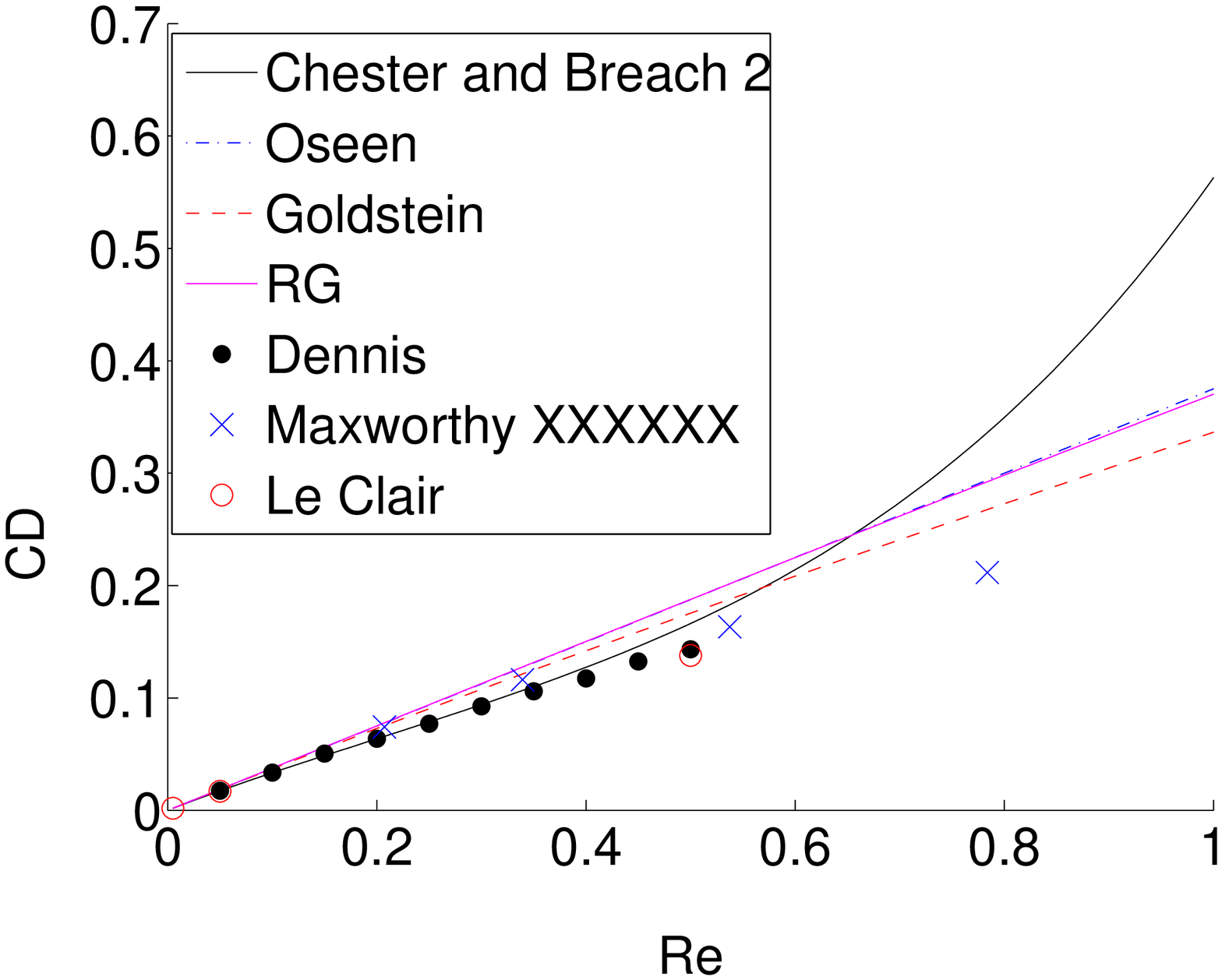}
\caption{(Color online) Drag on a sphere, comparing RG to other theories \cite{Maxworthy65,LC70,Den71}.}
\label{fig:sphere:RG}
\end{center}
\end{figure}

This result is plotted in Figure \ref{fig:sphere:RG}, where it is
compared against the principal results of Oseen theory, matched
asymptotic theory, numerical results, and experiments. As $R
\rightarrow 0$, there is excellent agreement. At small but
non-infinitesimal Reynolds numbers, RG is nearly identical to Oseen's
prediction (Eqn. \ref{Oseen:dragsphere}), which is disappointing. It is
surprising that Goldstein's result is better than the RG result, as
they are calculations of the same order in $R$, and are a series
approximation. That the matched asymptotics predictions are superior is
not surprising; Chester and Breach's result began with a much higher
order perturbative approximation. If a higher order RG calculation were
possible, RG ought to be better than the same order matched asymptotics
prediction.

\begin{figure}[tb]
\psfrag{Re}{\mbox{\Large $R$}}
\psfrag{CD}{\mbox{\Large $C_D R/6\pi - 1$}}
\psfrag{CD2 XXXXXX}{\tiny{$C_D \frac{R}{6\pi} - 1$}}
\psfrag{Maxworthy XXXXXX}{Maxworthy}
\psfrag{Le Clair}{Le Clair}
\psfrag{Dennis}{Dennis}
\psfrag{Oseen}{Oseen, Eqn. \ref{matchedcd}}
\psfrag{Proudman}{Proudman, Eqn. \ref{matchedcd}}
\psfrag{Chester and Breach 2}{Chester, Eqn. \ref{matchedcd}}
\psfrag{Goldstein}{Goldstein, Eqn. \ref{GoldsteinCD}}
\psfrag{RG}{RG, Eqn. \ref{sphere:cdresult}}
\begin{center}
\includegraphics[width=.9 \textwidth]{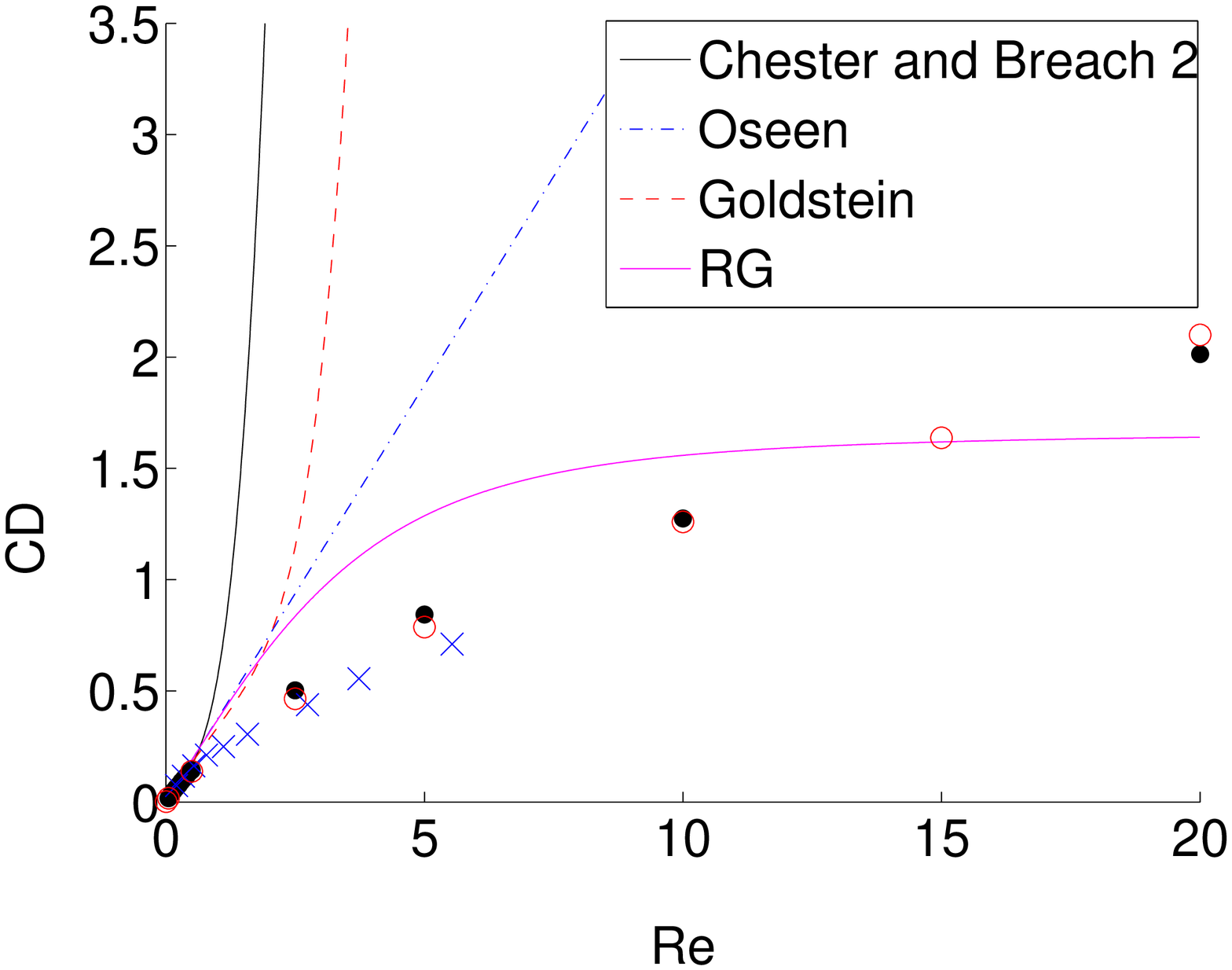}
\caption{(Color online) Drag on a sphere, comparing RG at larger $R$ \cite{Maxworthy65,LC70,Den71}.}
\label{fig:sphere:RG2}
\end{center}
\end{figure}

As in the case of the cylinder, the real strength of Eqn.
\ref{sphere:cdresult} can be seen as the Reynolds number increases.
Figure \ref{fig:sphere:RG2} demonstrates that all other theories
diverge from experimental measurements for $R\gtrsim1$. This is an
unavoidable aspect of their structure and derivation --- they are only
valid asymptotically. The RG prediction suffers from none of these
problems. Eqn. \ref{sphere:cdresult} is well behaved for all $R$,
although it does become less accurate at larger Reynolds numbers.

\section{CONCLUSIONS}

We have devoted a substantial effort to the historical problem of
calculating the drag coefficient for flow around a cylinder and a
sphere at low Reynolds number. We report four principal
accomplishments. First, we have untangled over 150 years of diffuse,
confusing, and sometimes contradictory experimental, numerical, and
theoretical results. We have expressed all important previous work
within a consistent mathematical framework, and explained the
approximations and assumptions which have gone into previous
calculations. Moreover, by plotting experimental results and
theoretical predictions with the leading order divergence removed (an
idea originally due to Maxworthy), we have consistently and critically
compared all available measurements. There are no other such exhaustive
comparative reviews available in the existing literature.

Secondly, we have extended traditional matched asymptotics
calculations. We advance and justify the idea that \emph{uniformly
valid approximations}, not the Stokes or Oseen
expansions, should be used to calculate
derivative quantities such as $C_D$. By combining this approach with
previously published matched asymptotics results, we obtain new results
for the drag coefficients. These results \emph{systematically} improve
on published drag coefficients, which relied only on the Stokes
expansion. This methodology also resolved a problem in the existing
literature: the most accurate calculations for a cylinder, due to
Skinner, had failed to improve $C_D$ \cite{Ski75}. When treated via a
uniformly valid approximation, our new result based on Skinner's
solutions betters all matched asymptotics predictions.

We have also explored the structure and subtleties involved in applying
renormalization group techniques to the ``terrible'' problem posed by
Hinch and Lagerstrom \cite{Hinch91,Lag72}. This problem, previously
solved by Chen et al. \cite{CGO96}, contains a rich and henceforth
unexplored collection of hidden subtleties. We exhaustively examined
all possible complications which can arise while solving this problem
with the renormalization group. To treat some of these possibilities,
we identified and implemented a new constraint on the RG calculation;
the renormalized perturbation solution itself, not just the expansion
on which it is based, must satisfy the governing equations to the
appropriate order in $\epsilon$. While this had been done implicitly in
previous calculations, we had to deal with it explicitly (e.g., by
appropriate choices of homogeneous solutions). In the process of doing
so, we obtained several new second order approximate solutions to the
``terrible'' problem, and demonstrated their equivalence.

The work with the ``terrible'' problem laid the foundation for our most
significant new calculation. In close analogy with the ``terrible''
problem, we used the RG to derive new results for the drag coefficients
for both a sphere and a cylinder (Eqns. \ref{sphere:cdresult} and
\ref{cylinder:O1CD}, respectively). These new results agree
asymptotically with previous theoretical predictions, but greatly
surpass them at larger $R$. Other theories diverge pathologically,
while the results from the RG calculation remain well behaved.

We demonstrated that these new techniques could reproduce and improve
upon the results of matched asymptotics --- when applied to the very
problem which that discipline was created to solve! Matched asymptotics
requires the use of two ingenious and intricate expansions, replete
with strange terms (like $R \log{R}$) which must be introduced while
solving the problem via a painful iterative process. RG requires only a
single generic expansion, which can always be written down a priori,
even in complicated singular perturbation problems with boundary
layers. It therefore gives rise to a much more economical solution,
requiring half the work and yielding a superior result. It is hoped
that demonstrating of the utility of these techniques on this
historical problem will result in increased interest and further
application of renormalization group techniques in fluid mechanics.

\section*{ACKNOWLEDGMENTS}
The authors are grateful, in some sense, to Charlie Doering for his
initial suggestion to consider the tortuous problems discussed here
using RG techniques.  This work was partially supported by the National
Science Foundation through Grant No. NSF-DMR-99-70690.

\bibliography{lowRbib}

\end{document}